\documentclass[iop,apj]{emulateapj}
\usepackage{amsmath,amssymb,color,verbatim}
\newcommand{\ie}{i.e.\/}
\newcommand{\eg}{e.g.\/}
\newcommand{\etc}{etc.\/}
\newcommand{\kms}{km\,s$^{-1}$}

\shorttitle{How to Calculate Molecular Column Density}
\shortauthors{Mangum \& Shirley}

\begin{document}
\title{How to Calculate Molecular Column Density}

\author{Jeffrey G.~Mangum}
\affil{National Radio Astronomy Observatory, 520 Edgemont Road,
  Charlottesville, VA  22903, USA}
\email{jmangum@nrao.edu}

\and

\author{Yancy L.~Shirley}
\affil{Steward Observatory, University of Arizona, 933 North Cherry
  Avenue, Tucson, AZ 85721, USA}
\email{yshirley@as.arizona.edu}

\begin{abstract}
The calculation of the molecular column density from molecular
spectral (rotational or ro-vibrational) transition measurements is one
of the most basic quantities 
derived from molecular spectroscopy.  Starting from first principles
where we describe the basic physics behind the radiative and
collisional excitation of molecules and the radiative transfer of
their emission, we derive a general expression for the molecular
column density.  As the calculation of the molecular column density
involves a knowledge of the molecular energy level degeneracies,
rotational partition functions, dipole moment matrix elements, and
line strengths, we include generalized derivations of these
molecule-specific quantities.  Given that approximations to the column
density equation are often useful, we explore the optically thin,
optically thick, and low-frequency limits to our derived general
molecular column density relation.  We also evaluate the limitations
of the common assumption that the molecular excitation temperature is
constant, and address the distinction between beam- and
source-averaged column densities.  As non-LTE approaches to the
calculation of molecular spectral line column density have become
quite common, we summarize non-LTE models which calculate molecular
cloud volume densities, kinetic temperatures, and molecular column
densities.  We conclude our discussion of the molecular
column density with worked examples for C$^{18}$O, C$^{17}$O,
N$_2$H$^+$, NH$_3$, and H$_2$CO.  Ancillary information on some
subtleties involving line profile functions, conversion between
integrated flux and brightness temperature, the calculation of the
uncertainty associated with an integrated intensity measurement, the
calculation of spectral line optical depth using hyperfine or
isotopologue measurements, the calculation of the kinetic temperature
from a symmetric molecule excitation temperature measurement, and
relative hyperfine intensity calculations for NH$_3$ are presented in
appendices.  The intent of this document is to provide a reference for
researchers studying astrophysical molecular spectroscopic
measurements.
\end{abstract}

\keywords{ISM: molecules}


\section{Introduction}
\label{Intro}

Molecular spectral line column densities are
fundamental to the study of the physical and chemical conditions
within dense molecular clouds.  Chemical composition, molecular
isomer and isotopomer ratios, molecular hydrogen column densities,
and molecular volume densities are all based on a measure of the
molecular column density.  Past reviews and tutorials on this
subject, including \cite{Genzel1991}, \cite{Evans1991},
\cite{Evans1999}, and \cite{Goldsmith1999}, along with spectroscopic
textbooks such as \cite{Townes1975}, \cite{Gordy1984}, and \cite{Draine2011} have
provided an excellent basis upon which one can derive the molecular
spectral line column density for a variety of dense molecular cloud
environments.  The intent of this tutorial is to provide students
and scientists a general single reference point for the 
calculation of the molecular spectral line column density.

We have organized the document such that some basic background information is
first provided to allow a contextural foundation to be lain for
further calculations.  This foundation includes a basic understanding
of radiative transfer and statistical equilibrium
(Sections~\ref{RadCol} and \ref{RadTrans}), molecular degeneracy
(Sections~\ref{Degeneracies} and \ref{Sym}), partition functions
(Section~\ref{Qrot}), line strength (Sections~\ref{LineStrength} and
\ref{SymtopsandLin}), and hyperfine structure
(Section~\ref{Hyperfine}).  The derivation of the most general form of
the molecular spectral line column density is presented in
Section~\ref{Colden}.  We also explore a number of commonly-used
practical approximations to the calculation of the molecular spectral line
column density in Section~\ref{ColDenApprox}.  As the 
calculation of the molecular spectral line column density normally
assumes constant excitation temperature, we address the consequences
of this assumption when non-LTE conditions are present in
Section~\ref{Tex}.  We also summarize non-LTE approaches to the
calculation of molecular cloud volume density, kinetic temperature,
and molecular column density in Section~\ref{NonLTE}. 

Molecular cloud structure and its relationship to the spatial
resolution of a measurement system affects the derivation of
molecular spectral line column density.  To quantify these effects
we address the distinction between beam- and source-averaged column
density in Section~\ref{BeamSourceAverage}.

As worked examples are often illustrative for many astrophysical
calculations, we provide step-by-step derivations of the molecular
column density for several representative molecules; C$^{18}$O,
C$^{17}$O, N$_2$H$^+$, NH$_3$, and H$_2$CO in Section~\ref{Examples}.
Appendices provide 
ancillary information on some subtleties often encountered when
calculating molecular spectral line column density, including line
profile functions (Appendix~\ref{Lineprofile}), conversion between
integrated flux and brightness temperature (Appendix~\ref{Intflux}),
the calculation of the uncertainty associated with an integrated
intensity measurement (Appendix~\ref{IntUncert}), the calculation of
spectral line optical depth using hyperfine or isotopologue
measurements (Appendix~\ref{HyperTau}), the calculation of the kinetic
temperature from a symmetric molecule excitation temperature
measurement (Appendix~\ref{AmonTexTk}), and relative hyperfine intensity 
calculations for NH$_3$ (Appendix~\ref{amontabs}).

\section{Radiative and Collisional Excitation of Molecules}
\label{RadCol}

When the energy levels of a molecule are in statistical equilibrium,
the rate of transitions populating a given energy level is balanced by
the rate of transitions which depopulate that energy level.  For a
molecule with multiple energy levels statistical equilibrium can be
written as:
\begin{equation}
n_i\sum_j{R_{ij}} = \sum_j{n_j R_{ji}},
\label{eq:stateq}
\end{equation}
where $n_i$ and $n_j$ are the populations of the energy
levels $i$ and $j$ and $R_{ij}$ and $R_{ji}$ are transition rates
between levels $i$ and $j$.  The transition rates contain
contributions from:

\begin{itemize}
\item Spontaneous radiative excitation ($A_{ij}$)
\item Stimulated radiative excitation and de-excitation ($R_{ij}
  \equiv n_iB_{ij}\int_0^\infty J_\nu \phi_{ij}(\nu) d\nu$)
\item Collisional excitation and de-excitation ($n_{collider}C_{ij}$)
\end{itemize}

\noindent{where} $\phi_\nu(\nu)$ is the line profile function and $J_\nu$
is defined as the integral of the specific intensity $I_\nu$ over the
source of emission: 
\begin{equation}
J_\nu \equiv \frac{1}{4\pi}\int I_\nu d\Omega.
\label{eq:jnu}
\end{equation}
Our statistical equilibrium equation then becomes:
\begin{multline}
n_i\Biggl[\sum_j\left(n_{collider}C_{ij} + B_{ij}\int_0^\infty J_\nu
\phi_{ij}(\nu) d\nu\right) + \sum_{j<i}A_{ij}\Biggr] \\
= \sum_j n_j\left(n_{collider}C_{ij} + B_{ji}\int_0^\infty J_\nu \phi_{ji}(\nu)
d\nu\right) \\
+ \sum_{j>i}n_jA_{ji}.
\label{eq:stateqmulti}
\end{multline}
For a two-level system with $i$ defined as the lower energy
level $l$ and $j$ defined as the upper energy level $u$,
$\sum_{j<i}A_{ij} = 0$ and the statistical equilibrium equation
(Equation~\ref{eq:stateqmulti}) becomes:
\begin{multline}
n_l\left(n_{collider}C_{lu} + B_{lu}\int_0^\infty J_\nu
\phi_{lu}(\nu) d\nu\right) = \\ n_u\left(n_{collider}C_{ul} +
B_{ul}\int_0^\infty J_\nu \phi_{ul}(\nu) d\nu + A_{ul}\right).
\label{eq:stateqtwolev}
\end{multline}
As hydrogen ($H$) is the most abundant atom in the interstellar
medium, and in dense molecular clouds most of the $H$ is in molecular
form ($H_2$), $n_{collider}$ is usually assumed to be $n(H_2)$.

At this point we can derive the Einstein relations $A_{ul}$, $B_{ul}$,
and $B_{lu}$ by considering only radiative excitation ($C_{lu} = C_{ul} = 0$)
and complete redistribution over the line profile ($\phi_{ul}(\nu) =
\phi_{lu}(\nu)$).  Physically, this means that emitted and absorbed photons 
are completely independent.  Equation~\ref{eq:stateqtwolev} then becomes:
\begin{eqnarray}
n_lB_{lu}\int_0^\infty J_\nu \phi_{lu}(\nu) &=&
n_uB_{ul}\int_0^\infty J_\nu \phi_{ul}(\nu) + n_uA_{ul}
\nonumber \\
\int_0^\infty \left[n_lB_{lu}J_\nu\phi_{lu}(\nu)\right] d\nu &=&
\int_0^\infty \left[n_uB_{ul}J_\nu\phi_{lu}(\nu) + n_uA_{ul}\right] d\nu
\nonumber \\
n_lB_{lu}J_\nu\phi_{lu} &=& n_uB_{ul}J_\nu\phi_{lu} + n_uA_{ul}.
\label{eq:einstein}
\end{eqnarray}

For a system in thermal equilibrium, the relative level populations
follow the Boltzmann distribution:
\begin{equation}
\frac{n_u}{n_l} \equiv \frac{g_u}{g_l}\exp\left(-\frac{h\nu}{kT}\right),
\label{eq:boltzmann}
\end{equation}
and the radiation field $J_\nu$ is described by the Planck
Function $B_\nu(T)$:
\begin{equation}
B_\nu(T) \equiv
  \frac{2h\nu^3}{c^2}\left[\exp\left(\frac{h\nu}{kT}\right) -
  1\right]^{-1}.
\label{eq:planckdef}
\end{equation}
Substituting Equations~\ref{eq:boltzmann} and
\ref{eq:planckdef} into Equation~\ref{eq:einstein} (which eliminates
the line profile function $\phi_{lu}$) yields, after some
rearrangement:
\begin{equation}
\left(\frac{c^2}{2h\nu^3}A_{ul} -
\frac{g_l}{g_u}B_{lu}\right)\left[\exp\left(\frac{h\nu}{kT}\right) -
1\right] = \frac{g_l}{g_u}B_{lu} + B_{ul},
\label{eq:planckdef2}
\end{equation}
which implies that:
\begin{eqnarray}
g_lB_{lu} &=& g_uB_{ul} \\
A_{ul} &=& \frac{2h\nu^3}{c^2} B_{ul}.
\label{eq:planckdef3}
\end{eqnarray}
For dipole emission, the spontaneous emission coefficient $A_{ul}$ can
be written in terms of the dipole matrix element $|\mu_{lu}|^2$ as:
\begin{equation}
A_{ul} \equiv \frac{64 \pi^4 \nu^3}{3hc^3}|{\mathbf \mu_{lu}}|^2.
\label{eq:adef}
\end{equation}

\section{Radiative Transfer}
\label{RadTrans}

The radiative transfer equation, ignoring scattering processes (such
as those involving electrons or dust) is defined as follows (see
\cite{Spitzer1978} or \cite{Draine2011} for details):
\begin{equation}
\frac{dI_\nu}{ds} = -\kappa_\nu I_\nu + j_\nu
\label{eq:radtransdef}
\end{equation}
where
\begin{align}
s &\equiv \textrm{Path of propagation along the line of sight} \nonumber \\
I_\nu &\equiv \textrm{Specific Intensity} \nonumber \\
\kappa_\nu &\equiv \textrm{Absorption Coefficient} \nonumber \\
           &= \frac{h\nu}{4\pi}\left(n_lB_{lu} - n_uB_{ul}\right)\phi_\nu \\
           &= \frac{c^2}{8\pi\nu^2}\frac{g_u}{g_l}n_lA_{ul}\left(1 -
\frac{g_ln_u}{g_un_l}\right)\phi_\nu \label{eq:kappa}\\
j_\nu &\equiv \textrm{Emission Coefficient} \nonumber \\
           &= \frac{h\nu}{4\pi} A_{ul} n_u \label{eq:j},
\end{align}

Since we generally do not know what the propagation path is for our
measured radiation it is convenient to change independent variables
from pathlength $s$ to ``optical depth'' $\tau_\nu$, which is defined
as:
\begin{equation}
d\tau_\nu \equiv \kappa_\nu ds,
\label{eq:taudef}
\end{equation}
where we use the convention adopted by
\cite{Draine2011}\footnote{\cite{Draine2011} points out that 
  \cite{Spitzer1978} uses the opposite convention, that radiation
  propagates in the direction of \textit{decreasing} optical depth.}
that the radiation propagates in the direction of \textit{increasing}
optical depth.  Switching variables from $s$ to $\tau_\nu$ in our
radiative transfer equation (Equation~\ref{eq:radtransdef}) results in
the following radiative transfer equation:
\begin{equation}
dI_\nu = S_\nu d\tau_\nu - I_\nu d\tau_\nu,
\label{eq:radtransdeftau}
\end{equation}
where we define the \textsl{Source Function} $S_\nu$:
\begin{equation}
S_\nu \equiv \frac{j_\nu}{\kappa_\nu}.
\label{eq:sourcefunc}
\end{equation}

By multiplying both sides of Equation~\ref{eq:radtransdeftau} by the
``integrating factor'' $e^{\tau_\nu}$, we can integrate the radiative
transfer equation from a starting point where $\tau_\nu = 0$ and
$I_\nu = I_\nu(0)$ to find that:
\begin{eqnarray}
&&e^{\tau_\nu} \left(dI_\nu + I_\nu d\tau_\nu\right) = e^{\tau_\nu}
S_\nu d\tau_\nu \nonumber \\
&&e^{\tau_\nu}I_\nu - I_\nu(0) = \int_0^{\tau^{\prime}} S_\nu d\tau^\prime \nonumber \\
&&I_\nu = I_\nu(0)e^{-\tau_\nu} + \int_0^{\tau^{\prime}}
\exp\left[-\left(\tau_\nu - \tau^\prime\right)\right] S_\nu d\tau^\prime.
\label{eq:radtransdefintegral}
\end{eqnarray}

Equation~\ref{eq:radtransdefintegral} is a completely general solution
to the equation of radiative transfer (again, assuming that scattering
is neglected).  It defines the intensity measured by the observer
($I_\nu$) as the sum of the background intensity ($I_\nu(0)$)
attenuated by the interstellar medium ($\exp\left(-\tau_\nu\right)$)
plus the integrated emission ($S_\nu d\tau^\prime$) attenuated by the
effective absorption due to the interstellar medium between the point
of emission and the observer \{$\exp\left[-\left(\tau_\nu -
  \tau^\prime\right)\right]$\}.

For an infinitely large medium the radiation field would be defined as
blackbody: $I_\nu = B_\nu$.  We further assume that the medium
through which the radiation is traveling is uniform at an excitation
temperature $T_{ex}$, defined as:
\begin{equation}
\label{eq:tex}
T_{ex} = \frac{h\nu / k}{\ln \left( \frac{n_l \, g_u}{n_u \, g_l} \right) },
\end{equation}
where $n$ is the density (cm$^{-3}$) in the upper ($u$) or lower ($l$)
energy level for a transition with frequency $\nu$.  The source
function $S_\nu$ is then equivalent to the Planck Function at
temperature $T_{ex}$ (Equation~\ref{eq:planckdef}): $S_\nu =
B_\nu(T_{ex})$.  This condition is sometimes referred to as ``local
thermodynamic equilibrium'' (LTE). 
Equation~\ref{eq:radtransdefintegral} becomes:
\begin{equation}
I_\nu = I_\nu(0)e^{-\tau_\nu} + \int_0^{\tau^{\prime}}
\exp\left[-\left(\tau_\nu - \tau^\prime\right)\right] B_\nu(T_{ex}) d\tau^\prime.
\label{eq:radtransdefintegraluniform}
\end{equation}
If we further assume that $T_{ex}$ is a constant,
Equation~\ref{eq:radtransdefintegraluniform} becomes:
\begin{equation}
I_\nu = I_\nu(0)\exp(-\tau_\nu) + B_\nu(T_{ex})\left[1-\exp(-\tau_\nu)\right].
\label{eq:radtransdef3}
\end{equation}
Molecular spectral line measurements involve differencing the measured
intensity toward a reference position which contains only background
intensity from that measured towards a molecular line source position:
\begin{eqnarray}
\Delta I_\nu &\equiv& I_\nu - I_\nu(0) \nonumber \\
&=& I_\nu(0)\exp(-\tau_\nu) +
B_\nu(T_{ex})\left[1-\exp(-\tau_\nu)\right] - I_\nu(0) \nonumber \\
&=& \left[B_\nu(T_{ex}) - B_\nu(T_{bg})\right]\left[1-\exp(-\tau_\nu)\right].
\label{eq:radtransdef4}
\end{eqnarray}

In many cases the specific intensity I$_\nu$ is replaced by the
\textsl{Rayleigh-Jeans Equivalent Temperature}, which is the
equivalent temperature of a black body at temperature T:
\begin{equation}
J_\nu(T) \equiv \frac{\frac{h\nu}{k}}{\exp\left(\frac{h\nu}{kT}\right)-1}.
\label{eq:rjtdef}
\end{equation}
If we further define the \textsl{Radiation Temperature
  $T_R$} as follows:
\begin{equation}
T_R \equiv \frac{c^2}{2k\nu^2} \Delta I_\nu,
\label{eq:trdef}
\end{equation}
we can relate $B_\nu(T)$ and $J_\nu(T)$ as follows:
\begin{equation}
\frac{c^2}{2k\nu^2} B_\nu(T) = J_\nu(T).
\label{eq:bnujnu}
\end{equation}
We can write our radiative transfer equation in a form
which involves the observable \textsl{Source Radiation Temperature 
  T$_R$} derived from a differencing measurement:
\begin{equation}
T_R = f\left[J_\nu(T_{ex}) - J_\nu(T_{bg})\right]\left[1 - \exp(-\tau_\nu)\right],
\label{eq:tr}
\end{equation}
where we have introduced an extra factor $f$ which is the
fraction of the spatial resolution of the measurement filled by the
source (sometimes called the ``filling factor'').  See
\cite{Ulich1976} for a complete derivation of the source radiation
temperature for the case of a single antenna position switched
measurement.

\section{Column Density}
\label{Colden}

In order to derive physical conditions in the interstellar medium it
is often useful to measure the number of molecules per unit area along
the line of sight.  This quantity, called the ``column density'', is
basically a first-step to deriving basic physical quantities such as
spatial density, molecular abundance, and kinetic temperature.
Using the two-energy level system defined above, we can express the
column density as the number of molecules in energy level $u$
integrated over the pathlength $ds$: 
\begin{equation}
N_u \equiv \int n_u ds.
\label{eq:nudef}
\end{equation}

Since we want to use our molecular spectral line measurements to
calculate the molecular column density, which will ultimately involve
the radiative transfer properties of the molecular spectral line
measured, we can use the definition of the optical depth
(Equation~\ref{eq:taudef}), the definition of the absorption
coefficient $\kappa$ (Equation~\ref{eq:kappa}), the Boltzmann
equation for statistical equilibrium (Equation~\ref{eq:boltzmann}), the
definition of the spontaneous emission coefficient $A_{ul}$
(Equation~\ref{eq:adef}), and our definition of the column density
(Equation~\ref{eq:nudef}) to relate $\tau_\nu$ to the number of
molecules in the upper energy state $N_u$:
\begin{eqnarray}
\tau_\nu  &=&
\frac{c^2}{8\pi\nu^2}\frac{g_u}{g_l}A_{ul}\phi_\nu\int ds^\prime
n_l(s^\prime) \left(1 - \frac{g_ln_u(s^\prime)}{g_un_l(s^\prime)}\right)
\nonumber \\
&=& \frac{c^2}{8\pi\nu^2}\frac{g_u}{g_l}A_{ul}\phi_\nu
N_l \left(1 - \frac{g_lN_u}{g_uN_l}\right) \nonumber \\
&=& \frac{c^2}{8\pi\nu^2}
\left[\exp\left(\frac{h\nu}{kT}\right) - 1 \right] A_{ul}\phi_\nu N_u \nonumber \\
&=& \frac{8\pi^3\nu|{\mathbf \mu_{lu}}|^2}{3hc}
\left[\exp\left(\frac{h\nu}{kT}\right) - 1 \right] \phi_\nu N_u
\nonumber \\
\int \tau_\nu d\nu &=& \frac{8\pi^3\nu|{\mathbf \mu_{lu}}|^2}{3hc}
\left[\exp\left(\frac{h\nu}{kT}\right) - 1 \right] N_u,
\label{eq:taucolden}
\end{eqnarray}
where in the last step we have integrated over the line
profile such that $\int\phi_\nu = 1$.  Rearranging and converting our
frequency axis to velocity ($\frac{d\nu}{\nu} = \frac{dv}{c}$) in
Equation~\ref{eq:taucolden}, we get an expression for the column
density of molecules in the upper transition state ($N_u$):
\begin{equation}
N_u = \frac{3h}{8\pi^3|{\mathbf \mu_{lu}}|^2}
\left[\exp\left(\frac{h\nu}{kT}\right) - 1 \right]^{-1} \int \tau_\nu dv.
\label{eq:colden}
\end{equation}

At this point we have our basic equation for the number of molecules
in the upper energy state $u$ of our two-level system.  In order to
relate this to the total molecular column density as measured by the
intensity of a transition at frequency $\nu$, we need to relate the
number of molecules in the upper energy level $u$ ($N_u$) to the total
population of all energy levels in the molecule $N_{tot}$.  Assuming
detailed balance at a constant temperature defined by the excitation
temperature $T_{ex}$, we can relate these two quantities as follows:
\begin{equation}
\frac{N_{tot}}{N_u} = \frac{Q_{rot}}{g_u}\exp\left(\frac{E_u}{kT_{ex}}\right),
\label{eq:ntot}
\end{equation}
where we have introduced the ``rotational partition function''
$Q_{rot}$, a quantity that represents a statistical sum over all
rotational energy levels in the molecule (see Section~\ref{Qrot}) and $g_u$,
the degeneracy of the energy level $u$.  Substituting for $N_u$ in
Equation~\ref{eq:ntot}, the total molecular column density becomes:
\begin{multline}
N_{tot} = \frac{3h}{8 \pi^3 |{\mathbf \mu_{lu}}|^2}
  \frac{Q_{rot}}{g_u} \exp\left(\frac{E_u}{kT_{ex}}\right)\\
  \times\left[\exp\left(\frac{h\nu}{kT_{ex}}\right) - 1\right]^{-1}
  \int\tau_\nu dv.
\label{eq:ntotfinal}
\end{multline}

In the following we show how to calculate the level degeneracy $g_u$
(Section~\ref{Degeneracies}), the rotational partition function $Q_{rot}
\equiv \sum_i g_i \exp\left(-\frac{E_i}{kT}\right)$ (Section~\ref{Qrot}),
the dipole matrix element $|{\mathbf \mu_{lu}}|^2$ and associated line
strength $S$ and dipole moment $\mu$ (such that $|{\mathbf
  \mu_{jk}}|^2 \equiv S\mu^2$; Section~\ref{LineStrength}).  For
absorption lines where the integral of the optical depth over velocity
may be derived from a spectrum, Equation~\ref{eq:ntotfinal} may be
used directly to determine the total column density.  For emission
lines, the integral over optical depth is typically converted to an
integrated intensity $\int T_R dv$ (see Section~\ref{ColDenApprox}).
We derive several commonly-used approximations to
Equation~\ref{eq:ntotfinal}, including optically thin, optically
thick, and the Rayleigh-Jeans approximation in
Section~\ref{ColDenApprox}.  We also evaluate the 
limitations of the common assumption that the molecular excitation
temperature is constant (Section~\ref{Tex}), summarize non-LTE models
of the molecular column density (Section~\ref{NonLTE}), and the
distinction between beam- and source-averaged column density
(Section~\ref{BeamSourceAverage}).  We then close this
discussion of molecular column density by working through several
example calculations (Section~\ref{Examples}).  We also discuss 
some minor issues related to the assumed line profile function
(Appendix~\ref{Lineprofile}), the relationship between integrated
fluxes and brightness temperatures (Appendix~\ref{Intflux}), and the
uncertainty associated with an integrated intensity measurement
(Appendix~\ref{IntUncert}) in the appendices.  In a calculation which
utilizes the column density calculation formalism presented below,
Appendix~\ref{AmonTexTk} describes the standard three-level model for 
low-temperature NH$_3$ excitation often used to derive the kinetic
temperature from measurements of the (1,1) and (2,2) inversion
transitions of that molecule.

\section{Degeneracies}
\label{Degeneracies}

For rotational molecular transitions the total degeneracy for an
an energy level of a transition is given by the product of rotational
($g_J$ and $g_K$) and spin ($g_I$) degeneracies:
\begin{equation}
g_u \equiv g_J g_K g_I.
\label{eq:degeneracy}
\end{equation}
In the following we derive the expressions for these three
contributions to the degeneracy of a molecular energy level.

\subsection{Rotational Degeneracy ($g_J$)}
\label{RotDegen}

The rotational degeneracy due to the projection of the angular
momentum on the spatial axis $z$, $g_J$, exists in all molecules and
is given by:
\begin{equation}
g_J = 2J_u + 1.
\label{eq:gj}
\end{equation}

\subsection{K Degeneracy ($g_K$)}
\label{KDegen}

The K degeneracy ($g_K$) describes the degeneracy associated with the
internal quantum number K in symmetric and asymmetric top molecules
due to projections of the total angular momentum onto a molecular axis.
Because of the opposite symmetry of the doubly-degenerate levels for
which $K\ne 0$, $g_K$ is defined as follows:
\begin{eqnarray}
g_K &=& 1 \textrm{ for K=0 and all linear and asymmetric top}
\nonumber \\
    && \textrm{molecules,} \\ 
    &=& 2 \textrm{ for K$\neq$ 0 in symmetric top molecules.}
\label{eq:kdegen}
\end{eqnarray}
K-level doubling is due to the asymmetry in the molecule
about the molecular axes and removes the K degeneracy in asymmetric
top molecules.

\subsection{Nuclear Spin Degeneracy ($g_I$)}
\label{NucDegen}

The nuclear spin degeneracy $g_I$ takes account of the
statistical weights associated with identical nuclei in a nonlinear
molecule with symmetry (which most nonlinear molecules have).  For a
molecule with \textit{no symmetry or hyperfine splitting}, each
rotational level will have a nuclear spin degeneracy given by:
\begin{eqnarray}
g_n &=& \prod_i (2I_i + 1) = \left(2I + 1\right)^\sigma \\
g_I &\equiv& \frac{g_{nuclear}}{g_n},
\label{eq:nucdegen}
\end{eqnarray}
where $I_i$ represents the spin of the $i$th nucleus,
$\sigma$ is the number of identical nuclei, and $g_{nuclear}$ 
is given in Table~\ref{tab:nucdegen} for two of the largest classes of
molecules found in the interstellar medium medium: those with two
(C$_{2v}$ symmetry\footnote{The number of symmetry states for a
  molecule are determined by the number of configurations within which
  the wavefunction of the molecule is unchanged with a rotation of
  $\pi$ about a symmetry axis and a reflection of $\pi$ through that
  symmetry plane.  For a nice tutorial on molecular symmetry see
  Stefan Immel's extensive discussion at \textit{http://csi.chemie.tu-darmstadt.de/ak/immel/tutorials/symmetry/index7.html}}) and three (C$_{3v}$ symmetry) identical nuclei.
Symmetry and hyperfine splitting changes $g_n$ for all practical
cases.  Hyperfine splitting is covered elsewhere in this 
document, while symmetry considerations are well-covered by
\cite{Gordy1984} (Chapter III.4).  See also \cite{Turner1991} for a
general discussion applicable to the high-temperature limit.
\textit{Keep in mind that if you are only interested in studying one symmetry
state (\ie\ the para species) of a molecule, $g_I = 1$.}  In the
following we list some examples of $g_{nuclear}$ calculations for
several molecules.
\begin{deluxetable*}{lllll}
\tablewidth{0pt}
\tablecolumns{5}
\tablecaption{Nuclear Statistical Weight Factors for C$_{2v}$ and C$_{3v}$ Molecules\tablenotemark{a}\label{tab:nucdegen}}
\tablehead{
\colhead{Identical Nuclei ($\sigma$)\tablenotemark{b}} & 
\colhead{Spin} & 
\colhead{$J$} & 
\colhead{$K$\tablenotemark{c}} & 
\colhead{$g_{nuclear}$}
}
\startdata
2 & $\frac{1}{2},\frac{3}{2},\ldots$ & Any & Even & $(2I+1)I$ \\
2 & $\frac{1}{2},\frac{3}{2},\ldots$ & Any & Odd & $(2I+1)(I+1)$ \\
2 & $0,1,2,\ldots$ & Any & Odd & $(2I+1)(I+1)$ \\
2 & $0,1,2,\ldots$ & Any & Even & $(2I+1)I$ \\
3 & Any & Even or Odd & 3n, $\ne 0$ & $\frac{1}{3}(2I+1)(4I^2+4I+3)$ \\
3 & Any & Even or Odd & $\ne 3$n & $\frac{1}{3}(2I+1)(4I^2+4I)$
\enddata
\tablenotetext{a}{Derived from \cite{Gordy1984}, Table~3.2 and 3.3.}
\tablenotetext{b}{$g_I \equiv \frac{g_{nuclear}}{(2I + 1)^\sigma}$}
\tablenotetext{c}{Where n is an integer.}
\end{deluxetable*}

\subsubsection{H$_2$CO and c-C$_3$H$_2$}
\label{GnucAsym}

Formaldehyde (H$_2$CO) is a (slightly; $\kappa = -0.96$; see
Equation~\ref{eq:raysasymmetry}) prolate (see Section~\ref{Sym})
asymmetric top molecule, while Cyclopropenylidene 
(c-C$_3$H$_2$) is a cyclic (slightly; $\kappa = +0.69$) oblate (see
Section~\ref{Sym}) asymmetric top molecule.  Both have 
two opposing identical H (spin=$\frac{1}{2}$) nuclei.  The coordinate
wavefunction is symmetric/asymmetric for K$_{-1}$ even/odd,
respectively.  Therefore, from the two identical spin $\frac{1}{2}$
nuclei cases in Table~\ref{tab:nucdegen}:
\begin{eqnarray}
g_{nuclear} &=& (2I+1)I = 1 \textrm{ for K$_{-1}$ even,} \\
           &=& (2I+1)(I+1) = 3 \textrm{ for K$_{-1}$ odd.}
\label{eq:gnucasym}
\end{eqnarray}

\subsubsection{NH$_3$ and CH$_3$CN}
\label{GnucSym}

Ammonia (NH$_3$) and Acetonitrile (CH$_3$CN) are symmetric top
molecules with three opposing identical H (spin=$\frac{1}{2}$) nuclei.
Therefore, from the three identical spin $\frac{1}{2}$ nuclei cases in
Table~\ref{tab:nucdegen}:
\begin{eqnarray}
g_{nuclear} &=& \frac{1}{3}(2I+1)(4I^2+4I+3) = 4 \textrm{ for K=3n,} \\
            &=& \frac{1}{3}(2I+1)(4I^2+4I) = 2 \textrm{ for K$\ne$3n.}
\label{eq:gnucsym}
\end{eqnarray}

\subsubsection{c--C$_3$H and SO$_2$}
\label{GnucCyclic}

Cyclopropynylidyne (c--C$_3$H) is an oblate ($\kappa = +0.17$)
asymmetric top molecule with two opposing identical carbon (spin=0) nuclei,
while SO$_2$ is a prolate ($\kappa = -0.94$) asymmetric top molecule
with two opposing identical oxygen (spin=0) nuclei.  For both
molecules the coordinate wavefunction is symmetric/asymmetric for K$_{-1}$ even/odd,
respectively.  Therefore, from the two identical spin 0 nuclei
cases in Table~\ref{tab:nucdegen}:
\begin{eqnarray}
g_{nuclear} &=& (2I+1)I = 0 \textrm{ for K$_{-1}$ even,} \\
           &=& (2I+1)(I+1) = 1 \textrm{ for K$_{-1}$ odd.}
\label{eq:gnuccyclic}
\end{eqnarray}
This indicates that half of the levels are missing (those
for which K$_{-1}$ is even).

\section{Symmetry Considerations for Asymmetric Rotor Molecules}
\label{Sym}

The symmetry of the total wavefunction $\psi$ for a given rotational
transition is determined by the product of the coordinate wavefunction
$\psi_e \psi_v \psi_r$ and the nuclear spin wavefunction $\psi_n$.
It is common to refer to a symmetric nuclear spin state as
``ortho'' and an anti-symmetric nuclear spin state as ``para''.
These wavefunctions are of two types: Fermions and Bosons.
Table~\ref{tab:sym} lists the symmetries for the various wavefunctions
in both cases for exchange of two identical nuclei.
\begin{deluxetable*}{llcccl}
\tablewidth{0pt}
\tablecolumns{6}
\tablecaption{Eigenfunction Symmetries for Exchange of Two Identical
  Nuclei\tablenotemark{a} \label{tab:sym}}
\tablehead{
&& \multicolumn{3}{c}{Wavefunction\tablenotemark{b}} & \\
\cline{3-5}
\colhead{Statistics} & \colhead{Spin ($I$)} & \colhead{Total ($\psi$)}
& \colhead{Coordinate ($\psi_e \psi_v \psi_r$)} & \colhead{Spin
  ($\psi_n$)} & \colhead{$g_{nuclear}$}
}
\startdata
Fermi & $\frac{1}{2},\frac{3}{2},\ldots$ & A & S & A & $(2I+1)I$ \\
Fermi & $\frac{1}{2},\frac{3}{2},\ldots$ & A & A & S & $(2I+1)(I+1)$ \\
Bose  & $0,1,2,\ldots$ & S & S & S & $(2I+1)(I+1)$ \\
Bose  & $0,1,2,\ldots$ & S & A & A & $(2I+1)I$
\enddata
\tablenotetext{a}{From \cite{Gordy1984}, Table~3.2.}
\tablenotetext{b}{Key: A = Asymmetric; S = Symmetric.  For nuclear
  spin states asymmetric = ``para'' while symmetric = ``ortho''.}
\end{deluxetable*}
Since an asymmetric top can be thought of as belonging to
one of two limiting cases, prolate or oblate symmetric, we need to
consider these two cases in the context of the coordinate wavefunction
$\psi_e \psi_v \psi_r$.
\begin{description}
\item[\textit{Limiting Prolate:}] We consider the symmetry of the
  coordinate wavefunctions with respect to rotation of 180$^\circ$
  about the axis of \textit{least} moment of inertia.  Since the
  coordinate wavefunction $\psi_e \psi_v \psi_r$ depends on this
  rotation angle $\xi$ as $\exp\left(\pm i K_{-1} \xi\right)$, it is
  \textit{symmetric when K$_{-1}$ is even and antisymmetric when
  K$_{-1}$ is odd}.  H$_2$CO is a limiting prolate asymmetric top
  molecule (though it is only \textit{slightly} asymmetric).
\item[\textit{Limiting Oblate:}] We consider the symmetry of the
  coordinate wavefunctions with respect to rotation of 180$^\circ$
  about the axis of \textit{greatest} moment of inertia.  Since the
  coordinate wavefunction $\psi_e \psi_v \psi_r$ depends on this
  rotation angle $\xi$ as $\exp\left(\pm i K_{+1} \xi\right)$, it is
  \textit{symmetric when K$_{+1}$ is even and antisymmetric when
  K$_{+1}$ is odd}.  NH$_2$D is a limiting oblate asymmetric top
  molecule.
\end{description}
The level of asymmetry in molecules is often described in terms of
Ray's asymmetry parameter $\kappa$ \citep{Ray1932}:
\begin{equation}
\kappa \equiv \frac{2B-A-C}{A-C},
\label{eq:raysasymmetry}
\end{equation}
where A, B, and C are constants which represent the three
principle moments of inertia for an asymmetric rotor molecule (see
\cite{Gordy1984} for more information), usually
expressed in megahertz (MHz).  The limits to $\kappa$ are:
\begin{itemize}
\item B = A: $\kappa = +1$ and the molecule is an oblate symmetric
  rotor.  
\item B = C: $\kappa = -1$ and the molecule is a prolate symmetric
  rotor.  
\end{itemize}
Taking a more general example, the rotational angular momentum
constants for H$_2$CO are A = 281970.37\,MHz, B = 38835.42558\,MHz,
and C = 34005.73031\,MHz.  Equation~\ref{eq:raysasymmetry} then yields
$\kappa \simeq -0.961$, which means that H$_2$CO is nearly a prolate
symmetric rotor.

\section{Rotational Partition Functions ($Q_{rot}$)}
\label{Qrot}

For a parcel of gas that exchanges energy with the ambient medium,
statistical mechanics states that the partition function $Q$ which
describes the relative population of states in the gas is given by:
\begin{equation}
Q = \sum_i g_i \exp\left(-\frac{E_i}{kT}\right).
\label{eq:q}
\end{equation}
Following \cite{Gordy1984} (chapter 3, section 3), the partition function
for molecules in a gaseous state is a function of the electronic,
vibrational, rotational, and nuclear spin states of the molecule.
Assuming that there are no interactions between these states, the
total partition function for the molecule can be expressed as the
product of the partition functions of these four types of energy
states:
\begin{equation}
Q = Q_e Q_v Q_r Q_n.
\label{eq:qall}
\end{equation}
The ground electronic state of most molecules observed in the
interstellar medium are stable, $^{1}\Sigma$ electronic states,
although there are a few notable exceptions (\ie\ OH $^{2}\Pi$, SO
$^{3}\Sigma$, CN $^{2}\Sigma$, \etc).  For these molecules $Q_e = 1$.
For simplicity we will also assume that the molecules are in their
ground vibrational state ($Q_v = 1$).  This 
leaves us with rotational and nuclear partition functions comprising
the total molecular partition function, which we can write as:
\begin{eqnarray}
Q_{rot} &\equiv& Q_r Q_n \nonumber \\
&=& \sum_{J,K,I} g_J g_K g_I \exp\left(-\frac{E_{JK}}{kT}\right),
\label{eq:qrot}
\end{eqnarray}
where the degeneracies $g_J$, $g_K$, and $g_I$ are
described in Section~\ref{RotDegen}, Section~\ref{KDegen}, Section~\ref{NucDegen},
respectively.  See \cite{Turner1991} for a nice general discussion
listing expressions for $Q_{rot}$ in the high-temperature limit for a
variety of molecules.  In the following we derive the rotational
partition function $Q_{rot}$ for linear, symmetric, and asymmetric
rotor molecules.

\subsection{Linear Molecule Rotational Partition Function}
\label{LinQrot}

\begin{figure}
\centering
\includegraphics[trim=10mm 0mm 0mm 0mm, clip, scale=0.45]{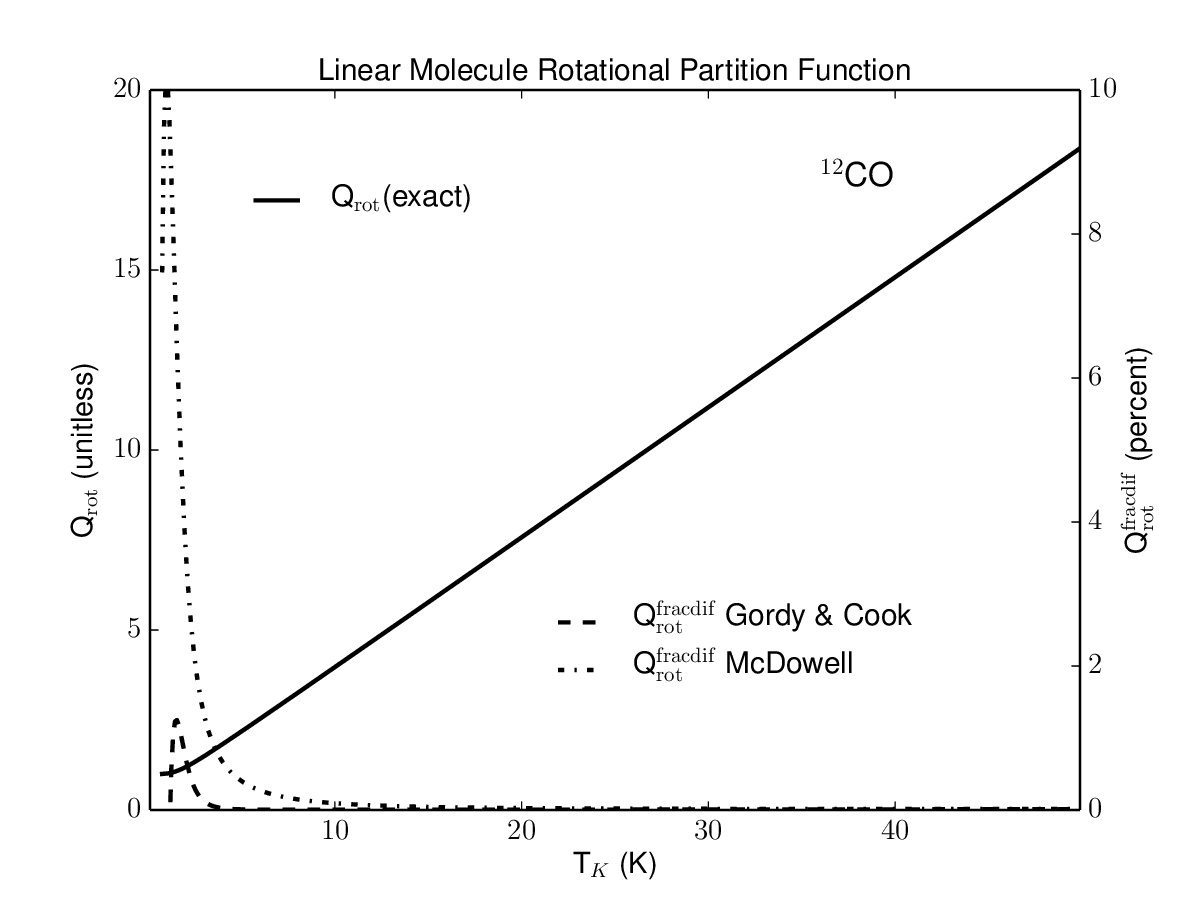}
\caption{Rotational partition function calculations for CO using the
  lowest 51 levels of the molecule.  Shown are $Q_{rot}$
  (Equation~\ref{eq:qrotlinear}; vertical scale on left) and the
  percentage differences between $Q_{rot}$ and the expansions of
  Equation~\ref{eq:qrotlinear} provided by
  Equation~\ref{eq:qrotlineardiatom} (using terms to third-order in
  $\frac{hB_0}{kT}$) and Equation~\ref{eq:qrotlinearpolyatom}
  (vertical scale on right).}
\label{fig:qrotco}
\end{figure}

For linear molecules:
\begin{itemize}
\item $g_J = 2J + 1$ (Section~\ref{RotDegen})
\item $g_K = 1$ (Section~\ref{KDegen})
\item $g_I = 1$ (since linear molecules with allowed electric dipole
  transitions are polar and have no center of symmetry)
\end{itemize}
which implies that Equation~\ref{eq:qrot} becomes:
\begin{equation}
Q_{rot} = \sum^\infty_{J=0} (2J + 1)\exp\left(-\frac{E_J}{kT}\right).
\label{eq:qrotlinear}
\end{equation}
The energy levels for a linear molecule can be described by
a multi-term expansion as a function of $J(J+1)$ \citep{Jennings1987}:
\begin{multline}
E_J = h(B_0J(J+1) - D_0J^2(J+1)^2 + H_0J^3(J+1)^3 \\
- L_0J^4(J+1)^4 + M_0J^5(J+1)^5 + ...),
\label{eq:energylinear}
\end{multline}
where $B_0$ is the rigid rotor rotation constant and $D_0$,
$H_0$, $L_0$, and $M_0$ are the first- through fourth-order
centrifugal distortion constants for the molecule, respectively, all
in megahertz (MHz).  Using the rigid rotor approximation to the level energies,
thus ignoring all terms other than those linear in J(J+1),
Equation~\ref{eq:energylinear} becomes:
\begin{equation}
E_J = hB_0J(J+1).
\label{eq:energylinearrigidrotor}
\end{equation}
This allows us to approximate $Q_{rot}$ for diatomic linear
molecules as follows:
\begin{align}
Q_{rot} \simeq& \sum^\infty_{J=0} (2J +
1)\exp\left(-\frac{hB_0J(J+1)}{kT}\right) \nonumber \\
\simeq& \frac{kT}{hB_0} + \frac{1}{3} +
\frac{1}{15}\left(\frac{hB_0}{kT}\right) +
\frac{4}{315}\left(\frac{hB_0}{kT}\right)^2 \nonumber \\
&+ \frac{1}{315}\left(\frac{hB_0}{kT}\right)^3 + \ldots
\label{eq:qrotlineardiatom}
\end{align}
(from \cite{Gordy1984} Chapter 3, Equation~3.64 and \cite{McDowell1988},
where application of the Euler-MacLaurin summation formula has been
used).  This approximate form is good to $<1$\% for $T > 2$\,K
(Figure~\ref{fig:qrotco}).  This level of accuracy is maintained even
if one includes only the first two terms in
Equation~\ref{eq:qrotlineardiatom}.

An alternate approximation for linear polyatomic molecules is derived
by \cite{McDowell1988}:
\begin{equation}
Q_{rot} \simeq \frac{kT}{hB_0}\exp\left(\frac{hB_0}{3kT}\right),
\label{eq:qrotlinearpolyatom}
\end{equation}
which is reported to be good to 0.01\% for $\frac{hB_0}{kT}
\lesssim 0.2$ ($T > 13.7$\,K for CO) and is good to better than 1\%
for $T > 3.5$\,K (Figure~\ref{fig:qrotco}).  Note that
Equation~\ref{eq:qrotlinearpolyatom} reduces to
Equation~\ref{eq:qrotlineardiatom} when expanded using a Taylor Series.

\vfill\eject
\subsection{Symmetric and Slightly-Asymmetric Rotor Molecule Rotational
  Partition Function}
\label{SymQrot}

\begin{figure}
\centering
\includegraphics[trim=10mm 0mm 0mm 0mm, clip, scale=0.45]{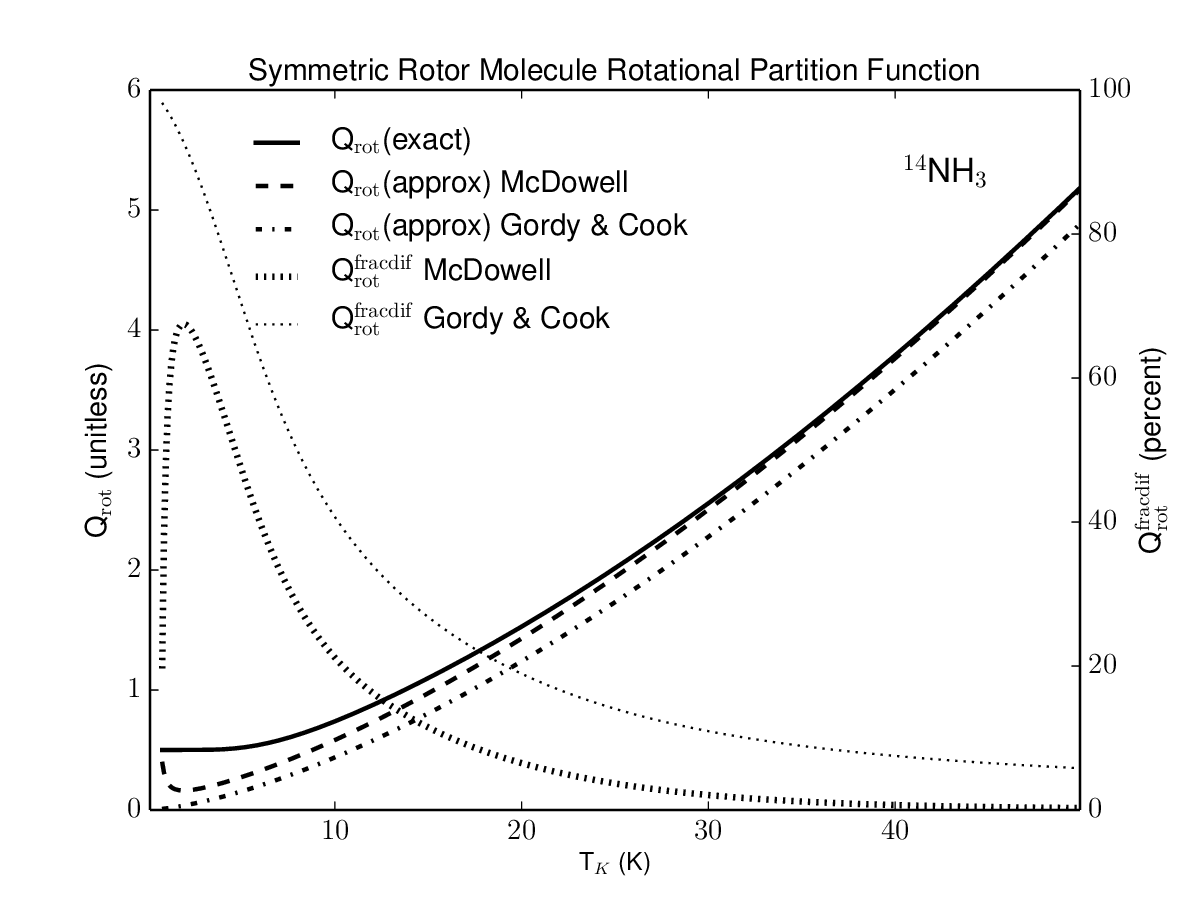}
\caption{Rotational partition function calculations for NH$_3$ using
  the lowest 51 levels of the molecule.  Shown are the exact summation
  for $Q_{rot}$ (Equation~\ref{eq:qrotsymmetric}) and the approximate
  forms given by Equations~\ref{eq:qrotsymmetricapprox} and
  \ref{eq:qrotsymmetricapprox2} (vertical scale on left).  Shown also
  are the fractional percentage differences (vertical scale on right), given by
  $100*\frac{Q_{rot}(exact) - Q_{rot}(approx)}{Q_{rot}(exact)}$, of
  these two approximations relative to $Q_{rot}$
  (Equation~\ref{eq:qrotsymmetric}).} 
\label{fig:qrotnh3}
\vspace{2mm}
\end{figure}

For symmetric and asymmetric rotor molecules:
\begin{itemize}
\item $g_J = 2J + 1$ (Section~\ref{RotDegen})
\item $g_K = 1$ for K = 0 and 2 for K $\neq$ 0 in symmetric rotors
  (Section~\ref{KDegen}) 
\item $g_K = 1$ for all K in asymmetric rotors
\item $g_I = \frac{g_{nuclear}}{\left(2I+1\right)^\sigma}$ (See Table~\ref{tab:nucdegen})
\end{itemize}
which implies that Equation~\ref{eq:qrot} becomes:
\begin{equation}
Q_{rot} = \sum^\infty_{J=0}\sum^J_{K=-J} g_K g_I (2J + 1)\exp\left(-\frac{E_{JK}}{kT}\right).
\label{eq:qrotsymmetric}
\end{equation}
Like the energy levels for a linear molecule, the energy
levels for a symmetric or slightly-asymmetric\footnote{Slightly-asymmetric
    rotor molecules are defined such that $B_0\simeq C_0$. In
    Equation~\ref{eq:energysymmetric} replace $B$ with $\sqrt{B_0C_0}$.}
rotor molecule can be described by a multi-term expansion as a function of $J(J+1)$:
\begin{multline}
E_{JK} = h(B_0J(J+1) + s_0K^2 + D_jJ^2(J+1)^2 \\
+ D_{jk}J(J+1)K^2 + D_kK^4 + H_{jkk}J(J+1)K^4 \\
+ H_{jjk}J^2(J+1)^2K^2 + H_{j6}J^3(J+1)^3 \\
+ H_{k6}K^6 +...),
\label{eq:energysymmetric}
\end{multline}
where $s_0 \equiv A_0 - B_0$ for a prolate symmetric
rotor molecule and $s_0 \equiv C_0 - B_0$ for an oblate symmetric
rotor, and the other constants represent various terms in the
centrifugal distortion of the molecule.  All constants are in
megahertz (MHz).
For rigid symmetric or slightly-asymmetric rotor molecules, using the
rigid rotor approximation to the level energies:
\begin{equation}
E_{JK} = h\left(B_0J(J+1) + s_0K^2\right).
\label{eq:energysymmetricsymmetricrigidrotor}
\end{equation}
From \cite{McDowell1990} we can then approximate $Q_{rot}$
for a symmetric or slightly-asymmetric rotor molecule as
follows:
\begin{multline}
Q_{rot} \simeq
\frac{\sqrt{m\pi}}{\sigma}\exp\left(\frac{hB_0(4-m)}{12kT}\right)\left(\frac{kT}{hB_0}\right)^{3/2} \\
\times\left[1 + \frac{1}{90}\left(\frac{hB_0(1-m)}{kT}\right)^2 + ...\right],
\label{eq:qrotsymmetricapprox}
\end{multline}
where
\begin{align}
m =& \frac{B_0}{A_0} \textrm{~for a prolate symmetric rotor molecule,} \nonumber \\
  =& \frac{B_0}{C_0} \textrm{~for an oblate symmetric rotor molecule,} \nonumber \\
  =& \frac{B^2_0}{A_0C_0} \textrm{~for a slightly-asymmetric rotor molecule} \nonumber \\
&\textrm{(see \cite{Herzberg1945}).} \nonumber
\label{eq:m}
\end{align}
If we expand the exponential and take only up to first order
terms in the expansion in Equation~\ref{eq:qrotsymmetricapprox}:
\begin{eqnarray}
Q_{rot} &\simeq&
\frac{\sqrt{m\pi}}{\sigma}\left(1+\frac{hB_0(4-m)}{12kT}+...\right)\left(\frac{kT}{hB_0}\right)^{3/2}
\nonumber \\
&\simeq& \frac{\sqrt{m\pi}}{\sigma}\left(\frac{kT}{hB_0}\right)^{3/2} \nonumber \\
&\simeq& \frac{1}{\sigma}\left[m\pi\left(\frac{kT}{hB_0}\right)^3\right]^{1/2},
\label{eq:qrotsymmetricapprox2}
\end{eqnarray}
which is the equation for symmetric rotor partition functions quoted
by \cite{Gordy1984} (Chapter 3, Equations~3.68 and 3.69).
Figure~\ref{fig:qrotnh3} compares $Q_{rot}$ calculated using
Equation~\ref{eq:qrotsymmetric} and the approximate forms given by
Equation~\ref{eq:qrotsymmetricapprox} and
Equation~\ref{eq:qrotsymmetricapprox2} for NH$_3$.  In this example
Equation~\ref{eq:qrotsymmetricapprox} is good to $\lesssim 20\%$ 
for $T_K > 10$\,K and $\lesssim 0.25\%$ for $T_K > 50$\,K, while
Equation~\ref{eq:qrotsymmetricapprox2} is much less accurate, good to
$\lesssim 40\%$ for $T_K > 10$\,K and $\lesssim 6\%$ for $T_K > 50$\,K.

\section{Dipole Moment Matrix Elements ($|\mu_{jk}|^2$) and
  Line Strengths ($S$)}
\label{LineStrength}

The following discussion is derived from the excellent discussion
given in \cite{Gordy1984}, Chapter II.6.  A detailed discussion of
line strengths for diatomic molecules can be found in \cite{Tatum1986}.
Spectral transitions are induced by interaction of the electric or
magnetic components of the radiation field in space with the electric
or magnetic dipole components fixed in the rotating molecule.  The
strength of this interaction is called the \textit{line strength $S$}.  
The matrix elements of the dipole moment with reference to the
space-fixed axes (X,Y,Z) for the rotational eigenfunctions $\psi_r$
can be written as follows:
\begin{equation}
\int \psi_r^\ast \mu_F \psi_r^\prime d\tau = 
\sum_g \mu_g \int \psi_r^\ast \Phi_{Fg} \psi_r^\prime d\tau,
\label{eq:dipole}
\end{equation}
where $\Phi_{Fg}$ is the direction cosine between the
space-fixed axes F=(X,Y,Z) and the molecule-fixed axes g=(x,y,z).
The matrix elements required to calculate line strengths for linear
and symmetric top molecules are known and can be evaluated in a
straightforward manner, but these calculations are rather tedious
because of the complex form of the eigenfunction.  Using commutation
rules between the angular momentum operators and the direction cosines
$\Phi_{Fg}$, \cite{Cross1944} derive the nonvanishing direction cosine
matrix elements in the symmetric top representation (J,K,M):
\begin{multline}
\langle J,K,M | \Phi_{Fg} | J^\prime, K^\prime, M^\prime \rangle = 
\langle J | \Phi_{Fg} | J^\prime \rangle \\
\times\langle J,K | \Phi_{Fg} | J^\prime, K^\prime \rangle
\langle J,M | \Phi_{Fg} | J^\prime, M^\prime \rangle.
\label{eq:matrixelementterms}
\end{multline}
where $M$ is the projection of the total angular momentum $J$ on a
fixed space axis.  There are $2J+1$ projections of $M$ allowed.  The
dipole moment matrix element $|\mu_{lu}|^2$ can then be written as:
\begin{equation}
|\mu_{lu}|^2 = \sum_{F=X,Y,Z}\sum_{M^\prime}|\langle
  J,K,M | \mu_F |J^\prime, K^\prime, M^\prime \rangle|^2,
\label{eq:matrixelement}
\end{equation}
where the sum over $g=x,y,z$ is contained in the expression
for $\mu_F$ (Equation \ref{eq:dipole}).
Table~\ref{tab:matrixelementterms} lists the direction cosine matrix
element factors in Equation \ref{eq:matrixelementterms} for symmetric
rotor and linear molecules.  In the following we give examples of the
use of the matrix elements in line strength calculations
\begin{deluxetable*}{llll}
\tablewidth{\textwidth}
\tablecolumns{4}
\tabletypesize{\footnotesize}
\tablecaption{Direction Cosine Matrix Element Factors\tablenotemark{a}
  for Linear\tablenotemark{b} and Symmetric Top
  Molecules\label{tab:matrixelementterms}}
\tablehead{
& \multicolumn{3}{c}{$J^\prime$ Value} \\
\cline{2-4}
\multicolumn{1}{c}{Matrix Element Term} &
\multicolumn{1}{c}{$J+1$}  & 
\multicolumn{1}{c}{$J$} & \multicolumn{1}{c}{$J-1$}
}
\startdata
$\langle J | \Phi_{Fg} | J^\prime \rangle$ &
	$\left\{4(J+1)\left[(2J+1)(2J+3)\right]^\frac{1}{2}\right\}^{-1}$ &
	$\left[4J(J+1)\right]^{-1}$ &
	$\left[4J(4J^2+1)^\frac{1}{2}\right]^{-1}$ \\
$\langle J,K | \Phi_{Fz} | J^\prime, K \rangle$ &
	$2\left[(J+1)^2-K^2\right]^\frac{1}{2}$ & 
	$2K$ &
	$-2(J^2-K^2)^\frac{1}{2}$ \\
$\langle J,K | \Phi_{Fy} | J^\prime, K\pm1 \rangle$ = 
$\mp i \langle J,K | \Phi_{Fx} | J^\prime, K\pm1 \rangle$ & 
	$\mp \left[(J \pm K + 1)(J \pm K + 2)\right]^\frac{1}{2}$ &
	$\left[J(J+1)-K(K+1)\right]^\frac{1}{2}$ &
	$\mp \left[(J \mp K)(J \mp K - 1)\right]^\frac{1}{2}$ \\
$\langle J,M | \Phi_{Zg} | J^\prime, M \rangle$ &
	$2\left[(J+1)^2-M^2\right]^\frac{1}{2}$ & 
	$2M$ &
	$-2(J^2-M^2)^\frac{1}{2}$ \\
$\langle J,M | \Phi_{Yg} | J^\prime, M\pm1 \rangle$ = 
$\pm i \langle J,M | \Phi_{Xg} | J^\prime, M\pm1 \rangle$ & 
	$\mp \left[(J \pm M + 1)(J \pm M + 2)\right]^\frac{1}{2}$ &
	$\left[J(J+1)-M(M+1)\right]^\frac{1}{2}$ &
	$\mp \left[(J \mp M)(J \mp M - 1)\right]^\frac{1}{2}$
\enddata
\tablenotetext{a}{Derived from \cite{Gordy1984}, Table~2.1, which is
  itself derived from \cite{Cross1944}.}
\tablenotetext{b}{For linear molecules, set K=0 in the terms listed.}
\end{deluxetable*}

\section{Linear and Symmetric Rotor Line Strengths}
\label{SymtopsandLin}

For all linear and most symmetric top molecules, the permanent dipole
moment of the molecule lies completely along the axis of symmetry of
the molecule ($\mu = \mu_z$).  This general rule is only violated for
the extremely-rare ``accidentally symmetric top'' molecule (where $I_x
= I_y$).  For all practical cases, then, Equation
\ref{eq:matrixelement} becomes:
\begin{eqnarray}
|\mu_{lu}|^2 &=& \mu^2 \sum_{F=X,Y,Z}\sum_{M^\prime}|\langle J,K,M | \Phi_{Fg}
|J^\prime, K^\prime, M^\prime \rangle|^2 \nonumber \\
&=& \mu^2 S,
\label{eq:matrixelement2}
\end{eqnarray}
where we have defined the line strength $S$ as:
\begin{equation}
S \equiv \sum_{F=X,Y,Z}\sum_{M^\prime}|\langle J,K,M | \Phi_{Fg}
|J^\prime, K^\prime, M^\prime \rangle|^2.
\label{eq:linestrengthdef}
\end{equation}

\subsection{$(J,K)\rightarrow (J-1,K)$ Transitions}
\label{SymtopsandLinP}

Using the matrix element terms listed in the fourth column of
Table~\ref{tab:matrixelementterms} with the definition of the
direction cosine matrix elements
(Equation~\ref{eq:matrixelementterms}), we can write the terms which
make-up Equation~\ref{eq:matrixelement2} for the case
$(J,K)\rightarrow (J-1,K)$\footnote{Also referred to as ``P-branch
  transitions''.} as follows:
\begin{multline}
|\mu_{lu}|^2 = \mu^2
  \left[\frac{(J^2-K^2)^{1/2}}{J(4J^2-1)^{1/2}}\right]
  \Biggl\{(J^2-M^2)^{1/2} \\
  +\left(\frac{i\pm1}{2}\right)\left[(J\mp M)(J\mp
  M-1)\right]^{1/2}\Biggr\}.
\label{eq:smatrixp}
\end{multline}
Applying these terms to the dipole moment matrix element
(Equation \ref{eq:matrixelement}, which simply entails squaring each
of the three terms in Equation \ref{eq:smatrixp} and expanding the
$\pm$ terms) and using the definition of $|\mu_{lu}|^2$
(Equation~\ref{eq:linestrengthdef}):
\begin{multline}
S = \left[\frac{(J^2-K^2)}{J^2(4J^2-1)}\right]
  \Biggl[(J^2-M^2) \\
+ \frac{1}{2}\left[(J-M)(J-M-1) +
  (J+M)(J+M-1)\right]\Biggr].
\label{eq:smatrixp2}
\end{multline}
Reducing Equation \ref{eq:smatrixp2} results in the
following for a symmetric top transition $(J,K)\rightarrow (J-1,K)$:
\begin{equation}
S = \frac{J^2-K^2}{J(2J+1)} \textrm{ for $(J,K)\rightarrow (J-1,K)$}.
\label{eq:linestrengthlp}
\end{equation}
To derive the equation for a linear molecule transition
$J\rightarrow J-1$, simply set $K=0$ in Equation~\ref{eq:linestrengthlp}.

\subsection{$(J,K)\rightarrow (J,K)$ Transitions}
\label{SymtopsandLinQ}

Using the matrix element terms listed in the fourth column of
Table~\ref{tab:matrixelementterms} with the definition of the
direction cosine matrix elements
(Equation~\ref{eq:matrixelementterms}), we can write the terms which
make-up Equation~\ref{eq:matrixelement2} for the case
$(J,K)\rightarrow (J,K)$\footnote{Also referred to as ``Q-branch
  transitions''.} as follows: 
\begin{multline}
|\mu_{lu}|^2 = \mu^2 \left[\frac{2K}{4J(J+1)}\right] \\
  \times\left\{2M \pm 2\left[J(J+1)-M(M\pm1)\right]^{1/2}\right\}.
\label{eq:smatrixq}
\end{multline}
Applying these terms to the dipole moment matrix element
(Equation \ref{eq:matrixelement}) and using the definition of
$|\mu_{lu}|^2$ (Section~\ref{LineStrength}):
\begin{multline}
S = \left[\frac{K^2}{4J^2(J+1)^2}\right] \Biggl[4M^2 \\
+ 2\left[J(J+1)-M(M+1) + J(J+1)-M(M-1)\right]\Biggr].
\label{eq:smatrixq2}
\end{multline}
Reducing Equation \ref{eq:smatrixq2} results in the
following for a symmetric top transition $(J,K)\rightarrow (J,K)$:
\begin{equation}
S = \frac{K^2}{J(J+1)} \textrm{ for $(J,K)\rightarrow (J,K)$}.
\label{eq:linestrengthlsq}
\end{equation}

\subsection{$(J,K)\rightarrow (J+1,K)$ Transitions}
\label{SymtopsandLinR}

Using the matrix element terms listed in the second column of
Table~\ref{tab:matrixelementterms} with the definition of the
direction cosine matrix elements
(Equation~\ref{eq:matrixelementterms}), we can write the terms which
make-up Equation~\ref{eq:matrixelement2} for the case
$(J,K)\rightarrow (J+1,K)$\footnote{Also referred to as ``R-branch
  transitions''.} as follows:
\begin{multline}
|\mu_{lu}|^2 = \mu^2
  \frac{\left[(J+1)^2-K^2\right]^{1/2}}{(J+1)\left[(2J+1)(2J+3)\right]^{1/2}}
  \\
\times\Biggl\{\left[(J+1)^2-M^2\right]^{1/2} \\
-\left(\frac{i\pm1}{2}\right)\left[(J\pm
    M+1)(J\pm M+2)\right]^{1/2}\Biggr\}.
\label{eq:smatrixr}
\end{multline}
Applying these terms to the dipole moment matrix element
(Equation \ref{eq:matrixelement}, which simply entails squaring each
of the three terms in Equation~\ref{eq:smatrixr} and expanding the
$\pm$ terms) and using the definition of $|\mu_{lu}|^2$
(Equation~\ref{eq:linestrengthdef}):
\begin{multline}
S = \frac{\left[(J+1)^2-K^2\right]^{1/2}}{(J+1)\left[(2J+1)(2J+3)\right]^{1/2}}
\Biggl\{(J+1)^2 - M^2 \\
+ \frac{1}{2}\Biggl[(J+M+1)(J+M+2) \\
+ (J-M+1)(J-M+2)\Biggr]\Biggr\}.
\label{eq:smatrixr2}
\end{multline}
Reducing Equation \ref{eq:smatrixr2} results in the
following for a symmetric top transition $(J,K)\rightarrow (J+1,K)$:
\begin{equation}
S = \frac{(J+1)^2-K^2}{(J+1)(2J+1)} \textrm{ for $(J,K)\rightarrow (J+1,K)$}.
\label{eq:linestrengthlr}
\end{equation}
To derive the equation for a linear molecule transition
$J\rightarrow J+1$, simply set $K=0$ in Equation~\ref{eq:linestrengthlr}.

%
\section{Hyperfine Structure and Relative Intensities}
\label{Hyperfine}

Hyperfine splitting in molecular spectra occurs when a nucleus in the molecule
has non-zero nuclear spin.  Hyperfine splitting is commonly observed in
molecules that contain $^1$H (I = 1/2), $^2$D (I = 1), $^{13}$C (I = 1/2),
$^{14}$N (I = 1), $^{15}$N (I = 1/2), $^{17}$O (I = 5/2), and $^{33}$S (I = 3/2) nuclei.  
For a molecule with a single coupling nucleus, the vector addition of
angular momenta is used to define a new hyperfine quantum number, F, such that 
$\vec{F} = \vec{I} + \vec{J}$
is the total molecular
angular momentum, where $\vec{I}$ and $\vec{J}$ are the nuclear spin and rotational
angular momentum of the molecule respectively
($\vec{I}$ and $\vec{J}$ precess around $\vec{F}$).  
The allowed scalar values of $\vec{F}$ are given by
\begin{equation}
F = J+I, J+I-1, ... , |J-I| \;\; .
\label{eq:cg}
\end{equation}
Selection rules stipulate that $\Delta F = 0,\pm 1$ (but $F = 0$
cannot go to $0$).  
For example, the $^{14}$N nucleus in HCN is responsible for strong
hyperfine coupling because the spin of $^{14}$N is I = 1.  
Equation~\ref{eq:cg} shows that the J = 0 level remains unsplit 
(F = 1), the J = 1 level splits into a triplet (F = 2, 1, 0),
and the J = 2 level also splits into a triplet (F = 3, 2, 1).
The selection rule allows 3 hyperfine transitions from $J=1-0$ and
six hyperfine transitions from $J=2-1$.  Since I = 1 and
the Clebsch-Gordon series (Equation~\ref{eq:cg}) terminates with F = $|$J $-$ I$|$, all of
the levels with J $>$ 2 will also be split into triplets.  

The relative strengths can be calculated by using irreducible
tensor methods (see \cite{Gordy1984} Chapter 15).  
We define the relative strength such that the sum of the relative 
strength, $R_i$, of all
transitions from $F^\prime \rightarrow F$ for a given
$J^\prime \rightarrow J$ are
equal to one:
\begin{equation}
\sum_{F^\prime F}^{} R_i(J^\prime F^\prime \rightarrow\ J F) = 1 .
\end{equation}
The relative line strengths are 
calculated in terms of a $6-j$ symbol,
\begin{equation}
R_i(J^\prime F^\prime \rightarrow\ J F) = \frac{(2F+1)(2F^\prime +1)}{(2I+1)}
\begin{Bmatrix}
I & F^\prime & J^\prime \\ 
1 & J        & F 
\end{Bmatrix}^2.
\end{equation}
With the aid of $6-j$ Tables \citep{Edmonds1974}\footnote{Many online calculation tools
are available that will calculate $6j$ symbols.  For example, see 
http://www.svengato.com/sixj.html.}, 
and the property that
$6-j$ symbols are invariant with permutation of the columns, we find
the appropriate $6-j$ symbol for each transition:
\begin{multline}
\begin{Bmatrix}
a & b & c \\ 
1 & c-1 & b-1  
\end{Bmatrix}^2 = \\ \frac{s(s + 1)(s - 2a - 1)(s - 2a)}{(2b - 1)2b(2b +
  1)(2c - 1)2c(2c + 1)} \\
\shoveleft{\begin{Bmatrix}
a & b & c \\ 
1 & c-1 & b  
\end{Bmatrix}^2 = }\\ \frac{2(s + 1)(s - 2a)(s - 2b)(s - 2c + 1)}{2b(2b +
  1)(2b + 2)(2c - 1)2c(2c + 1)} \\
\shoveleft{\begin{Bmatrix}
a & b & c \\ 
1 & c-1 & b+1 
\end{Bmatrix}^2 = }\\ \frac{(s - 2b - 1)(s - 2b)(s - 2c +1)(s - 2c + 2)}
{(2b + 1)(2b + 2)(2b + 3)(2c - 1)2c(2c + 1)} \\
\shoveleft{\begin{Bmatrix}
a & b & c \\ 
1 & c & b  
\end{Bmatrix}^2 = }\\ \frac{[2b(b + 1) + 2c(c + 1) - 2a(a + 1)]^2}{2b(2b +
  1)(2b + 2)2c(2c + 1)(2c + 2)} \;\;, \\
\label{eq:6j}
\end{multline}
where $s = a + b +c$.  For example, in the case of the $J=1-0$ $F=2-1$
transition of HC$^{14}$N, the first $6-j$ symbol in
Equation~\ref{eq:6j} should be used with $a = I 
= 1$, $b = F^{\prime} = 2$, and $c = J^{\prime} = 1$.

The formalism above may be generalized when multiple coupling nuclei are
present.  By extension, an arbitrary number, $n$, of coupling nuclei may be included in
the nested vector addition of angular momenta (\ie\  
$\vec{F}_i = \vec{J} + \vec{I}_1$, $\vec{F}_{2} = \vec{F}_1 + \vec{I}_2$,\nodata ,  $\vec{F} = \vec{F}_{n-1} + \vec{I}_n$) and the relative
intensities can be calculated by multiplying the relative strength equation
$n$ times with appropriate quantum numbers for each step in the nested
vector addition.
For example, in the case of two coupling nuclei with unequal strength,
such as the outer ($I_1$) and inner ($I_2$) $^{14}$N of N$_2$H$^+$, the nested 
coupling scheme $\vec{F}_1 = \vec{J} + \vec{I}_1$ and $\vec{F} = \vec{F}_1 + \vec{I}_2$
is used.  The relative strength of the hyperfine transitions are then given by
\begin{multline}
R_i(J^\prime F_1^\prime F^\prime \rightarrow\ J F_1 F) = \\
\frac{(2F_1^\prime+1)(2F_1+1)
(2F^\prime +1)(2F +1)}{(2I_1+1)(2I_2+1)}\\
\times \begin{Bmatrix}
I_1 & F_1^\prime & J^\prime \\ 
1   & J          & F_1 
\end{Bmatrix}^2
\begin{Bmatrix}
I_2 & F^\prime & F_1^\prime \\ 
1   & F_1      & F
\end{Bmatrix}^2 .
\label{eq:hyperrel}
\end{multline}
While we have specifically discussed hyperfine splitting from nuclei with
unequal coupling in this section, the addition of vector angular momenta
in quantum mechanics is a general problem and the formalism above may be
extended to problems with equal coupling nuclei (i.e. H$_2$CO) or for fine structure
splitting (\ie\ in molecules with electronic states with total electronic spin $S > 0$)
with an appropriate mapping of the quantum numbers.

We end this section by noting an important caveat with regards
the interpretation of the relative intensities of hyperfine
transitions.  Anomalies between observed hyperfine transition
intensities and those predicted by the quantum mechanics described
in this section have been observed (see \cite{Kwan1974},
\cite{Guilloteau1981}, \cite{Walmsley1982}, \cite{Stutzki1984},
\cite{Stutzki1985}, \cite{Bachiller1997}, \cite{Daniel2006}, and
\cite{Hily-Blant2010} for examples).  These anomalies are likely due
to spectral line overlap between hyperfine transitions and can
result in non-LTE hyperfine ratios.

\section{Approximations to the Column Density Equation}
\label{ColDenApprox}

In the following we derive several commonly-used approximations to the
column density Equation~\ref{eq:ntotfinal}: Optically thin ($\tau\ll
1$), optically thin when in the Rayleigh-Jeans limit ($h\nu\ll
kT_{ex}$), and optically thin when in the Rayleigh-Jeans limit and
negligible background emission ($T_{bg}\ll T_{ex}$).

\subsection{Optically Thin Approximation}
\label{Thin}

If we assume that $\tau_\nu\ll 1$, the radiative transfer equation
(Equation~\ref{eq:tr}) becomes:
\begin{eqnarray}
T_R &=& f\left[J(T_{ex}) - J(T_{bg})\right]\tau,
\label{eq:trthin}
\end{eqnarray}
and the column density equation (Equation
\ref{eq:ntotfinal}) becomes:
\begin{multline}
N^{thin}_{tot} = \left(\frac{3h}{8 \pi^3 S \mu^2
  R_i}\right)\left(\frac{Q_{rot}}{g_J g_K g_I}\right)
\frac{\exp\left(\frac{E_u}{kT_{ex}}\right)}{\exp\left(\frac{h\nu}{kT_{ex}}\right)
  - 1}\\
  \times\int\frac{T_R dv}{f\left(J_\nu(T_{ex}) - J_\nu(T_{bg})\right)}.
\label{eq:ntotthin0}
\end{multline}
The $J_{\nu}(T_{ex}) - J_{\nu}(T_{bg})$ term may be removed from the
integral because the Planck function does not vary substantially
across the frequency extent of a typical spectral line.  For example,
if we take the CO $J=1-0$ transition at 115\,GHz, then for a spectral line with
a full width zero intensity of 200\,MHz (corresponding to an extent of
520\,\kms), the fractional change in the Planck function across the
spectral line is less than 0.3\% for all excitation temperatures above the
CMB.  For most sources within the Milky Way Galaxy, the extent of a
spectral line is an order of magnitude smaller and the corresponding
change in the Planck function across the transition is also smaller.  The
fractional uncertainty in the Planck function increases as the
frequency of the transition increases; however, even at 1\,THz, the
fractional uncertainty is still less than $3$\% for a 500\,km/s wide
transition.  As a result, the $J_{\nu}(T_{ex}) - J_{\nu}(T_{bg})$ term is
approximately constant in all realistic cases and the integral term in
Equation~\ref{eq:ntotthin0} reduces to just the integrated intensity of
the spectral line:
\begin{multline}
N^{thin}_{tot} = \left(\frac{3h}{8 \pi^3 S \mu^2
  R_i}\right)\left(\frac{Q_{rot}}{g_J g_K g_I}\right)
\frac{\exp\left(\frac{E_u}{kT_{ex}}\right)}{\exp\left(\frac{h\nu}{kT_{ex}}\right)
  - 1} \\
\times\frac{1}{\left(J_\nu(T_{ex}) - J_\nu(T_{bg})\right)}
  \int\frac{T_R dv}{f}.
\label{eq:ntotthin}
\end{multline}

\vfill\eject
\subsection{Optically Thin with Rayleigh-Jeans Approximation}
\label{ThinRayleigh}

\begin{figure}
\centering
\includegraphics[scale=0.43]{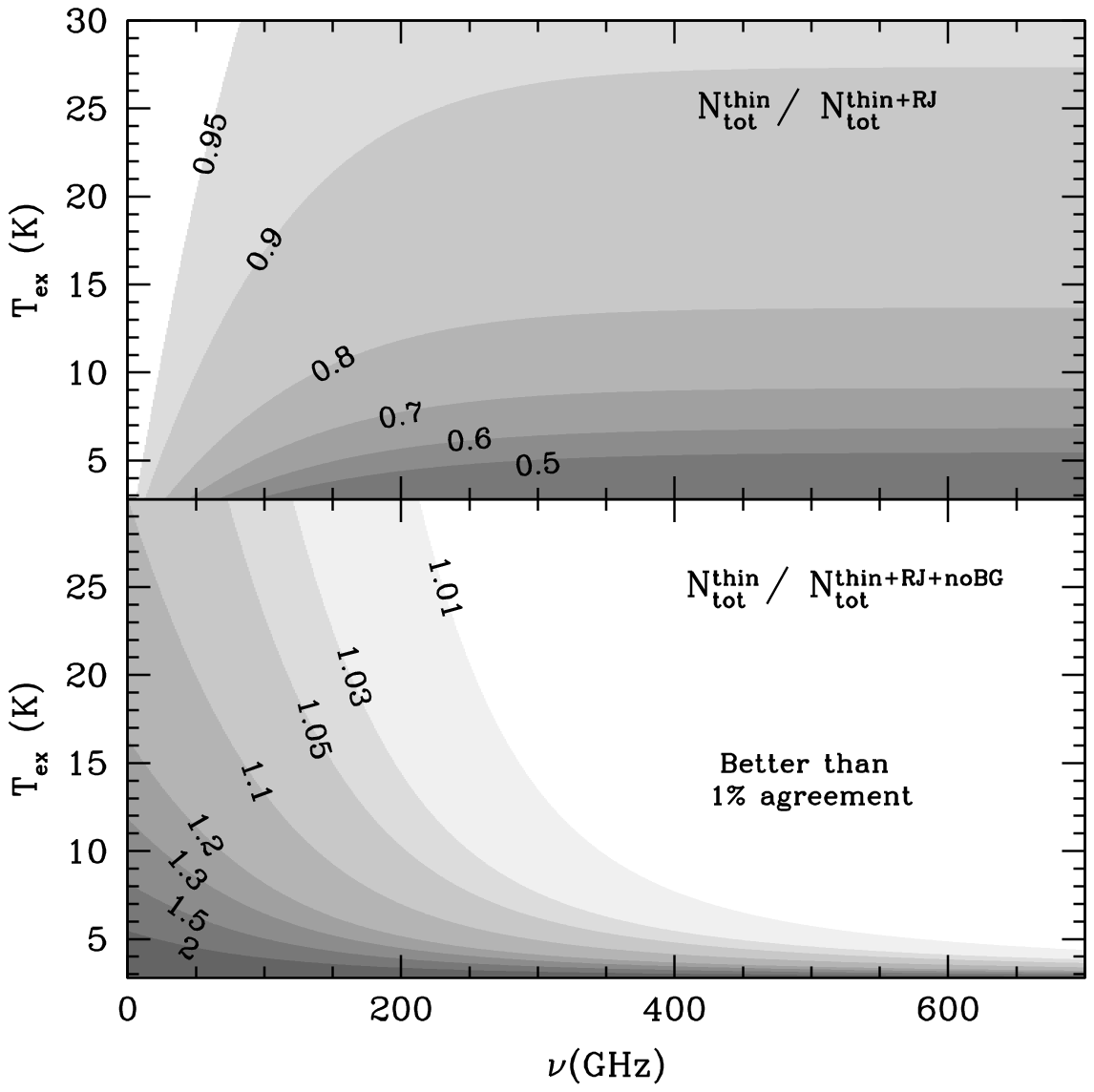}
\caption{In the top panel we compare the molecular column density
  calculated assuming optically-thin ($N^{thin}_{tot}$;
  Equation~\ref{eq:ntotthin}) and optically-thin within the
  Rayleigh-Jeans approximation ($N^{thin+RJ}_{tot}$;
  Equation~\ref{eq:ntotthinrayleigh}).  In the bottom panel we show
  the comparison between molecular column density calculated assuming
  optically-thin and optically-thin within the Rayleigh-Jeans
  approximation and with the assumption of negligible background
  emission ($N^{thin+RJ+noBG}_{tot}$: Equation~\ref{eq:ntotthinrayleighnobg}).}
\label{fig:ColumnDensityThinComparison}
\end{figure}

Assume that $\tau\ll 1$ and $h\nu\ll kT_{ex}$.  This reduces the term
in $[~]$ in Equation \ref{eq:ntotfinal} to $\frac{h\nu}{kT_{ex}}$,
$J_\nu(T)$ to $T$, and reduces the radiative transfer equation to that
derived in Equation~\ref{eq:trthin}. Equation~\ref{eq:ntotfinal} then
reduces to
\begin{multline}
N^{thin+RJ}_{tot} = \left(\frac{3k}{8 \pi^3 \nu S \mu^2
    R_i}\right)\left(\frac{Q_{rot}}{g_J g_K g_I}\right)
  \exp\left(\frac{E_u}{kT_{ex}}\right) \\
  \times\int\frac{T_R T_{ex} dv}{f\left(T_{ex} - T_{bg}\right)}.
\label{eq:ntotthinrayleigh}
\end{multline}
We test the validity of these approximations by comparing
the ratio of Equation~\ref{eq:ntotthin} ($N_{tot}^{thin}$) to
Equation~\ref{eq:ntotthinrayleigh} ($N_{tot}^{thin+RJ}$) in
Figure~\ref{fig:ColumnDensityThinComparison}.  The column density 
calculated from Equation~\ref{eq:ntotthin} is always less than the column
density calculated from Equation~\ref{eq:ntotthinrayleigh} in the
Rayleigh-Jeans limit.   Darker shading corresponds to worse agreement
between the equations. For example, the Rayleigh-Jeans approximation
is a good approximation (less than 5\% disagreement between equations)
for the low frequency (4.8\,GHz) H$_2$CO K-doublet transition for all
excitation temperatures.   The Rayleigh-Jeans approximation is still
a good approximation for the NH$_3$ inversion transitions
(\eg\. (1,1), (2,2), \etc) at 24\,GHz with less than 10\% disagreement
between the equations for $T_{ex} > 5$\,K.   Rayleigh-Jeans failure is
more apparent for transitions at millimeter and shorter wavelengths.
For instance, the C$^{18}$O $J=1-0$ line at 109.7\,GHz has more than
15\% disagreement between the optically-thin and optically-thin plus
Rayleigh-Jeans approximations for $T_{ex} < 9$\,K that rapidly
increases toward lower excitation temperatures.  It is possible to use
Equation~\ref{eq:ntotthinrayleigh} at submillimeter wavelengths $\nu >
300$\,GHz if the excitation temperature is large (\ie\ less than 10\%
disagreement at 300\,GHz if $T_{ex} > 26$\,K); however, for most cases
in dense regions of molecular clouds where the tracer is sub-thermally
populated with low excitation temperatures, Equation~\ref{eq:ntotthin}
should be used.

%
\subsection{Optically Thin with Rayleigh-Jeans Approximation and
  Negligible Background}
\label{ThinRayleighNoBG}

Assuming that the temperature of the background source
(\ie\ the cosmic microwave background radiation) is small in
comparison to the molecular excitation temperature ($T_{bg}\ll
T_{ex}$), Equation \ref{eq:ntotthinrayleigh} becomes:
\begin{multline}
N^{thin+RJ+noBG}_{tot} = \left(\frac{3k}{8 \pi^3 \nu S \mu^2
  R_i}\right)\left(\frac{Q_{rot}}{g_J g_K g_I}\right) \\
  \times\exp\left(\frac{E_u}{kT_{ex}}\right)
  \int \frac{T_R dv}{f}.
\label{eq:ntotthinrayleighnobg}
\end{multline}
We test when it is appropriate to use this approximation by comparing
the column density calculated with Equation~\ref{eq:ntotthin}
($N_{tot}^{thin}$) versus Equation~\ref{eq:ntotthinrayleighnobg}
($N_{tot}^{thin+RJ+noBG}$).  The ratio of these two equations is given by
$N_{tot}^{thin}/N_{tot}^{thin+RJ+noBG} = J_{\nu}(T_{ex}) /
[J_{\nu}(T_{ex}) - J_{\nu}(T_{bg})]$.
Figure~\ref{fig:ColumnDensityThinComparison} shows contours of this
ratio as a function of $T_{ex}$ and the transition frequency.   The
general trend is such that the thin$+$RJ$+$noBG approximation is
poor for low frequencies and for low $T_{ex}$.  At first, it seems
counterintuitive that agreement between the two approximations would
be better at high frequencies when failure of the Rayleigh-Jeans
limit is more likely to occur (see Section~\ref{ThinRayleigh});
however, ignoring the background term in
Equation~\ref{eq:ntotthinrayleighnobg} compensates for
Rayleigh-Jeans failure.  For high frequency and for $T_{ex} >>
T_{bg}$, the $J_{\nu}(T_{bg})$ term in the equation ratio becomes
small such that $N_{tot}^{thin}/N_{tot}^{thin+RJ+noBG} \rightarrow 1$.
The disagreement between the two equations is worse at smaller $T_{ex}$.

As a first example, we consider the NH$_3$ (1,1) lowest inversion
transition at 23.7\,GHz.  At $T_{ex} = 10$\,K, a common value found in
NH$_3$ studies \citep[see][]{Rosolowsky2008},  the thin$+$RJ$+$noBG
approximation results in a systematic 30\% underestimate of the column
density.  At these frequencies, the CMB cannot be ignored.  This
result indicates that a column density expression with background
(Equation~\ref{eq:ntotthin} or \ref{eq:ntotthinrayleigh}) should be
used to calculate column densities from NH$_3$ (1,1) observations.  In
contrast, the thin+RJ+noBG approximation is more applicable at higher
frequencies.  For the HCO$^+$ $J=3-2$ transition at 267.5\,GHz,
Equation~\ref{eq:ntotthinrayleighnobg} (thin+RJ+noBG approximation)
has better than 10\% accuracy for $T_{ex} > 5.4$\,K.
Equation~\ref{eq:ntotthinrayleighnobg} may be used for submillimeter
transitions ($\nu > 300$\,GHz) unless the level populations for
the transition are severely sub-thermally populated ($T_{ex} < 5$\,K).

\vfill\eject
\subsection{Not Optically Thin Approximation\protect\footnote{The
    condition $\tau\gtrsim 1$ is often somewhat
  erroneously referred to as ``optically thick'', while in fact the
  analysis presented in this section is appropriate for all conditions
  \textit{excluding} $\tau\ll 1$.}}
\label{NotThin}

If we assume that $\tau_\nu\gtrsim 1$, we use the radiative transfer 
equation (Equation~\ref{eq:tr}) to solve for $\tau_\nu$:
\begin{equation}
\tau_\nu = -\ln\left[1 - \frac{T_R}{f\left[J(T_{ex}) - J(T_{bg})\right]}\right],
\label{eq:tauthick}
\end{equation}
and insert this expression for $\tau_\nu$ in the column
density equation (Equation \ref{eq:ntotfinal}):
\begin{multline}
N^{thick}_{tot} = \left(\frac{3h}{8 \pi^3 S \mu^2
  R_i}\right)\left(\frac{Q_{rot}}{g_J g_K g_I}\right)
\frac{\exp\left(\frac{E_u}{kT_{ex}}\right)}{\exp\left(\frac{h\nu}{kT_{ex}}\right)
  - 1} \\
  \times\int -\ln\left[1 - \frac{T_R}{f\left[J(T_{ex}) -
        J(T_{bg})\right]}\right] dv.
\label{eq:ntotgen}
\end{multline}
Note that Equation~\ref{eq:ntotgen} is quite general, applicable to
all values of $\nu$, $\tau$, $T_{ex}$, $T_{bg}$, and $T_R$.
Furthermore, one can derive a simple relationship between the column
density derived assuming $\tau\ll 1$ and those derived assuming
$\tau\nless 1$ by noting that one can write $N^{thin}_{tot}$
(Equation~\ref{eq:ntotthin}) using the radiative transfer equation
(Equation~\ref{eq:tr}) as follows:
\begin{multline}
N^{thin}_{tot} = \left(\frac{3h}{8 \pi^3 S \mu^2
  R_i}\right)\left(\frac{Q_{rot}}{g_J g_K g_I}\right)
\frac{\exp\left(\frac{E_u}{kT_{ex}}\right)}{\exp\left(\frac{h\nu}{kT_{ex}}\right)
  - 1} \\
  \times\int\left(1 - \exp(-\tau)\right) dv.
\label{eq:ntotthin2}
\end{multline}
Therefore, a column density calculated assuming $\tau\ll 1$ is related
to a column density calculated more generally by the
following:
\begin{equation}
N_{tot} = N^{thin}_{tot} \frac{\tau}{1 - \exp(-\tau)}.
\label{eq:ntotthingenrat}
\end{equation}
The factor $\tau/(1-\exp(-\tau))$ is the ``optical depth correction
factor'' derived by \cite{Goldsmith1999}.  Note that
Equation~\ref{eq:ntotthingenrat} also applies to the approximation cases
$h\nu\ll kT_{ex}$ (Section~\ref{ThinRayleigh}) and $T_{bg} \ll T_{ex}$
(\ref{ThinRayleighNoBG}).

\section{The Consequences of Assuming a Constant Excitation
  Temperature}
\label{Tex}

The calculation of column density requires a determination of the
gas excitation temperature (Equation~\ref{eq:tex}).  The excitation
temperature is a very general concept that describes an energy density
(in units of temperature), whether kinetic, radiative, rotational,
vibrational, spin, \etc 

Since molecules are fundamentally multi-level
systems, an excitation temperature 
may be defined for each transition allowed via electric dipole (or magnetic
dipole or electric quadrupole) selection rules\footnote{You can also 
define an ``excitation temperature'' between levels which are not
radiatively coupled - \eg\ the (1,1) and 
(2,2) inversion states of NH$_3$.  This is usually called a rotation
temperature T$_{rot}$.}.  Most molecules (in particular, dense gas tracers) in
the ISM are in non-LTE with different gas excitation temperatures for
different transitions.  However, to simplify the problem of
calculating the total molecular column density from a single
transition, the constant excitation temperature approximation (CTEX)
is usually assumed where all transitions have the same excitation
temperature 
\begin{equation}
T_{ex} \;\; = \;\; T_{ex}^{CTEX} \; \forall \; u \rightarrow l \;\;.
\end{equation}

\begin{figure}
\centering
\includegraphics[scale=0.43]{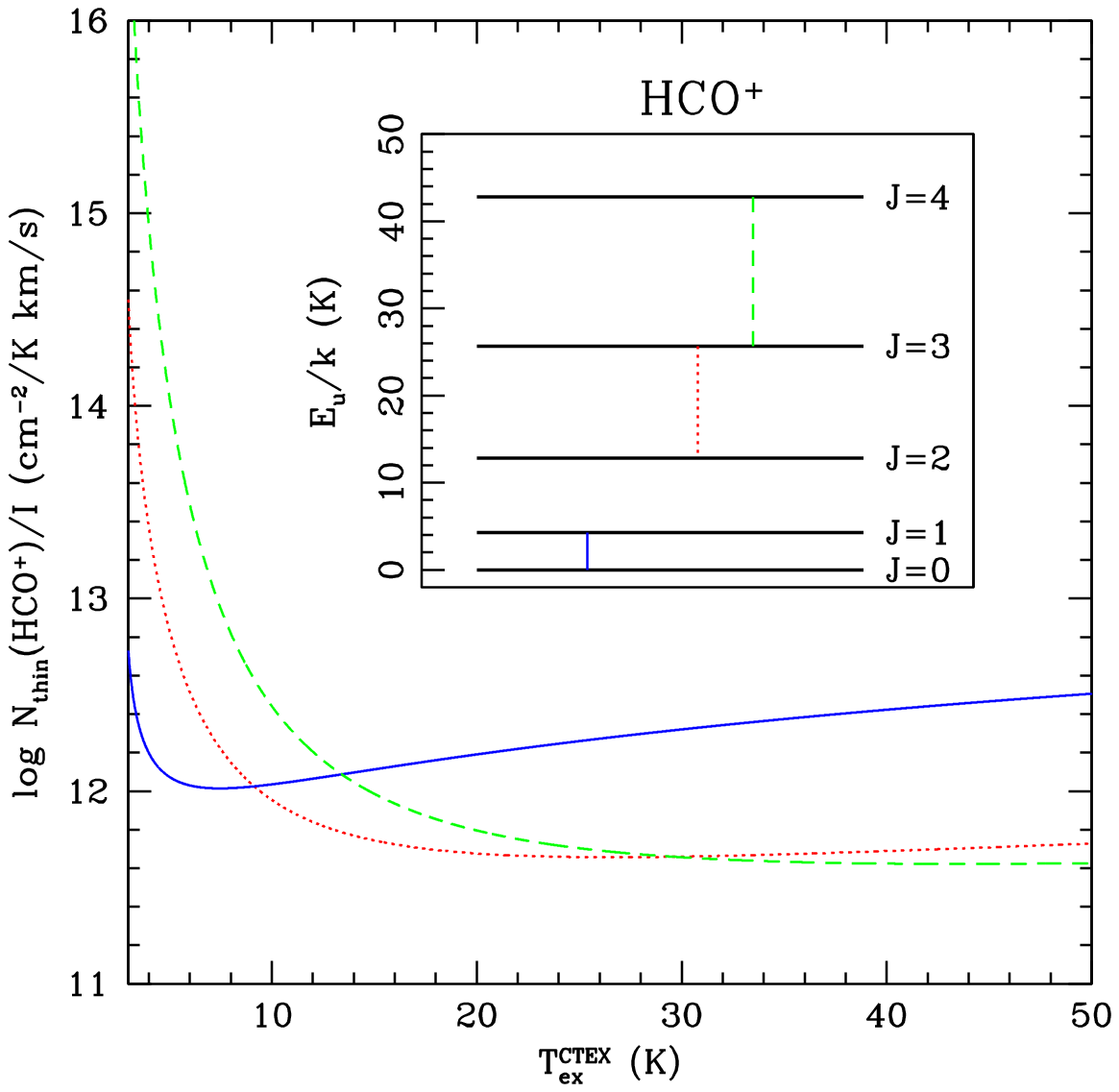}
\caption{The optically thin column density at $1$\,K\,km/s for transitions of
HCO$^+$.  The $J=1-0$ (blue solid line), $J=3-2$
(red dotted line), and $J=4-3$ (green dashed line) are plotted
for different excitation temperatures in the CTEX approximation.  The inset
shows the energy level diagram for the lowest $5$ energy levels of HCO$^+$
and the three transitions easily observable from ground-based observatories.}
\label{fig:HCOPTex}
\end{figure}

The optically thin column densities using Equation~\ref{eq:ntotthin}
are shown in Figure~\ref{fig:HCOPTex} for the three lowest energy
transitions of the HCO$^+$ molecule that are observable from the
ground in atmospheric windows.  Note that the column density equation
diverges for $T_{ex}^{CTEX} \rightarrow 2.735$\,K due to the
$J_{\nu}(T_{ex}) - J_{\nu}(T_{bg})$ 
term in the denominator.  This divergence has the consequence that as the level
populations approach equilibrium with the CMB, it becomes more difficult
to accurately determine the column density.  The plotted curves
show $N_{thin}/I$ or the column density in the optically thin limit
for an integrated intensity of $1$\,K\,km/s.  These curves should be
multiplied by the observed integrated intensity and corrected for
any optical depth effects by $\tau/(1 - \exp(-\tau))$.
The $1-0$ transition is the least sensitive transition to
changes in the excitation temperature above $3$\,K with
a minimum at $T_{ex}^{CTEX} = 7.46$\,K.  As a result, the difference
between assuming an excitation temperature between $5$\,K and $10$\,K
results in a less than $10$\% uncertainty in the calculated column
density assuming CTEX.  This statement is in general true for the
$3$\,mm transitions of several famous dense gas tracers such as
HCO$^+$, HCN, N$_2$H$^+$, and CS.  If a lower limit to the column
density is desirable, then the column density can be calculated at the
excitation temperature of the minimum excitation temperature curve
(\eg\ $N_{thin}/I > 1.03 \times 10^{12}$\,cm$^{-2}$ / K\,km/s for 
the HCO$^+$ $J=1-0$ transition at $T_{ex}^{CTEX} = 7.46$\,K.)

Estimating the total column density from higher energy transitions is
more uncertain for low excitation temperatures.  Unfortunately, this
situation does occur in observations of dense cores in the ISM where
the dense gas tracer is typically both optically thick and very sub-thermally
populated \citep[see][]{Shirley2013}.  For the HCO$^+$ $J=4-3$ 
transition, the uncertainty in the total column density calculation
between assuming an excitation temperature of $5$\,K versus $10$\,K is
a factor of $40$!  The $J=4-3$ curve does not reach its
minimum until $T_{ex}^{CTEX} = 43.55$\,K, close to the energy of the
$J = 4$ level above ground ($E_u/k = 42.80$\,K).  It is therefore
advisable to observe transitions with $E_u/k$ that are comparable to
the typical excitation temperature to minimize this uncertainty.  

Proper estimation of the gas excitation temperature usually requires
observations of multiple transitions coupled with modeling of the
statistical equilibrium and radiative transfer.  In general, the
excitation temperature is a sigmoid function that varies between the
background radiation temperature at low density and the gas kinetic
temperature at high density.  At low gas densities, collisions are not
as important for setting the level populations as interaction with the
background radiation field and the excitation temperature equilibrates
with the radiation temperature.  At high gas densities, collisions
dominate in setting the level populations and the excitation
temperature equilibrates to the gas kinetic temperature of the
dominant collisional partner.  A transition is said to be
``thermalized'' if T$_{ex}$ is equal to T$_K$.  Subthermal excitation
indicates that T$_{ex}$ is less than T$_K$.  In the general case of
non-LTE in the ISM, it is possible that some transitions for a
particular molecule may be thermalized (\ie\ CO $J=1-0$), while higher
energy levels are subthermally populated (\ie\ CO $J=6-5$).  Strong
non-LTE effects are also observed in some cases (\ie\ masers where
naturally occurring population inversions can occur).  Most molecular
transitions of dense gas tracers observed in the ISM are subthermally
populated.

Multi-level effects can cause departures
from the sigmoid shape.  For example, a bottleneck due to longer rates
of spontaneous decay, $A_{ul} \sim \nu^3$, for low J transitions can
cause super-thermal excitation in the $J = 1-0$ transition
where, over a narrow range of densities, $T_{ex} > T_{K}$.  A full
discussion of multi-level statistical equilibrium with radiative
transfer is beyond the scope of this tutorial.  There are situations
where the excitation temperature can be directly estimated from the
observations. If it is known that the line is very optically thick,
then the  $(1 - e^{-\tau})$ term in the radiative transfer equation
(Equation~\ref{eq:tr}) approaches 1 and the excitation temperature may
be solved analytically with an assumption of the gas filling factor:
\begin{equation}
T_{ex}^{obs} = \frac{h\nu /k}{\ln\left( 1 + \frac{h\nu /k}{T_{R}/f +
    J_{\nu}(T_{cmb})}\right)} \;\;\;\; (\rm{if} \; \tau >> 1) \;.
\label{eq:texobs}
\end{equation}
Molecular line ratios of well resolved hyperfine transitions or ratios of
isotopologues may be used to determine the optical depth in a line by
taking ratios of Equation~\ref{eq:tr} (see Appendix~\ref{HyperTau}),
although these methods usually assume that the excitation temperature
in each hyperfine line or among different isotopologue transitions are
the same which may not always be true.  

\begin{figure}
\centering
\includegraphics[scale=0.45]{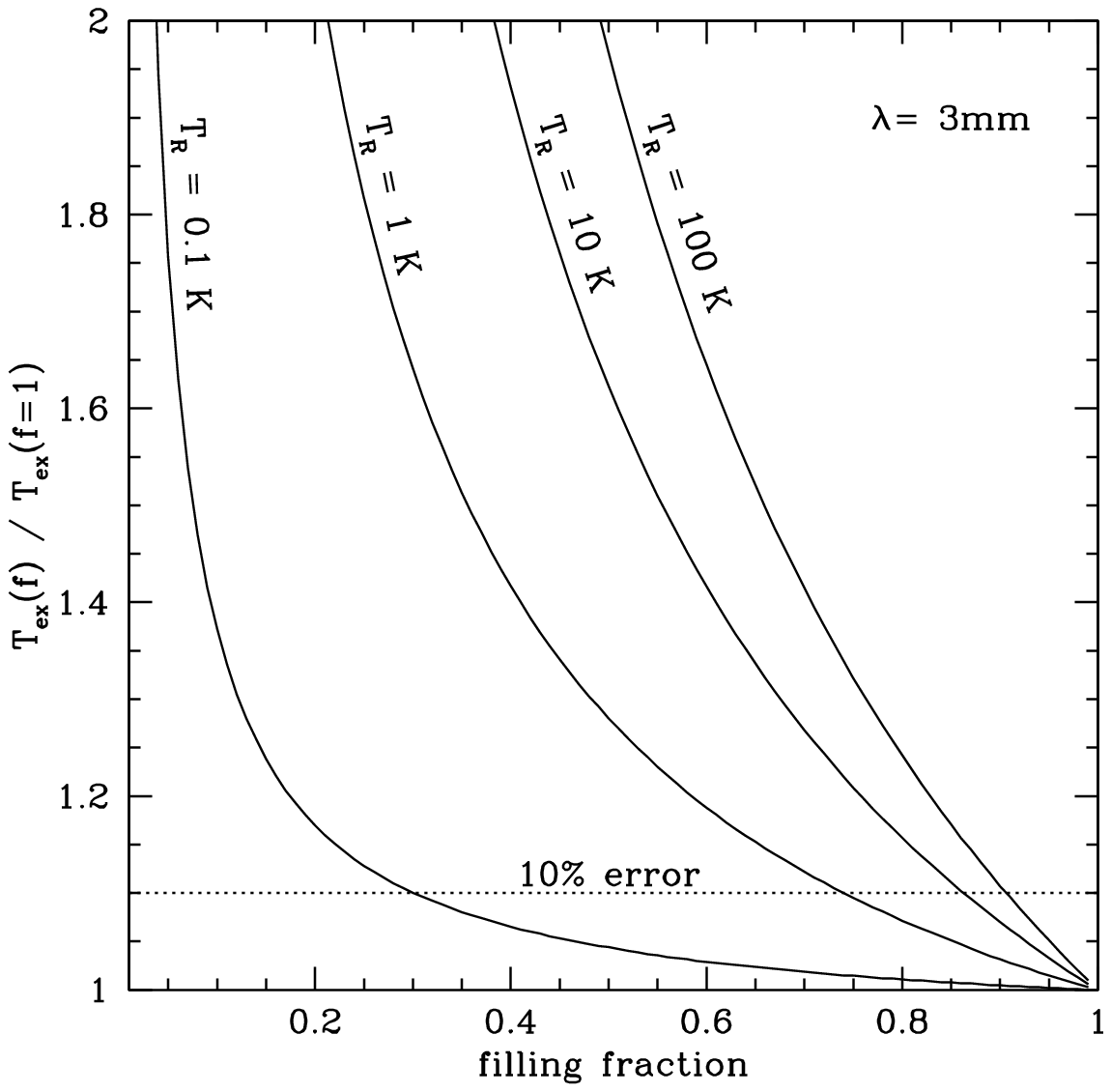}
\caption{The ratio of the excitation temperature at a given gas filling
fraction compared to the excitation temperature calculated assuming a
gas filling fraction $f = 1$.  The four curves correspond to spectral
lines with $T_R = 0.1, 1, 10, 100$\,K.  The curves are calculated for a
transition at 3 mm.  The curves are a weak function of the transition
frequency.  The horizontal dashed line correspond to an error in
$T_{ex}$ of 10\% compared to the $f = 1$ assumption.}
\label{fig:TexPlot}
\end{figure}

Figure~\ref{fig:TexPlot} shows the dependence of Equation~\ref{eq:tr}
on the filling fraction by plotting the ratio of the excitation
temperature at a filling fraction, $f$, divided by the excitation
temperature with a filling fraction of $f = 1$ (the usual assumption
when filling fraction is unknown).  For intense spectral lines $T_R >
1$\,K, calculation of $T_{ex}$ from Equation~\ref{eq:tr} is very
sensitive to the gas filing fraction.  For a line with $T_R = 1$\,K,
$T_{ex}$ is 28\% larger for $f = 0.5$.  $T_{ex}(f)$ quickly increases
toward lower gas filling fractions.  The dependence on gas filling
fraction is steeper for brighter lines and flatter for weaker lines.
The curves plotted in Figure~\ref{fig:TexPlot} are not a strong
function of the wavelength for transitions with $\lambda > 1$\,cm.  

If the line is optically thin, then it is not possible to determine
the excitation temperature from a single transition as the observed
line temperature, $T_R$, is degenerate with excitation temperature,
optical depth, and filling fraction.  However, in this case, the
excitation temperature can still be estimated from multiple
transitions using a rotation diagram.  A detailed discussion of the
limits of the CTEX approximation for HCO$^+$ observations may be found
in \cite{Shirley2013}.

\section{Non-LTE Approaches to the Derivation of Molecular Column
  Density}
\label{NonLTE}

Although beyond the scope of this tutorial, we should mention
that there are a number of non-LTE approaches to the solution of the
coupled multi-level statistical equilibrium and radiative transfer
equations.  These model-based solutions normally rely on the
measurements of a number of molecular spectral lines from a given
species to constrain the physical conditions within the dense gas
environment under study.  If one is interested in the global
physical conditions within a molecular cloud, the radiative transfer
solution for these models is simplified greatly by the introduction
of a geometrically-averaged escape probability (usually denoted by
$\beta$).  A number of escape probability expressions which depend
upon the assumed molecular cloud geometry and global velocity field
can be used in these global cloud models (see \cite{vanderTak2007}
for an excellent discussion).  One of the most popular of these
non-LTE escape probability models incorporates the Large Velocity
Gradient (LVG) approximation \citep{Sobolev1960} to the solution of
the radiative transfer and statistical equilibrium equations.  At
the heart of the LVG approximation is the assumption that photons
which are emitted within the modelled molecular cloud escape quickly
from the cloud due to the large velocity gradient (often assumed to
be a result of cloud collapse).  As described in \cite{Mangum1993}
and \cite{vanderTak2007}, in the LVG approximation the escape
probability is given by $\beta = \frac{1-e^{-\tau}}{\tau}$ within a
uniform collapsing spherical geometry.  The molecular cloud is
assumed to have uniform volume density, kinetic temperature, and
molecular column density.  Inputs to an LVG model include a
specification of the molecular spectrum and collisional excitation
rates involving the primary collision partner in the molecular cloud
(usually assumed to be $H_2$) and the molecular species under study.
Although the LVG approximation entails the use of a rather
simplified uniform physical model, it often produces plausible
physical conditions for environments where the transitions used to
constrain the model are not too optically thick.

Other non-LTE approaches include the generalized escape
probability code \textit{RADEX} \citep{vanderTak2007},
microturbulent velocity models \citep{Leung1976} and Monte Carlo
approaches \citep{Brinch2010}.  In the microturbulent model solution
the molecular cloud is assumed to have 
a power-law distribution in volume density and kinetic temperature
with a line profile that is produced due to a combination of thermal
and turbulent motions.  The radiative transfer equation is
iteratively solved using a quasi-diffusion method.  Monte Carlo
models are generalized solutions to the radiative transfer in
molecular clouds which incorporate up to three dimensions of
arbitrarily complex structure.  This complex structure is organized
into a grid of cells with uniform volume density, kinetic
temperature, and molecular abundance.  By ray tracing along random
paths through this grid the radiative transfer equation is
iteratively solved.

Even though non-LTE approaches result is a more accurate
representation of the physical conditions in a molecular cloud, they
must be constrained by multiple molecular spectral line
measurements from a given molecular species.  Non-LTE models also
require knowledge of the collisional excitation rates for the
molecular species under study.  Collisional excitation rates for a
wide variety of molecules have been calculated (see, for example, the
Leiden Atomic and Molecular Database (LAMDA); \cite{Schoier2005}).
In general, though, limitations to their accuracy and kinetic
temperature range can result in added uncertainty in non-LTE
modelling of radiative transfer.  Due to these limitations, 
LTE calculations of the molecular column density are often
satisfactory with limited measurements.

\section{Beam-Averaged Versus Source-Averaged Column Density}
\label{BeamSourceAverage}

An aspect of the calculation of molecular column density from a
spectral line measurement that is often overlooked is the
difference between beam- and source-averaged measurements.
Beam-averaged molecular column densities are derived from spectral
line measurements with a defined spatial resolution element (a radio
telescope ``beam'' $\theta_{beam}$).  Without an understanding of the actual spatial
extent of the measured emission (\ie\ $\theta_{beam} >
\theta_{source}$), one normally assumes that it is simply averaged
over the beam of the measurement, or ``beam-averaged''.  For
measurements which resolve the spatial extent of emission
(\ie\ $\theta_{beam} < \theta_{source}$), spectral line measurements
measured over a source, usually considered ``extended'', result in
``source-averaged'' molecular column densities.

\section{Molecular Column Density Calculation Examples}
\label{Examples}

In the following we describe in detail some illustrative calculations
of the molecular column density.  These example molecules were chosen
for their commonality (C$^{18}$O is an abundant, though often
optically-thin, isotopomer of CO), relatively simple hyperfine
structure (C$^{17}$O), somewhat more complex hyperfine structure
(N$_2$H$^+$), or general usage as kinetic temperature (NH$_3$) or
volume density (H$_2$CO) probes.  In these examples we adopt the more
conventional notation for a generic (emission or absorption)
transition, \ie\ ``$J=1-0$'', instead of the more explicit (and somewhat
less convenient) ``$J=1\rightarrow 0$'' notation.  We also address a
common point of confusion for students learning this material and its
reliance on CGS (Centimeter-Gram-Second) units.  For the calculations
presented below dipole moments are quoted in the CGS unit ``Debye'',
which is equal to $10^{-18}$ esu cm (electrostatic units times
centimeter).  A good source for molecular dipole moments ($\mu$) and
rotational constants (\ie\ $A_0$, $B_0$, $C_0$) is the Jet Propulsion
Laboratory (JPL) Molecular Spectroscopy database and spectral line
catalog \citep{Pickett1998}\footnote{Available
online at \textit{http://spec.jpl.nasa.gov} (use the ``catalog
directory with links'' to access dipole moment and molecular constants
information).}.

\subsection{C$^{18}$O}
\label{c18o}

To derive the column density for C$^{18}$O from a measurement of its
$J=1-0$ transition we use the general equation for molecular column
density (Equation~\ref{eq:ntotfinal}) with the following properties of
the C$^{18}$O $J=1-0$ transition:
\begin{align}
S &= \frac{J_u}{2J_u + 1} \nonumber \\
\mu &= 0.11079~\textrm{Debye} = 1.1079\times10^{-19}~\textrm{esu cm} \nonumber \\
B_0 &= 54891.420~\textrm{MHz} \nonumber \\
g_J &= 2J_u + 1 \nonumber \\
g_K &= 1~\textrm{(for linear molecules)} \nonumber \\
g_I &= 1~\textrm{(for linear molecules)} \nonumber \\
Q_{rot} &\simeq \frac{kT}{hB}+\frac{1}{3}~\mathrm{(Equation~\ref{eq:qrotlineardiatom})} \nonumber \\
        &\simeq 0.38\left(T+0.88\right) \nonumber \\
E_u &= 5.27~\textrm{K} \nonumber \\
\nu &= 109.782182~\textrm{GHz} \nonumber
\end{align}
which leads to:
\begin{multline}
N_{tot}(C^{18}O) = \frac{3h}{8 \pi^3 \mu^2 J_u R_i}
  \left(\frac{kT_{ex}}{hB} + \frac{1}{3}\right)
  \exp\left(\frac{E_u}{kT_{ex}}\right) \\
  \times\left[\exp\left(\frac{h\nu}{kT_{ex}}\right) - 1\right]^{-1}
  \int\tau_\nu dv.
\label{eq:ntotc18o1}
\end{multline}
Assuming that the emission is optically thin ($\tau_\nu \ll 1$;
Equation~\ref{eq:ntotthin}), Equation~\ref{eq:ntotc18o1} becomes:
\begin{multline}
N_{tot}(C^{18}O) = \frac{2.48\times 10^{14} \left(T_{ex}+0.88\right)
\exp\left(\frac{5.27}{T_{ex}}\right)}{\exp\left(\frac{5.27}{T_{ex}}\right)
  - 1} \\
\times\left[\frac{\int T_R dv(km/s)}{f\left(J_\nu(T_{ex}) - J_\nu(T_{bg})\right)}\right]~\textrm{cm}^{-2}.
\label{eq:ntotc18ot}
\end{multline}

If we are using integrated fluxes ($S_\nu \Delta v$) instead of
integrated brightness temperatures, we use Equation
\ref{eq:ntotfluxden} (see Appendix~\ref{Intflux}), with the assumption
that $\tau\ll 1$):
\begin{align}
N_{tot}(C^{18}O) =& \frac{3c^2}{16\pi^3 \Omega_s S \mu^2 \nu^3}
    \left(\frac{Q_{rot}}{g_J g_K g_I}\right) \nonumber \\
&   \times\exp\left(\frac{E_u}{kT}\right) \int S_\nu \Delta v \nonumber \\
=& \frac{4.77\times 10^{15} \left(T_{ex}+0.88\right)}{\theta_{maj}(asec)
    \theta_{min}(asec)}
\frac{\exp\left(\frac{5.27}{T_{ex}}\right)}{\exp\left(\frac{5.27}{T_{ex}}\right)
  - 1} \nonumber \\
& \times\left[\frac{\int S_\nu(Jy) dv(km/s)}{f\left(J_\nu(T_{ex}) - J_\nu(T_{bg})\right)}\right]~\textrm{cm}^{-2}.
\label{eq:ntotc18os}
\end{align}
where we have assumed a Gaussian source ($\Omega_s = 1.133\theta_{maj}\theta_{min}$)
with $\theta_{maj}$ and $\theta_{min}$
the major and minor source diameters (in arcseconds).

\subsection{C$^{17}$O}
\label{c17o}

\begin{figure}
\centering
\includegraphics[scale=0.60]{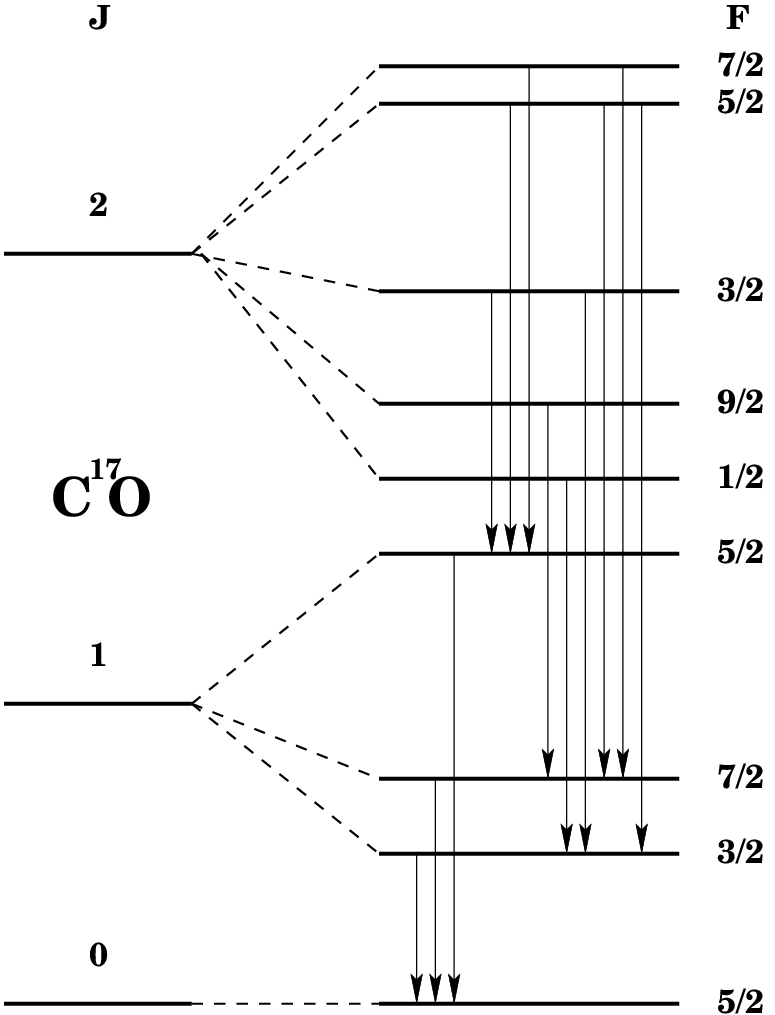}
\vspace{5mm}
\caption{Electric quadrupole hyperfine energy level structure for the
  J=0, 1, and 2 levels of C$^{17}$O.  Note that the 3
  ($J=1-0$) and 9 ($J=2-1$) allowed transitions
  are marked with arrows ordered by increasing frequency from left to
  right.}
\label{fig:c17olevs}
\end{figure}

C$^{17}$O is a linear molecule with hyperfine structure due to
interaction with the electric quadrupole moment of the $^{17}$O (I =
$\frac{5}{2}$) nucleus.  Using the selection rule:
\begin{equation}
F = J + I, J + I - 1, J + I - 2,...,|J - I|, \nonumber
\label{eq:fselect}
\end{equation}
we find that each J-level is split into the hyperfine
levels indicated in Table~\ref{tab:c17ohyperlevs} (for the first five
J-levels).  Since the selection rules for the single-spin coupling
case is, $\Delta F = 0,\pm1$, and $\Delta J = \pm1$, there are 3,
9, and 14 allowed hyperfine transitions for the $J = 1-0$,
$J = 2-1$, and $J = 3-2$ transitions,
respectively.  Figure~\ref{fig:c17olevs} shows the 
energy level structure for the $J=1-0$ and $J=2-1$
transitions.

\begin{deluxetable}{ccl}
\tablewidth{0pt}
\tablecaption{Allowed C$^{17}$O Hyperfine Energy Levels\label{tab:c17ohyperlevs}}
\tablehead{
\colhead{J} & \colhead{Number of Energy Levels} & \colhead{Allowed F}
}
\startdata
0 & 1 & $\frac{5}{2}$ \\
1 & 3 & $\frac{7}{2}, \frac{5}{2}, \frac{3}{2}$ \\
2 & 5 & $\frac{9}{2}, \frac{7}{2}, \frac{5}{2}, \frac{3}{2}, \frac{1}{2}$ \\
3 & 6 & $\frac{11}{2}, \frac{9}{2}, \frac{7}{2}, \frac{5}{2},
\frac{3}{2}, \frac{1}{2}$ \\
4 & 7 & $\frac{13}{2}, \frac{11}{2}, \frac{9}{2}, \frac{7}{2}, \frac{5}{2}, \frac{3}{2}, \frac{1}{2}$
\enddata
\end{deluxetable}

\begin{figure}
\centering
\includegraphics[scale=0.32]{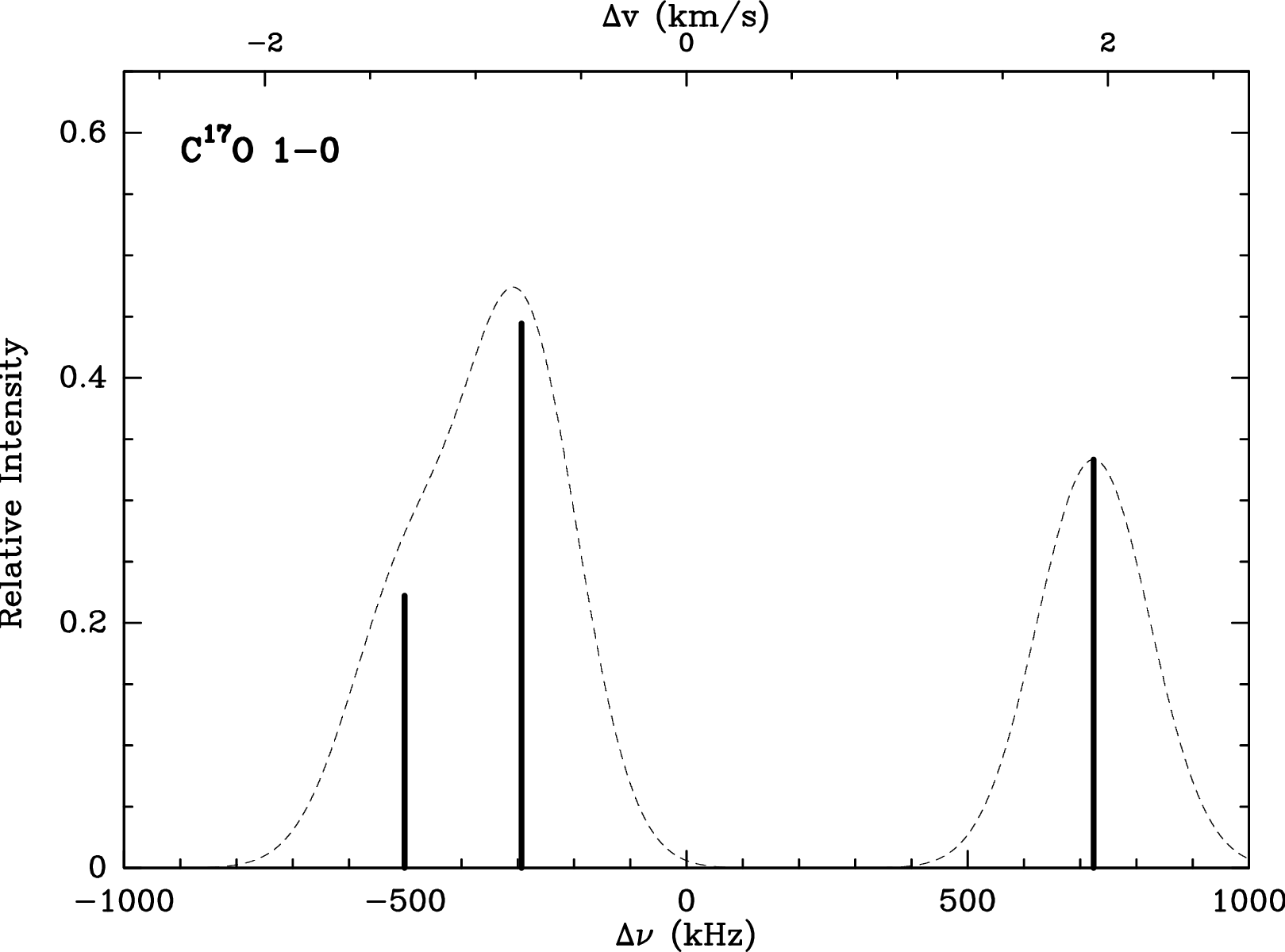} \\[10pt]
\includegraphics[scale=0.32]{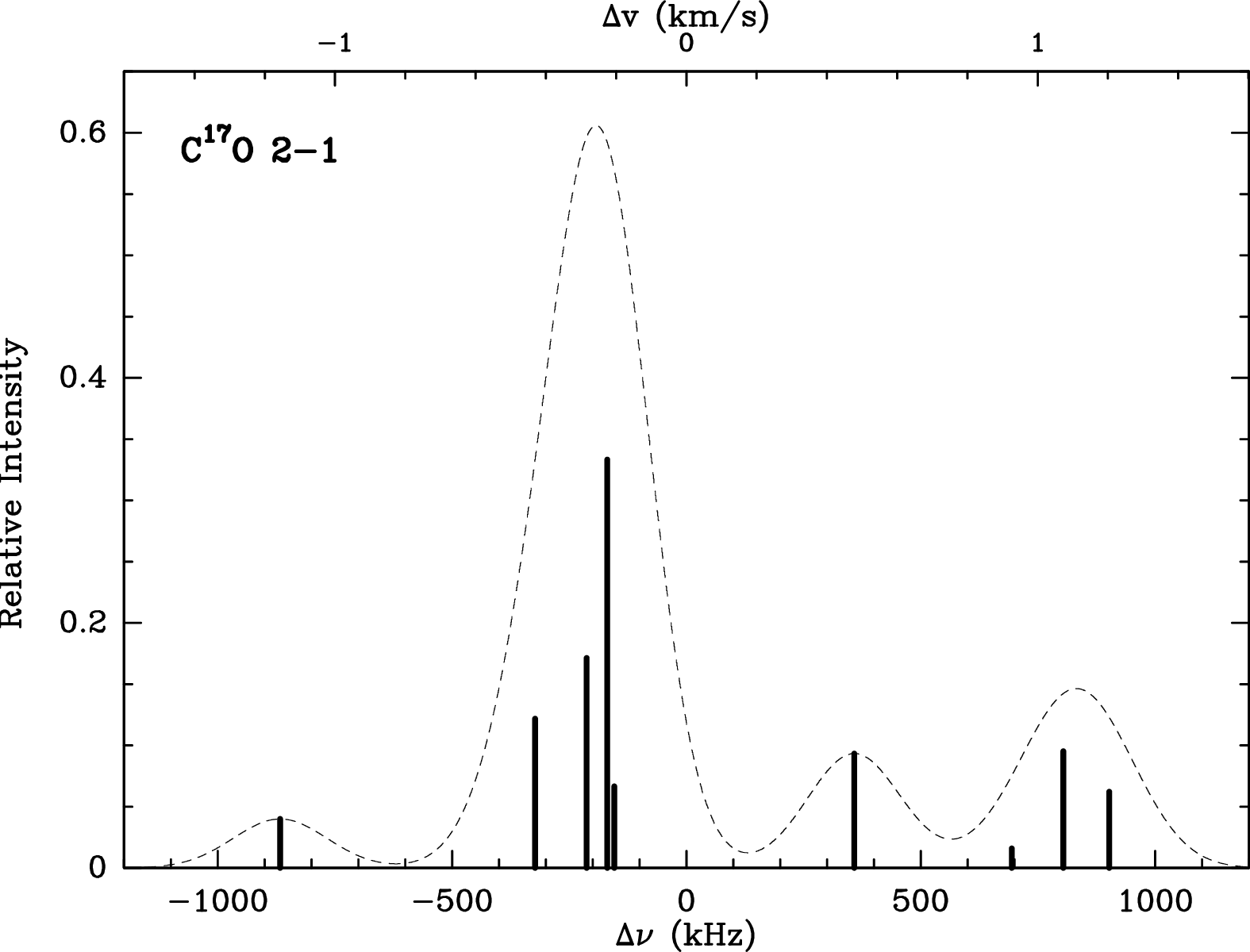}
\caption{Synthetic spectra for the C$^{17}$O $J=1-0$ (top)
  and $2-1$ (bottom) transitions.  Horizontal axes are
  offset velocity (top) and frequency (bottom) relative to 112359285.0
  and 224714386.0 kHz, respectively.  
  Overlain as a dashed line is a
  synthetic 100 kHz gaussian linewidth source spectrum.}
\label{fig:c17orelplt}
\end{figure}

We can calculate the relative hyperfine intensities ($R_i$) for the
$J=1-0$ and $J=2-1$ transitions using the
formalism derived in Section~\ref{Hyperfine}.  Using the formalism
described in Section~\ref{Hyperfine} we
can derive the relevant $R_i$ for the electric quadrupole hyperfine
coupling cases ($R_i(F,J), I=\frac{5}{2}$; Table~\ref{tab:hyperc17o}).
Figure~\ref{fig:c17orelplt} shows the synthetic spectra for the
C$^{17}$O $J=1-0$ and $J=2-1$ transitions.

\begin{deluxetable*}{cccccccccc}
\tablewidth{0pt}
\tablecolumns{10}
\tablecaption{Hyperfine Intensities\tablenotemark{a} for C$^{17}$O
  $J=1-0$ and $J=2-1$\label{tab:hyperc17o}}
\tablehead{
\colhead{$F^\prime \rightarrow F$} & 
\colhead{$J^\prime \rightarrow J$} &
\colhead{a} & \colhead{b} & \colhead{c} &
\colhead{$\frac{(2F^\prime + 1)(2F + 1)}{(2I + 1)}$} &  
\colhead{$\Delta\nu$\tablenotemark{b} (kHz)} & \colhead{6j} &
\colhead{R$_i(F,J)$}
}
\startdata
$(\frac{3}{2},\frac{5}{2})$ & (1,0) & $\frac{5}{2}$ & 
0 & $\frac{5}{2}$ & 4 & $-$501 & $-\frac{1}{3\sqrt{2}}$ &
$\frac{2}{9}$ \\
$(\frac{7}{2},\frac{5}{2})$ & (1,0) & $\frac{5}{2}$ & 
1 & $\frac{7}{2}$ & 8 & $-$293 & $-\frac{1}{3\sqrt{2}}$ &
$\frac{4}{9}$ \\
$(\frac{5}{2},\frac{5}{2})$ & (1,0) & $\frac{5}{2}$ & 
$\frac{5}{2}$ & 1 & 6 & $+$724 & $\frac{1}{3\sqrt{2}}$ &
$\frac{3}{9}$ \\
\tableline
$(\frac{3}{2},\frac{5}{2})$ & (2,1) & $\frac{5}{2}$ & 
1 & $\frac{5}{2}$ & 4 & $-$867 & $\frac{1}{10}$ &
$\frac{1}{25}$ \\
$(\frac{5}{2},\frac{5}{2})$ & (2,1) & $\frac{5}{2}$ & 
$\frac{5}{2}$ & 2 & 6 & $-$323 & $-\frac{4\sqrt{2}}{15\sqrt{7}}$ &
$\frac{64}{525}$ \\
$(\frac{7}{2},\frac{5}{2})$ & (2,1) & $\frac{5}{2}$ & 
2 & $\frac{7}{2}$ & 8 & $-$213 & $\frac{\sqrt{3}}{2\sqrt{35}}$ &
$\frac{6}{35}$ \\
$(\frac{9}{2},\frac{7}{2})$ & (2,1) & $\frac{5}{2}$ &
2 & $\frac{9}{2}$ & $\frac{40}{3}$ & $-$169 & $-\frac{1}{2\sqrt{10}}$ &
$\frac{1}{3}$ \\
$(\frac{1}{2},\frac{3}{2})$ & (2,1) & $\frac{5}{2}$ &
1 & $\frac{3}{2}$ & $\frac{4}{3}$ & $-$154 &
$-\frac{1}{2\sqrt{5}}$ & $\frac{1}{15}$ \\ 
$(\frac{3}{2},\frac{3}{2})$ & (2,1) & $\frac{5}{2}$ &
$\frac{3}{2}$ & 2 & $\frac{8}{3}$ & $+$358 &
$\frac{\sqrt{7}}{10\sqrt{2}}$ & $\frac{7}{75}$ \\
$(\frac{5}{2},\frac{7}{2})$ & (2,1) & $\frac{5}{2}$ &
1 & $\frac{7}{2}$ & 8 & $+$694 & $-\frac{1}{6\sqrt{14}}$ &
$\frac{1}{63}$ \\
$(\frac{7}{2},\frac{7}{2})$ & (2,1) & $\frac{5}{2}$ &
$\frac{7}{2}$ & 2 & $\frac{32}{3}$ & $+$804 & $\frac{1}{4\sqrt{7}}$ & 
$\frac{2}{21}$ \\
$(\frac{5}{2},\frac{3}{2})$ & (2,1) & $\frac{5}{2}$ &
2 & $\frac{5}{2}$ & 4 & $+$902 & $-\frac{\sqrt{7}}{15\sqrt{2}}$ &
$\frac{14}{225}$
\enddata
\tablenotetext{a}{The sum of the relative intensities 
  $\sum_i R_i = 1.0$ for each $\Delta J = 1$ transition.}
\tablenotetext{b}{Frequency offsets in kHz relative to 112359.285 and
224714.368\,MHz for $J=1-0$ and $J=2-1$, respectively (from the
spectroscopic constants in \cite{Klapper2003}).}
\end{deluxetable*}

To derive the column density for C$^{17}$O from a measurement of its
$J=1-0$ transition we use the general
equation for molecular column density (Equation~\ref{eq:ntotfinal})
with the following properties of the C$^{17}$O $J=1-0$
F=$\frac{5}{2}-\frac{5}{2}$ transition:
\begin{eqnarray}
S &=& \frac{J_u}{2J_u + 1} \nonumber \\
\mu &=& 0.11032~\textrm{Debye} = 1.1032\times10^{-19}~\textrm{esu cm} \nonumber \\
B_0 &=& 56179.9911~\textrm{MHz} \nonumber \\
g_J &=& 2J_u + 1 \nonumber \\
g_K &=& 1~\textrm{(for linear molecules)} \nonumber \\
g_I &=& 1~\textrm{(for linear molecules)} \nonumber \\
Q_{rot} &\simeq&
\frac{kT}{hB}+\frac{1}{3}~\textrm{(Equation~\ref{eq:qrotlineardiatom})}
\nonumber \\
        &\simeq& 0.37\left(T+0.90\right) \nonumber \\
E_u &=& 5.39~\textrm{K} \nonumber \\
R_i &=& \frac{3}{9} \nonumber
\end{eqnarray}
which leads to:
\begin{multline}
N_{tot}(C^{17}O) = \frac{3h}{8 \pi^3 \mu^2 J_u R_i}
  \left(\frac{kT_{ex}}{hB} + \frac{1}{3}\right)
  \exp\left(\frac{E_u}{kT_{ex}}\right) \\
  \times\left[\exp\left(\frac{h\nu}{kT_{ex}}\right) - 1\right]^{-1}
  \int\tau_\nu dv.
\label{eq:ntotc17o1}
\end{multline}
Assuming that the emission is optically thin ($\tau_\nu \ll 1$;
Equation~\ref{eq:ntotthin}), Equation~\ref{eq:ntotc17o1} becomes:
\begin{multline}
N_{tot}(C^{17}O) = \frac{2.44\times 10^{14}\left(T_{ex}+0.90\right)}{R_i}
\frac{\exp\left(\frac{E_u}{kT_{ex}}\right)}{\exp\left(\frac{h\nu}{kT_{ex}}\right)
  - 1} \\
\times\left[\frac{\int T_R dv(km/s)}{f\left(J_\nu(T_{ex}) -
    J_\nu(T_{bg})\right)}\right]~\textrm{cm}^{-2},
\label{eq:ntotc17ot}
\end{multline}
where $\nu$ is the frequency of the hyperfine transition
used.  For example, if the F=$\frac{5}{2}-\frac{5}{2}$ hyperfine was
chosen for this calculation, $R_i = \frac{3}{9}$ (See
Table~\ref{tab:hyperc17o}) and $\nu = 112359.285 + 0.724~\textrm{MHz}
= 112.360009$\,GHz.  Equation~\ref{eq:ntotc17ot} then becomes:
\begin{multline}
N_{tot}(C^{17}O) = 7.32\times 10^{14}\left(T_{ex}+0.90\right)
\frac{\exp\left(\frac{5.39}{T_{ex}}\right)}{\exp\left(\frac{5.39}{T_{ex}}\right)
  - 1} \\
\times\left[\frac{\int T_R dv(km/s)}{f\left(J_\nu(T_{ex}) -
    J_\nu(T_{bg})\right)}\right]~\textrm{cm}^{-2}.
\label{eq:ntotc17ot1}
\end{multline}

\subsection{N$_2$H$^+$}
\label{n2hp}

\begin{figure}
\centering
\includegraphics[scale=0.45]{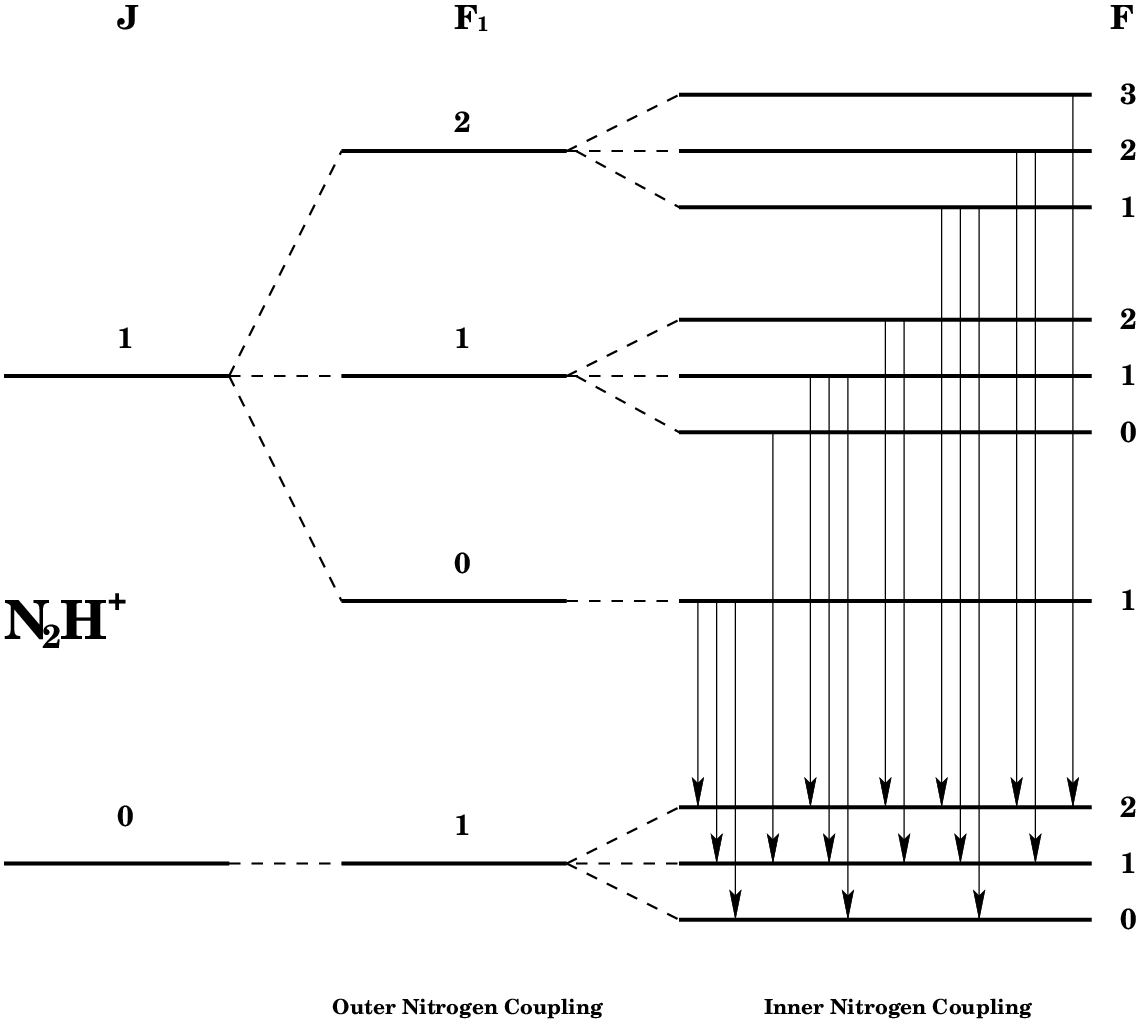}
\vspace{5mm}
\caption{Energy level structure for the $J=1-0$ transition
  of N$_2$H$^+$.  Note that of the 15 hyperfine split levels only 7
  are observed due to the fact that there is no hyperfine splitting of
  the J=0 level, but it is degenerate in energy.  Grouping of the
  indicated transitions show the 
  7 observed transitions.  Transitions are ordered by increasing
  frequency from left to right.}
\label{fig:n2hp10levs}
\end{figure}

N$_2$H$^+$ is a multiple spin coupling molecule due to the interaction
between its spin and the quadrupole moments of the two nitrogen
nuclei.  For a nice detailed description of the hyperfine levels of
the $J = 1-0$ transition see \cite{Shirley2005,Pagani2009}.  Since the
outer N nucleus has a much larger coupling strength than the inner N
nucleus, the hyperfine structure can be determined by a sequential
application of the spin coupling:
\begin{eqnarray}
\vec{F_1} &=& \vec{J} + \vec{I_N}, \nonumber \\
\vec{F} &=& \vec{F_1} + \vec{I_N}. \nonumber
\end{eqnarray}
When the coupling from both N nuclei is considered:
\begin{itemize}
\item The $J = 0$ level is split into 3 energy levels,
\item The $J = 1$ level is split into 7 energy levels,
\item The $J = 2$ and higher levels are split into nine energy levels.
\end{itemize}
Since the selection rules for the single-spin coupling
case apply, $\Delta F_1 = 0,\pm1$, $\Delta F = 0,\pm1$,
$0\nrightarrow0$, and 
$\Delta J = \pm1$, there are 15, 34, and 38 hyperfine
transitions for the $J = 1-0$, $J = 2-1$, and $J = 3-2$ transitions,
respectively.  Figure~\ref{fig:n2hp10levs} shows the energy level
structure for the $J=1-0$ transition. 

\begin{figure}
\centering
\includegraphics[scale=0.32]{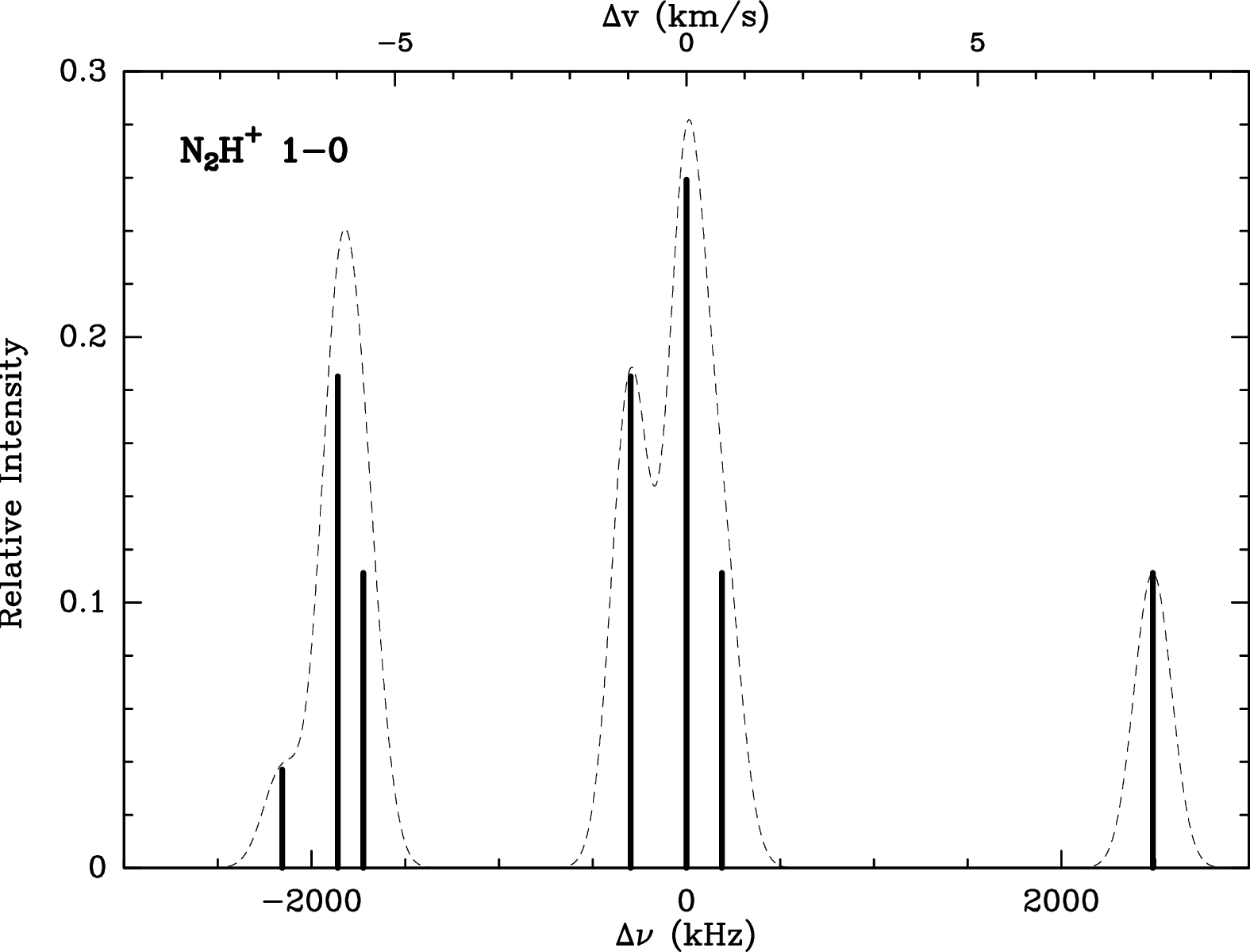}
\caption{Synthetic spectra for the N$_2$H$^+$ $J=1-0$
  transition.  Horizontal axes are offset velocity (top) and frequency
  (bottom) relative to 93173.7767\,MHz.  
  Overlain as a dashed line is a synthetic 100\,kHz ($\Delta v =
  0.3218$\,\kms) gaussian linewidth source spectrum.}
\label{fig:n2hp10relplt}
\end{figure}

To illustrate the hyperfine intensity calculation for N$_2$H$^+$, we
derive the relative intensities for the $J=1-0$ transition.
Relative intensities, derived using the formalism presented in
  Section~\ref{Hyperfine} and Equation~\ref{eq:hyperrel}, are listed in
Tables~\ref{tab:sixjn2hp10fp}, \ref{tab:sixjn2hp10f}, and
\ref{tab:hypern2hp10}. Figure~\ref{fig:n2hp10relplt} shows the
synthetic spectrum for the N$_2$H$^+$ $J=1-0$ transition.

\begin{deluxetable}{cccccccc}
\tablewidth{0pt}
\tablecolumns{8}
\tablecaption{Outer Nitrogen ($F_1$) Hyperfine Intensities for N$_2$H$^+$
  $J=1-0$\label{tab:sixjn2hp10fp}}
\tablehead{
\colhead{$F^\prime_1 \rightarrow F_1$\tablenotemark{a}} &
\colhead{$J^\prime \rightarrow J$\tablenotemark{a}} &
\colhead{a} & \colhead{b} & \colhead{c} &
\colhead{$\frac{(2F^\prime_1 + 1)(2F_1 + 1)}{(2I_N + 1)}$} &
\colhead{6j} & \colhead{R$_i(F_1,J)$}
}
\startdata
(0,1) & (1,0) & 1 & 0 & 1 & 1 & $\frac{1}{3}$ & $\frac{1}{9}$ \\
(1,1) & (1,0) & 1 & 1 & 1 & 3 & $-\frac{1}{3}$  & $\frac{1}{3}$ \\
(2,1) & (1,0) & 1 & 1 & 2 & 5 & $\frac{1}{3}$  & $\frac{5}{9}$
\enddata
\tablenotetext{a}{$I_N = 1$.}
\end{deluxetable}

\begin{deluxetable*}{cccccccc}
\tablewidth{0pt}
\tablecaption{Inner Nitrogen ($F$) Hyperfine Intensities for N$_2$H$^+$
  $J=1-0$\label{tab:sixjn2hp10f}}
\tablehead{
\colhead{$F^\prime \rightarrow F$\tablenotemark{a}} & 
\colhead{$F^\prime_1 \rightarrow F_1$\tablenotemark{a}} & 
\colhead{a} & \colhead{b} & \colhead{c} &
\colhead{$\frac{(2F^\prime + 1)(2F + 1)}{(2I_N + 1)}$} & 
\colhead{6j} & \colhead{R$_i(F,F_1)$}
}
\startdata
(1,0) & (0,1) & 1 & 0 & 1 & 1 & $\frac{1}{3}$ & $\frac{1}{9}$ \\

(1,1) & (0,1) & 1 & 1 & 1 & 3 & $-\frac{1}{3}$ & $\frac{1}{3}$ \\

(1,2) & (0,1) & 1 & 1 & 2 & 5 & $\frac{1}{3}$ & $\frac{5}{9}$ \\

(0,1) & (1,1) & 1 & 1 & 1 & 1 & $\frac{1}{3}$ & $\frac{1}{9}$ \\

(1,0) & (1,1) & 1 & 1 & 1 & 1 & $-\frac{1}{3}$ & $\frac{1}{9}$ \\

(1,1) & (1,1) & 1 & 1 & 1 & 3 & $\frac{1}{6}$ & $\frac{1}{12}$ \\

(1,2) & (1,1) & 1 & 1 & 2 & 5 & $\frac{1}{6}$ & $\frac{5}{36}$ \\

(2,1) & (1,1) & 1 & 1 & 2 & 5 & $\frac{1}{6}$ & $\frac{5}{36}$ \\

(2,2) & (1,1) & 1 & 2 & 1 & $\frac{25}{3}$ 
                              & $-\frac{1}{2\sqrt{5}}$ & $\frac{5}{12}$ \\

(1,0) & (2,1) & 1 & 2 & 1 & 1 & $\frac{1}{3}$ & $\frac{1}{9}$ \\

(1,1) & (2,1) & 1 & 1 & 2 & 3 & $\frac{1}{6}$ & $\frac{1}{12}$ \\

(1,2) & (2,1) & 1 & 1 & 2 & 5 & $\frac{1}{30}$ & $\frac{1}{180}$ \\

(2,1) & (2,1) & 1 & 2 & 2 & 5 & $-\frac{1}{2\sqrt{5}}$ & $\frac{1}{4}$ \\

(2,2) & (2,1) & 1 & 2 & 2 & $\frac{25}{3}$ 
                              & $-\frac{1}{10}$ & $\frac{1}{12}$ \\

(3,2) & (2,1) & 1 & 2 & 3 & $\frac{35}{3}$ 
                              & $\frac{1}{5}$ & $\frac{7}{15}$
\enddata
\tablenotetext{a}{$I_N = 1$.}
\end{deluxetable*}

\begin{deluxetable*}{cccccc}
\tablewidth{0pt}
\tablecaption{Hyperfine Intensities\tablenotemark{a} for N$_2$H$^+$ J=$1-0$\label{tab:hypern2hp10}}
\tablehead{
\colhead{$F^\prime \rightarrow F$\tablenotemark{b}} & 
\colhead{$F^\prime_1 \rightarrow F_1$\tablenotemark{b}} & 
\colhead{$J^\prime \rightarrow J$} &
\colhead{R$_i(F_1,J)$R$_i(F,F_1)$} &
\colhead{$\Delta\nu$\tablenotemark{b} (kHz)} &
\colhead{R$_i$(obs)\tablenotemark{c}}
}
\startdata
(0,1) & (1,1) & (1,0) & $\frac{1}{27}$ & $-2155.7$ & $\frac{1}{27}$ \\
\tableline
(2,2) & (1,1) & (1,0) & $\frac{5}{36}$ & $-1859.9$ & $\frac{5}{27}$ \\
(2,1) & (1,1) & (1,0) & $\frac{5}{108}$ & & \\
\tableline
(1,2) & (1,1) & (1,0) & $\frac{5}{108}$ & $-1723.4$ & $\frac{1}{9}$ \\
(1,1) & (1,1) & (1,0) & $\frac{1}{36}$ & & \\
(1,0) & (1,1) & (1,0) & $\frac{1}{27}$ & & \\
\tableline
(2,1) & (2,1) & (1,0) & $\frac{5}{36}$ & $-297.1$ & $\frac{5}{27}$ \\
(2,2) & (2,1) & (1,0) & $\frac{5}{108}$ & & \\
\tableline
(3,2) & (2,1) & (1,0) & $\frac{7}{27}$ & $+0.0$ & $\frac{7}{27}$ \\
\tableline
(1,1) & (2,1) & (1,0) & $\frac{5}{108}$ & $+189.9$ & $\frac{1}{9}$ \\
(1,2) & (2,1) & (1,0) & $\frac{1}{324}$ & & \\
(1,0) & (2,1) & (1,0) & $\frac{5}{81}$ & & \\
\tableline
(1,2) & (0,1) & (1,0) & $\frac{5}{81}$ & $+2488.3$ & $\frac{1}{9}$ \\
(1,1) & (0,1) & (1,0) & $\frac{1}{27}$ & & \\
(1,0) & (0,1) & (1,0) & $\frac{1}{81}$ & &
\enddata
\tablenotetext{a}{The sum of the relative intensities
  $\sum_i R_i = 1.0$.}
\tablenotetext{b}{Frequency offset in kHz relative to 93173.7767\,MHz \citep{Caselli1995}.} 
\tablenotetext{c}{Since the J=0 level has no hyperfine splitting, only
  the sum of all transitions into the J=0 is observed.}
\end{deluxetable*}

To derive the column density for N$_2$H$^+$ we start with
the general equation for the total molecular column density
(Equation~\ref{eq:ntotfinal}) with:
\begin{eqnarray}
S &=& \frac{J_u}{2J_u + 1} \textrm{ (see Section~\ref{SymtopsandLinP})}
\nonumber \\
\mu &=& 3.40~\textrm{Debye} = 3.40\times10^{-18}~\textrm{esu cm} \nonumber \\
B_0 &=& 46586.88~\textrm{MHz} \nonumber \\
R_i &=& \textrm{(see Section~\ref{Hyperfine} or, for J=1-0, see
  Table~\ref{tab:hypern2hp10})} \nonumber \\
g_J &=& 2J_u + 1 \nonumber \\
g_K &=& 1 \textrm{ (for linear molecules)} \nonumber \\
g_I &=& 1 \textrm{ (for linear molecules)} \nonumber \\
Q_{rot} &\simeq& \frac{kT}{hB}+\frac{1}{3}
~(\mathrm{Equation}~\ref{eq:qrotlineardiatom}) \nonumber \\ 
        &\simeq& 0.45\left(T+0.75\right) \nonumber \\
E_u &=& 4.4716~\textrm{K} \nonumber
\end{eqnarray}
which leads to:
\begin{multline}
N_{tot}(N_2H^+) = \frac{3h}{8 \pi^3 \mu^2}
  \frac{Q_{rot}}{J_uR_i} \exp\left(\frac{E_u}{kT_{ex}}\right) \\
  \times\left[\exp\left(\frac{h\nu}{kT_{ex}}\right) - 1\right]^{-1}
  \int\tau_\nu dv.
\label{eq:ntotn2hp}
\end{multline}
Assuming optically thin emission (Equation~\ref{eq:ntotthin}), we find
that Equation~\ref{eq:ntotn2hp} becomes\footnote{Alternatively, one
  can derive $\tau$ using the ratio of two hyperfine transition
  intensities as described in Appendix~\ref{HyperTau}.}:
\begin{multline}
N_{tot}(N_2H^+) = \frac{3.10\times10^{11}\left(T_{ex}+0.75\right)}{R_i}
\frac{\exp\left(\frac{E_u}{kT_{ex}}\right)}{\exp\left(\frac{h\nu}{kT_{ex}}\right)
  - 1} \\
\times\left[\frac{\int T_R dv(km/s)}{f\left(J_\nu(T_{ex}) -
    J_\nu(T_{bg})\right)}\right]~\textrm{cm}^{-2},
\label{eq:ntotn2hpfinal}
\end{multline}
where $\nu$ is the frequency of the hyperfine transition
used.  For example, if the F=(2,1), J=(1,0) hyperfine was
chosen for this calculation, $R_i = \frac{7}{27}$ (See
Table~\ref{tab:hypern2hp10}) and $\nu = 93.1737767$\,GHz.
Equation~\ref{eq:ntotn2hpfinal} then becomes:
\begin{multline}
N_{tot}(N_2H^+ (3,2;2,1)) =
1.20\times10^{12}\left(T_{ex}+0.75\right) \\
\frac{\exp\left(\frac{4.47}{T_{ex}}\right)}{\exp\left(\frac{4.47}{T_{ex}}\right) - 1}
\times\left[\frac{\int T_R dv(km/s)}{f\left(J_\nu(T_{ex}) -
    J_\nu(T_{bg})\right)}\right]~\textrm{cm}^{-2}. 
\label{eq:ntotn2hpfinal1}
\end{multline}
where the transition designation is in the
(F$^\prime$,F$^\prime_1$:F,F$_1$) format.  Note that this transition
is blended with the nearby F$^\prime$ = 1 and 2 (F$^\prime_1$,F$_1$) =
(2,1) hyperfine transitions under many molecular cloud physical
conditions.

\subsection{NH$_3$}
\label{Amon}

Ammonia (NH$_3$) is a symmetric top molecule with three opposing 
identical H (spin=$\frac{1}{2}$) nuclei.  Quantum mechanical tunneling
of the N nucleus through the potential energy plane formed by the H nuclei
leads to inversion splitting of each NH$_3$ energy level.  On top of
this inversion splitting the energy levels are split due to two
hyperfine interactions:
\begin{description}
\item[J--I$_N$:] Coupling between the quadrupole moment of the N
  nucleus (I$_N = 1$) and the electric field of the H atoms, which
  splits each energy level into three hyperfine states.  For this
  interaction the angular momentum vectors are defined as follows:
  $\vec{F_1} = \vec{J} + \vec{I_N}$.
\item[F$_1$--I$_H$:] Coupling between the magnetic dipole of the three
  H nuclei (I$_H = \frac{1}{2}$ for K$\ne 0$ or 3n (para), I$_H =
  \frac{3}{2}$ for K=3n (ortho)) with the weak current generated by
  the rotation of the molecule.  For this interaction the angular
  momentum vectors are defined as follows: $\vec{F} = \vec{F_1} +
  \vec{I_H}$. 
\end{description}
Weaker N-H spin-spin and H-H spin-spin interactions also
exist, but only represent small perturbations of the existing
hyperfine energy levels.  

\begin{figure*}
\centering
\includegraphics[scale=0.60,angle=-90]{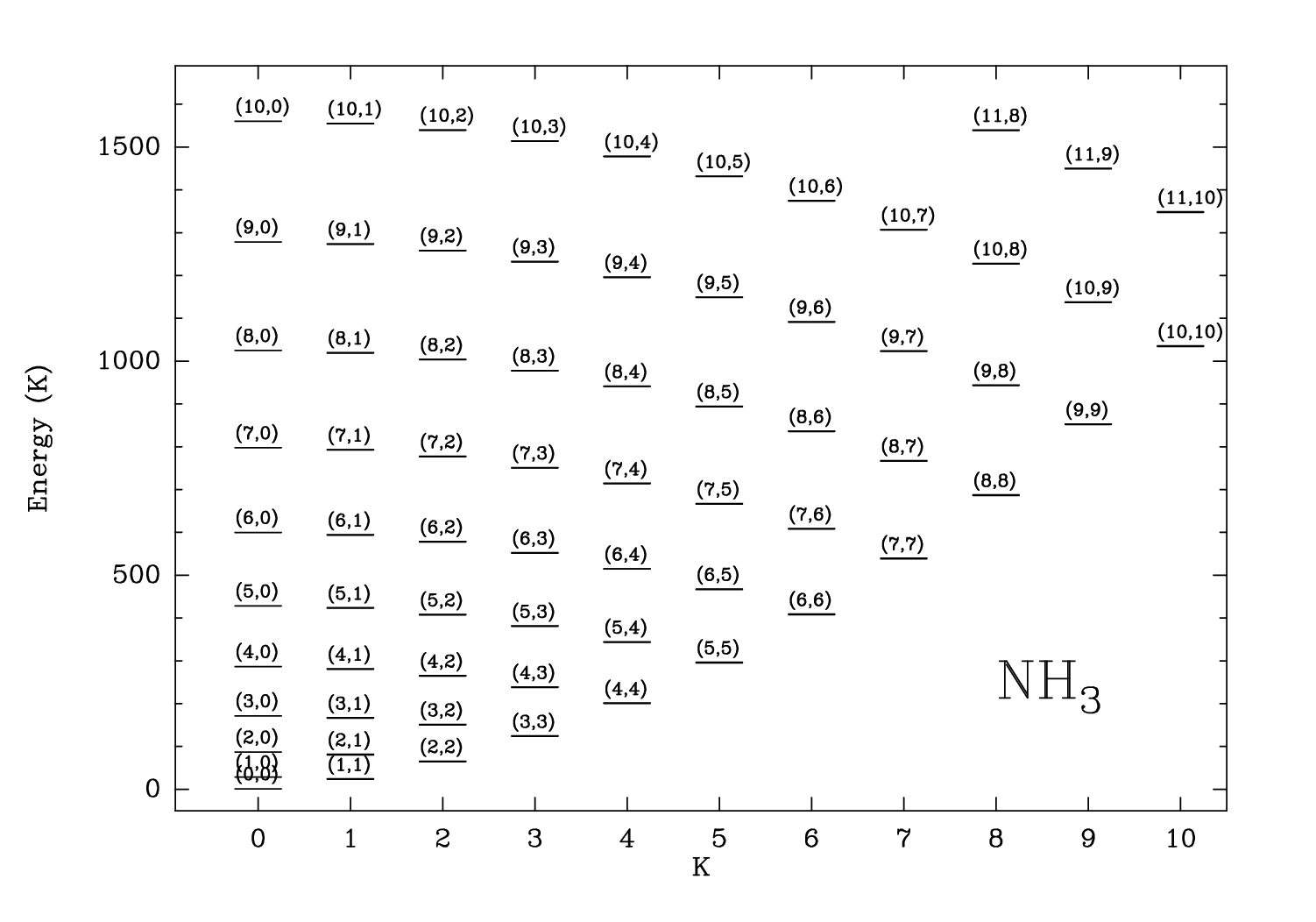}
\caption{Rotational energy level diagram for NH$_3$.  All levels with
  energy $<1600$\,K are shown.}  
\label{fig:NH3LevelPlot}
\end{figure*}

\begin{figure}
\centering
\includegraphics[scale=0.45]{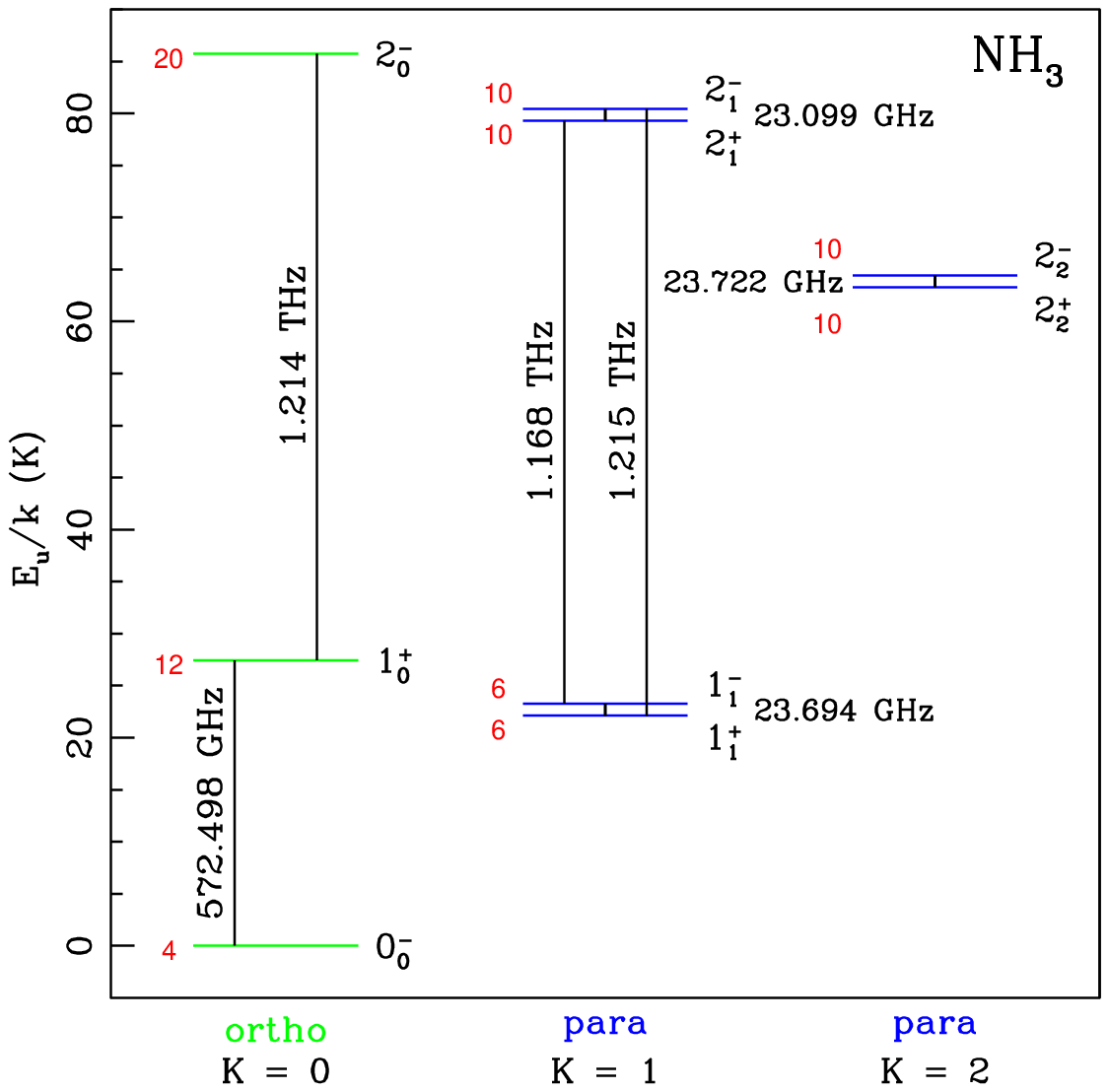}
\caption{Ammonia energy levels below 100\,K.  The notation
  $J_K^{\pm}$ is used where $\pm$ refers to the symmetry of the energy
  level.  Allowed electric dipole transitions are indicated with
  frequencies of the transitions.  The total statistical weight of
  each energy level is indicated in red to the left of each energy level.}
\label{fig:nh3lev}
\end{figure}

\begin{figure*}
\centering
\includegraphics[scale=0.55]{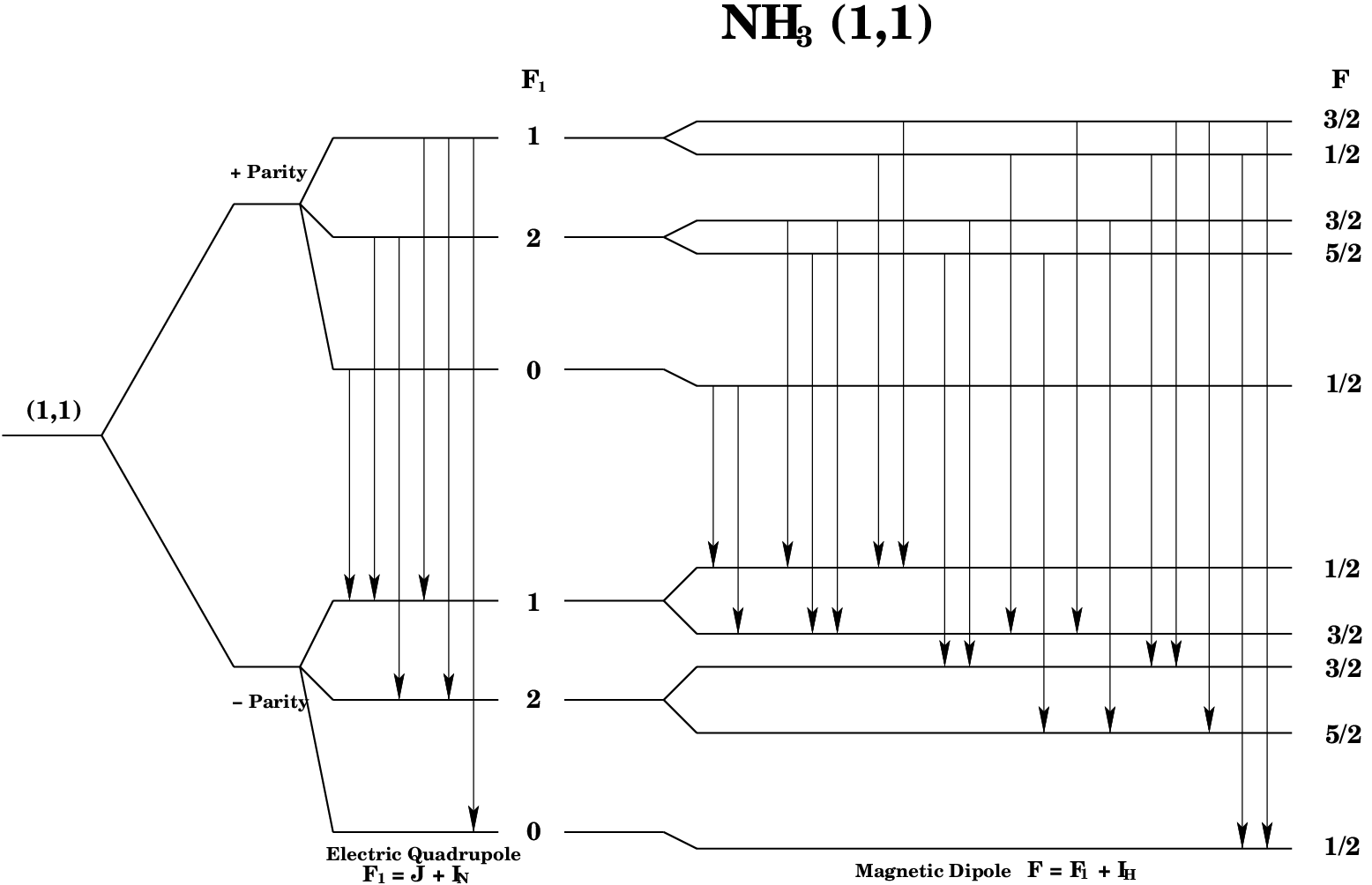} \\ [30pt]
\includegraphics[scale=0.55]{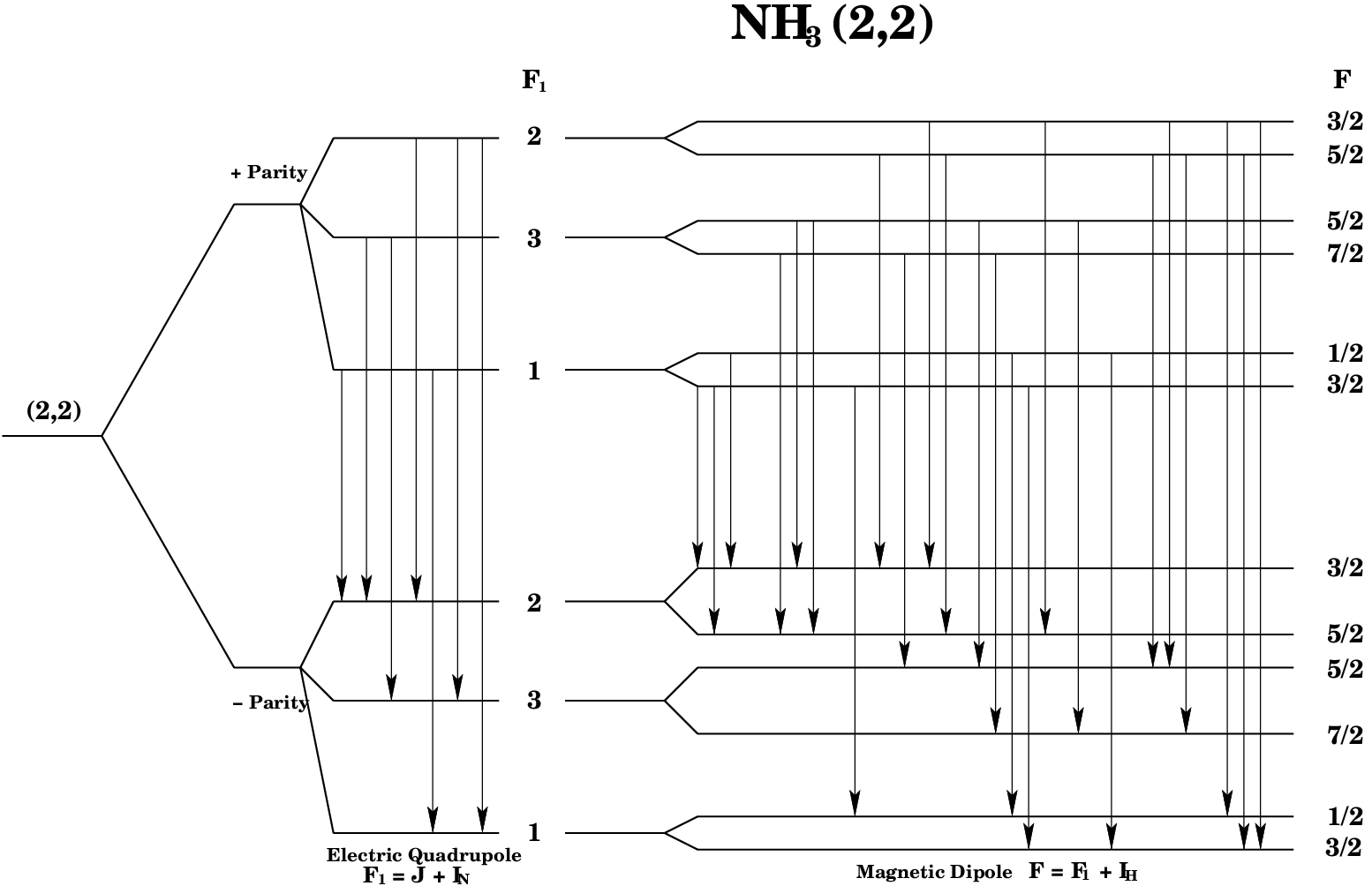}
\vspace{5mm}
\caption{Inversion and hyperfine energy level structure for the (1,1)
  (top) and (2,2) (bottom) transitions of NH$_3$.  Note that the 18
  (1,1) and 24 (2,2) allowed transitions are marked with arrows
  ordered by increasing frequency from left to right.   Adapted from
  \cite{Ho1983}.}
\label{fig:nh31122levs}
\end{figure*}

Figure~\ref{fig:NH3LevelPlot} shows all NH$_3$
energy levels below 1600 K, while Table~\ref{tab:nh3energy} lists the
level energies for $J\leq6$.  Figure~\ref{fig:nh3lev} shows the
rotational energy level diagram for the first four J-levels of NH$_3$,
while Figure~\ref{fig:nh31122levs} shows the inversion and hyperfine
level structure for the (1,1) and (2,2) transitions.  

\begin{deluxetable}{lclc}
\tablewidth{0pt}
\tablecaption{NH$_3$ Level Energies\tablenotemark{a,b}\label{tab:nh3energy}}
\tablehead{
\colhead{Level} & \colhead{Energy (K)} & \colhead{Level} &
\colhead{Energy (K)}
}
\startdata
(0,0,a) & 1.14 & \ldots & \ldots \\
(1,1,s) & 23.21 & (1,1,a) & 24.35 \\
(1,0,s) & 28.64 & \ldots & \ldots \\
(2,2,s) & 64.20 & (2,2,a) & 65.34 \\
(2,1,s) & 80.47 & (2,1,a) & 81.58 \\
(2,0,a) & 86.99 & \ldots & \ldots \\
(3,3,s) & 122.97 & (3,3,a) & 124.11 \\
(3,2,s) & 150.06 & (3,2,a) & 151.16 \\
(3,1,s) & 166.29 & (3,1,a) & 167.36 \\
(3,0,s) & 171.70 & \ldots & \ldots \\
(4,4,s) & 199.51 & (4,4,a) & 200.66 \\
(4,3,s) & 237.40 & (4,3,a) & 238.48 \\
(4,2,s) & 264.41 & (4,2,a) & 265.45 \\
(4,1,s) & 280.58 & (4,1,a) & 281.60 \\
(4,0,a) & 286.98 & \ldots & \ldots \\
(5,5,s) & 293.82 & (5,5,a) & 295.00 \\
(5,4,s) & 342.49 & (5,4,a) & 343.58 \\
(5,3,s) & 380.23 & (5,3,a) & 381.25 \\
(5,2,s) & 407.12 & (5,2,s) & 408.10 \\
(5,1,s) & 423.23 & (5,1,a) & 424.18 \\
(5,0,s) & 428.60 & \ldots & \ldots \\
(6,6,s) & 405.91 & (6,6,a) & 407.12 \\
(6,3,s) & 551.30 & (6,3,a) & 552.25 \\
(6,0,a) & 600.30 & \ldots & \ldots
\enddata
\tablenotetext{a}{{Listed} in level energy order per J and
  inversion-paired as appropriate.}
\tablenotetext{b}{See \cite{Poynter1975} for lower-state energy
  calculations.}
\end{deluxetable}

\begin{figure}
\centering
\includegraphics[scale=0.32]{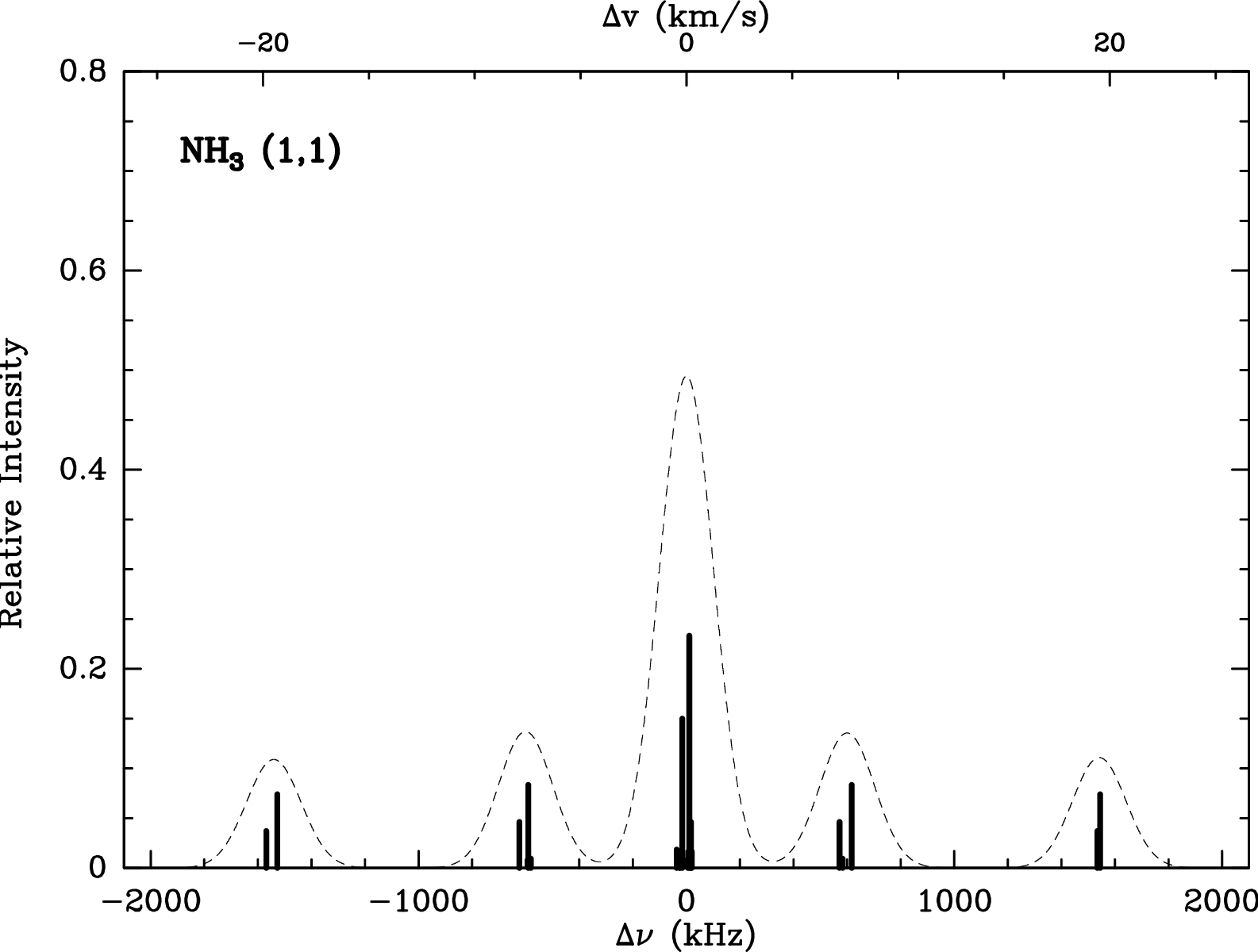} \\[10pt]
\includegraphics[scale=0.32]{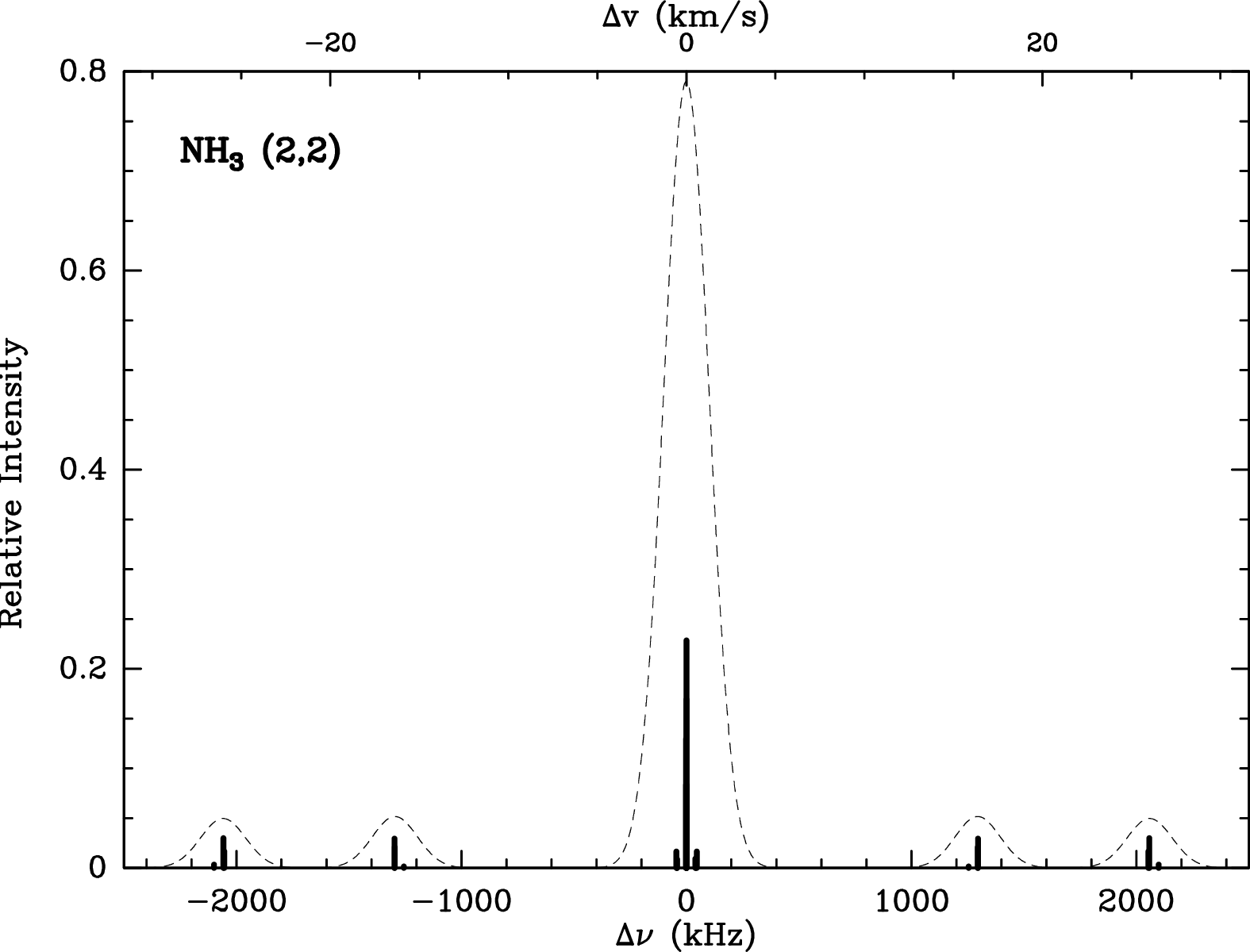}
\caption{Synthetic spectra for the NH$_3$ (1,1) (top) and (2,2)
  (bottom) transitions.  Horizontal axes are offset velocity (top) and
  frequency (bottom) relative to 23694.495487 and 23722.633335\,MHz,
  respectively.  
  Overlain as a dashed line is a synthetic 100\,kHz ($\Delta v =
  1.265$ and $1.264$\,\kms, respectively) gaussian linewidth source spectrum.}
\label{fig:nh3relplt}
\end{figure}

We can calculate the relative hyperfine intensities ($R_i$) for the
(1,1), (2,2), (3,3), and (4,4) transitions using the formalism derived in
Section~\ref{Hyperfine}.  Using the formalism presented in
  Section~\ref{Hyperfine} we can derive the 
relevant $R_i$ for the quadrupole hyperfine ($R_i(F_1,J)$, I$_N = 1$; 
Tables~\ref{tab:sixjji11}, \ref{tab:sixjji2k}, \ref{tab:sixjji3k}, and
\ref{tab:sixjji4k}) and magnetic
hyperfine ($R_i(F,F_1)$, I$_H = \frac{1}{2}$ (para, $K\ne 0$ or 3n) or
$\frac{3}{2}$ (ortho, $K=3n$); Tables~\ref{tab:sixjfi11},
\ref{tab:sixjfi22}, \ref{tab:sixjfi33}, and \ref{tab:sixjfi44})
coupling cases.  The resultant total hyperfine intensities are derived
by multiplying the magnetic hyperfine intensities ($R_i(F,F_1)$) by
the associated quadrupole hyperfine intensities ($R_i(F_1,J)$).
Tables~\ref{tab:hyper11} and \ref{tab:hyper2k} show the results of
this calculation for the NH$_3$ (1,1) and (2,2) transitions,
respectively.  Figure~\ref{fig:nh3relplt} shows the
synthetic spectra for the NH$_3$ (1,1) and (2,2) transitions.

For illustration we can derive the column density equation for a
para-NH$_3$ (K$\ne$0 or 3n) inversion ($\Delta K=0$) transition.  For
para-NH$_3$ inversion transitions:
\begin{align}
S &= \frac{K^2}{J_u(J_u + 1)} \nonumber \\
\mu &= 1.468~\textrm{Debye} = 1.468\times10^{-18}~\textrm{esu cm} \nonumber \\
R_i &= \textrm{(see Section~\ref{Hyperfine} or, for (1,1) and (2,2), see}
\nonumber \\
   & \textrm{Tables~\ref{tab:hyper11} and \ref{tab:hyper2k})} \nonumber \\
g_J &= 2J_u + 1 \nonumber \\
g_K &= 2 \textrm{ for K$\ne$ 0} \nonumber \\
g_I &= \frac{2}{8} \textrm{ for K$\ne$ 3n} \nonumber
\end{align}
Doing the calculation in two stages, we can first derive the
following equation for the molecular column density in NH$_3$ as
derived from a measurement of a (J,K) inversion ($\Delta J =
0$,$\Delta K = 0$) transition assuming optically thin emission (see
\cite{Rosolowsky2008} for the methodology to fit hyperfine
transitions in the optically thick limit) using Equation
\ref{eq:ntotthin}:
\begin{multline}
N_{tot}(NH_3) = \frac{3h}{8 \pi^3 \mu^2 R_i}
\frac{J_u(J_u + 1)}{K^2} \frac{Q_{rot}}{g_J g_K g_I} \\
\times\frac{\exp\left(\frac{E_u}{kT_{ex}}\right)}{\exp\left(\frac{h\nu}{kT_{ex}}\right) 
    - 1} \left[\frac{\int T_R dv}{f\left(J_\nu(T_{ex}) -
    J_\nu(T_{bg})\right)}\right].
\label{eq:ntotamon}
\end{multline}
Further assuming that we are measuring para-NH$_3$ (K$\ne$0 or 3n),
Equation~\ref{eq:ntotamon} becomes:
\begin{multline}
N_{tot}(NH_3) \simeq \frac{7.44\times 10^{12}
  J_u(J_u + 1) Q_{rot}}{R_i K^2 (2J_u + 1)} \\
\times\frac{\exp\left(\frac{E_u}{kT_{ex}}\right)}{\exp\left(\frac{h\nu}{kT_{ex}}\right)
  - 1} \left[\frac{\int T_R dv(km/s)}{f\left(J_\nu(T_{ex}) -
    J_\nu(T_{bg})\right)}\right] \textrm{ cm$^{-2}$}.
\label{eq:ntotparaamon}
\end{multline}

\subsection{H$_2$CO}
\label{Formaldehyde}

\begin{figure*}
\centering
\includegraphics[scale=0.60, angle=-90]{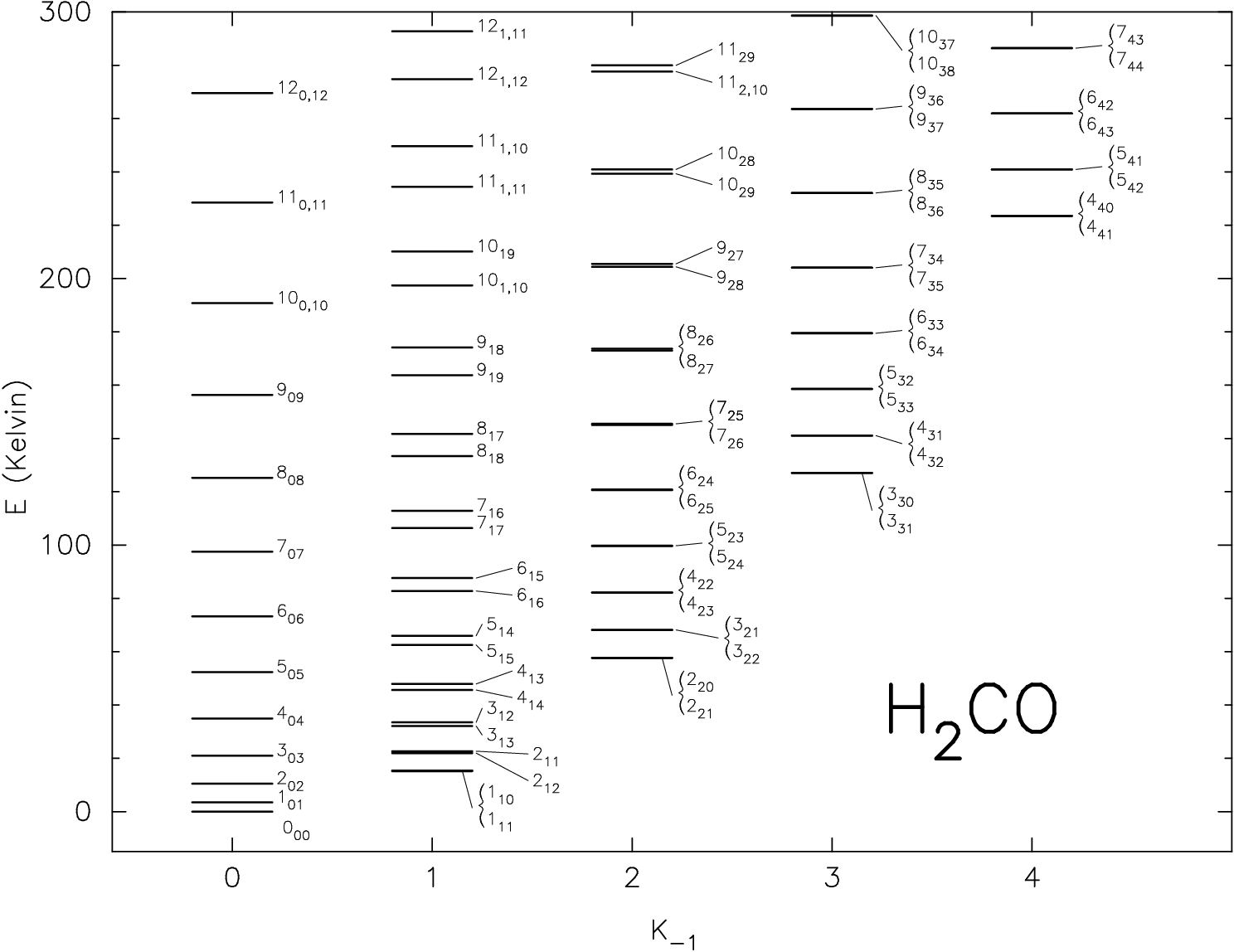}
\caption{Rotational energy level diagram for H$_2$CO including all energy levels
  with $E \leq 300$\,K.}
\label{fig:h2colev}
\end{figure*}

Formaldehyde (H$_2$CO) is a slightly asymmetric rotor molecule with
$\kappa \simeq -0.961$ (Equation~\ref{eq:raysasymmetry}), which means
that H$_2$CO is nearly a prolate symmetric rotor.  The slight
asymmetry in H$_2$CO 
results in limiting prolate (quantum number K$_{-1}$) and oblate
(quantum number K$_{+1}$) symmetric rotor energy levels that are
closely spaced in energy, a feature commonly referred to as
``K-doublet splitting''.  Figure~\ref{fig:h2colev} shows the energy
level diagram for H$_2$CO including all energy levels $E \leq
300$\,K.  In addition to the asymmetric rotor energy level structure
H$_2$CO possesses spin-rotation and spin-spin hyperfine energy level
structure.  Magnetic dipole interaction between the H nuclei (I$_H =
0/1$ for para-/ortho-H$_2$CO) and rotational motion of the
molecule result in spin-rotation hyperfine energy level
splitting. Since the two H nuclei have equal coupling, the two H
nuclei spins couple with each other first and the hyperfine coupling
scheme is $\vec{I}_H = \vec{I}_1 + \vec{I}_2$ and $\vec{F} = \vec{J} + \vec{I}_H$.
Since the interchange of the two $^{1}$H atoms ($I_1 = I_2 = 1/2$)
obey Fermi-Dirac statistics (total wave function is anti-symmetric to
this interchange) and alternating $K_a$ ladders in H$_2$CO have a
different rotational wavefunction symmetry, only ortho-H$_2$CO levels
($I_H = 1$) with $K_a =$ odd integers have hyperfine splitting.  For
the ortho-H$_2$CO $1_{10}-1_{11}$ transition, the frequency offsets of
these hyperfine transitions are $\Delta\nu \leq 18.5$\,kHz.  The
weaker spin-spin interactions between the nuclei are generally not
considered.

\begin{figure}
\centering
\includegraphics[scale=0.32]{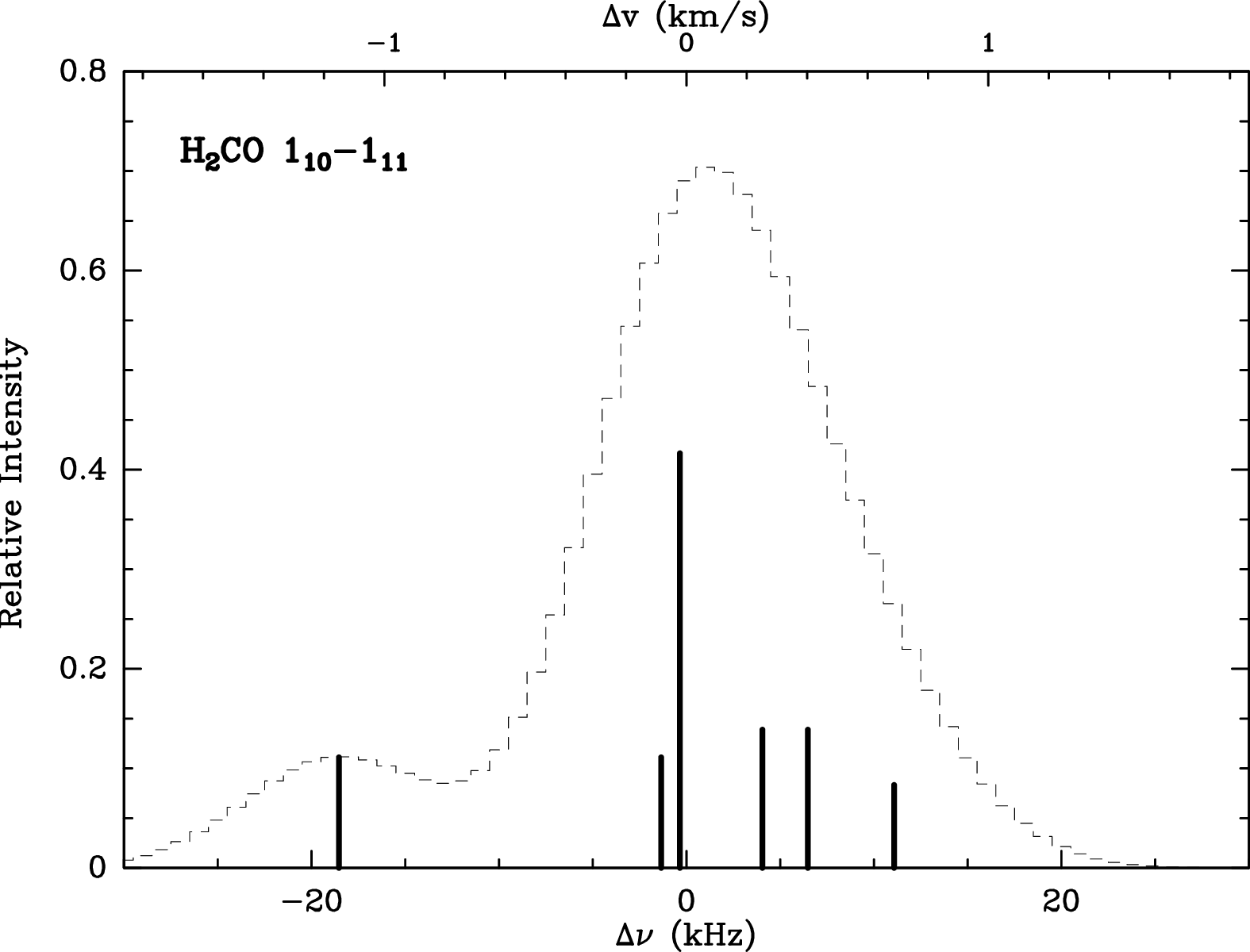} \\[10pt]
\includegraphics[scale=0.32]{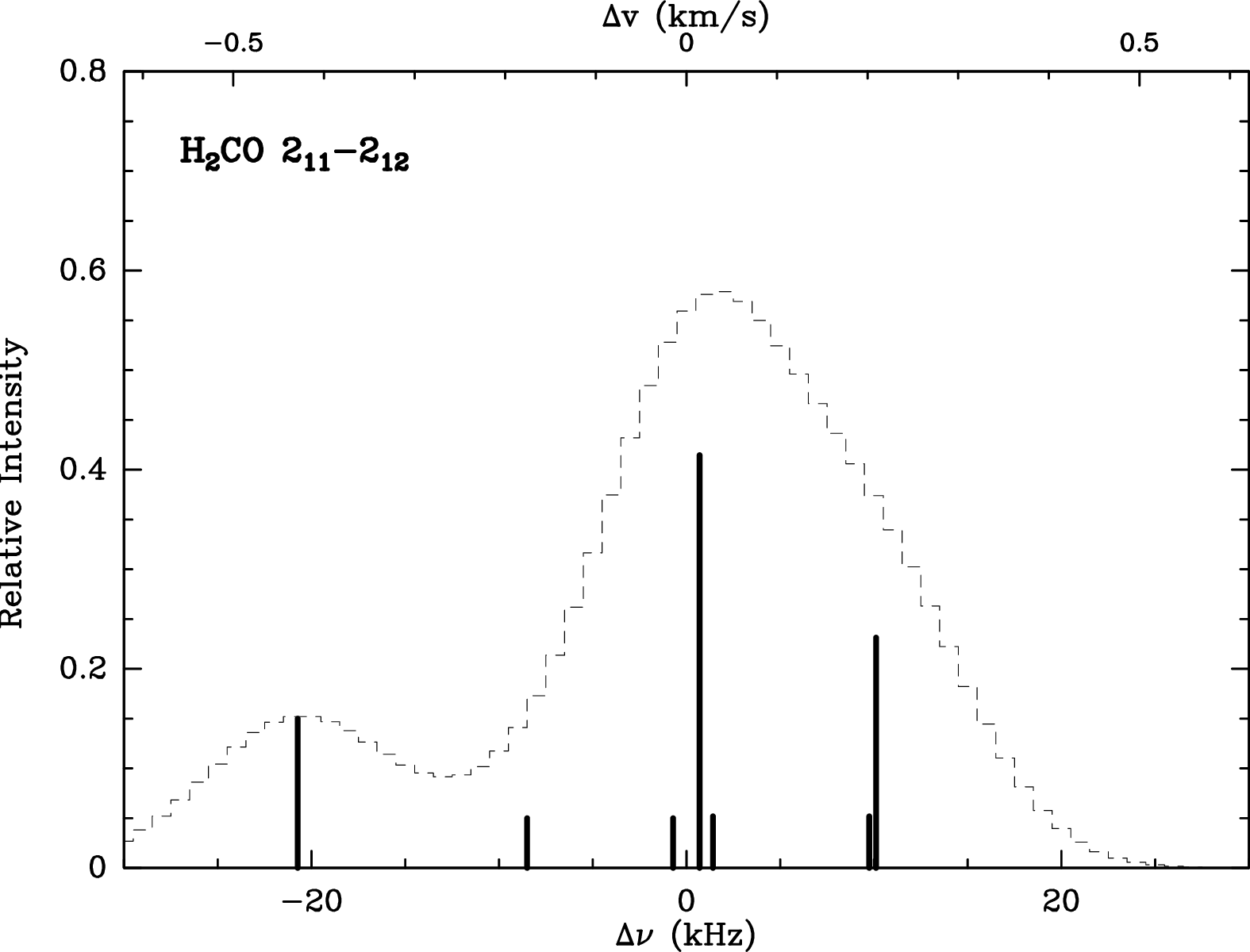} \\[10pt]
\includegraphics[scale=0.32]{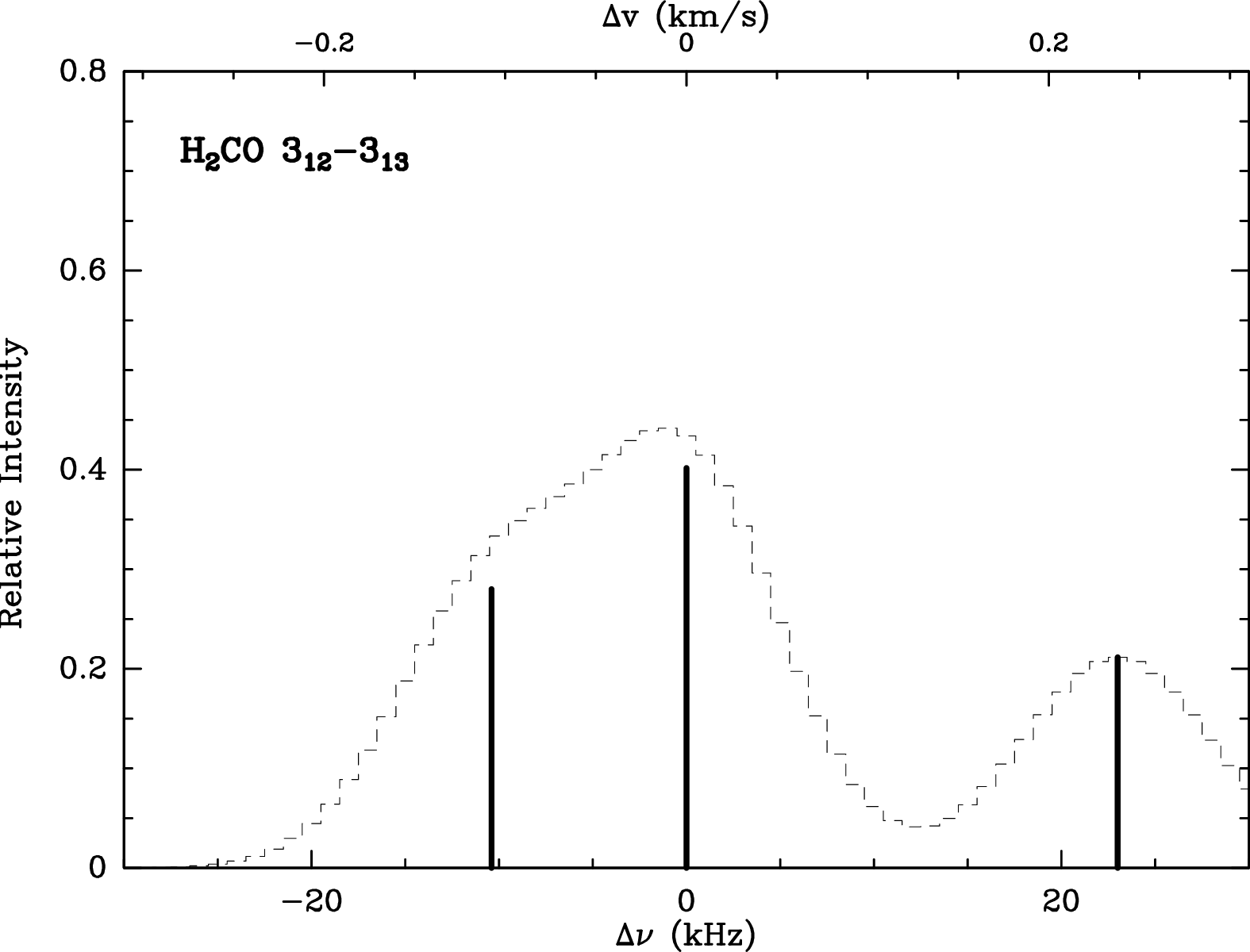}
\caption{Synthetic spectra for the H$_2$CO $1_{10}-1_{11}$,
  $2_{11}-2_{12}$, and $3_{12}-3_{13}$ transitions. Horizontal axes
  are offset velocity (top) and frequency (bottom) relative to 4829.6596\,MHz
  for $1_{10}-1_{11}$, 14488.650\,MHz for $2_{11}-2_{12}$, and
  28974.85\,MHz for $3_{12}-3_{13}$, respectively.  
  For the 
  $3_{12}-3_{13}$ transition only the $\Delta F = 0$ hyperfine
  transitions are shown.  Overlain as a dashed line is a synthetic 10\,kHz
  ($\Delta v = 0.621$, $0.207$, and $0.103$\,\kms, respectively) 
  gaussian linewidth source spectrum.} 
\label{fig:h2corelplt}
\vspace{2mm}
\end{figure}

Table~\ref{tab:h2cofreqint} lists the frequencies and relative
intensities for the spin-rotation hyperfine transitions of the H$_2$CO
$1_{10}-1_{11}$, $2_{11}-2_{12}$, and $3_{12}-3_{13}$ transitions.
Figure~\ref{fig:h2corelplt} shows the synthetic spectra for the H$_2$CO
$1_{10}-1_{11}$, $2_{11}-2_{12}$, and $3_{12}-3_{13}$ transitions.
Note that the hyperfine 
intensities are exactly equal to those calculated for the
spin-rotation hyperfine components of NH$_3$ (see Section~\ref{amontabs}).

\begin{deluxetable*}{ccccccccc}
\tablewidth{0pt}
\tablecaption{F$_1$-I$_H$ Hyperfine Frequencies and Intensities for
  H$_2$CO J=1, 2, and 3 K-Doublet Transitions\label{tab:h2cofreqint}}
\tablehead{
\colhead{$F^\prime \rightarrow F$\tablenotemark{a}} &
\colhead{$J^\prime \rightarrow J$\tablenotemark{a}} &
\colhead{$\Delta_{HF}$\tablenotemark{b} (kHz)} & 
\colhead{a} & \colhead{b} & \colhead{c} &
\colhead{$\frac{(2F^\prime_1 + 1)(2F_1 + 1)}{(2I_H + 1)}$}  
& \colhead{6j} & \colhead{R$_i(F_1,J)$}
}
\startdata
(1,0) & (1,1) & $-18.53$ & 1 & 1 & 1 & 1 & $-\frac{1}{3}$ & $\frac{1}{9}$ \\
(0,1) & (1,1) & $-1.34$ & 1 & 1 & 1 & 1 & $-\frac{1}{3}$ & $\frac{1}{9}$ \\
(2,2) & (1,1) & $-0.35$ & 1 & 2 & 1 & $\frac{25}{3}$ 
& $-\frac{1}{2\sqrt{5}}$ & $\frac{5}{12}$ \\
(2,1) & (1,1) & $+4.05$ & 1 & 1 & 2 & 5 & $\frac{1}{6}$  & $\frac{5}{36}$ \\
(1,2) & (1,1) & $+6.48$ & 1 & 1 & 2 & 5 & $\frac{1}{6}$  & $\frac{5}{36}$ \\
(1,1) & (1,1) & $+11.08$ & 1 & 1 & 1 & 3 & $\frac{1}{6}$  & $\frac{1}{12}$ \\

(1,1) & (2,2) & $-20.73$ & 1 & 1 & 2 & 3 & $\frac{1}{2\sqrt{5}}$  & $\frac{3}{20}$ \\
(1,2) & (2,2) & $-8.5$ & 1 & 2 & 2 & 5 & $-\frac{1}{10}$ & $\frac{1}{20}$ \\
(2,1) & (2,2) & $-0.71$ & 1 & 2 & 2 & 5 & $-\frac{1}{10}$ & $\frac{1}{20}$ \\
(3,3) & (2,2) & $+0.71$ & 1 & 3 & 2 & $\frac{49}{3}$ 
& $-\frac{2\sqrt{2}}{3\sqrt{35}}$ & $\frac{392}{945}$ \\
(3,2) & (2,2) & $+1.42$ & 1 & 2 & 3 & $\frac{35}{3}$ & $\frac{1}{15}$  & $\frac{7}{135}$ \\
(2,3) & (2,2) & $+9.76$ & 1 & 2 & 3 & $\frac{35}{3}$ & $\frac{1}{15}$  & $\frac{7}{135}$ \\
(2,2) & (2,2) & $+10.12$ & 1 & 2 & 2 & $\frac{25}{3}$ & $\frac{1}{6}$  & $\frac{25}{108}$ \\

(2,3) & (3,3) & \ldots & 1 & 3 & 3 & $\frac{35}{3}$ & $-\frac{1}{21}$ & $\frac{5}{189}$ \\
(4,3) & (3,3) & \ldots & 1 & 3 & 4 & 21 & $\frac{1}{28}$  & $\frac{3}{112}$ \\
(4,4) & (3,3) & $+0.00$ & 1 & 4 & 3 & 27 & $-\frac{\sqrt{5}}{4\sqrt{21}}$ &
                                                 $\frac{45}{112}$ \\
(3,3) & (3,3) & $-10.4$ & 1 & 3 & 3 & $\frac{49}{3}$ & $\frac{11}{84}$ &
$\frac{121}{432}$ \\
(2,2) & (3,3) & $+23.0$ & 1 & 2 & 3 & $\frac{25}{3}$ & $\frac{-2\sqrt{2}}{3\sqrt{35}}$ & $\frac{40}{189}$ \\
(3,4) & (3,3) & \ldots & 1 & 3 & 4 & 21 & $\frac{1}{28}$  & $\frac{3}{112}$ \\
(3,2) & (3,3) & \ldots & 1 & 3 & 3 & $\frac{35}{3}$ & $-\frac{1}{21}$
& $\frac{5}{189}$
\enddata
\tablenotetext{a}{$I_H = 1$ (ortho-H$_2$CO).}
\tablenotetext{b}{Frequency offset in kHz relative to 4829.6596\,MHz
  for $1_{10}-1_{11}$, 14488.65\,MHz for $2_{11}-2_{12}$, and
  28974.85\,MHz for $3_{12}-3_{13}$.}
\end{deluxetable*}

For illustration we can derive the column density equation for an
ortho-H$_2$CO ($K_{-1}$ odd) K-doublet ($\Delta K_{-1}= 0$)
transition.  For ortho-H$_2$CO transitions:
\begin{align}
S &= \frac{K^2}{J_u(J_u + 1)} \nonumber \\
\mu &= 2.331~\textrm{Debye} = 2.331\times10^{-18}~\textrm{esu cm} \nonumber \\
R_i &= \textrm{(see Section~\ref{Hyperfine} or, for $1_{10}-1_{11}$,
  $2_{11}-2_{12}$,} \nonumber \\
 & \textrm{or $3_{12}-3_{13}$ see Table~\ref{tab:h2cofreqint})} \nonumber \\ 
g_J &= 2J_u + 1 \nonumber \\
g_K &= 1 \textrm{ (for an asymmetric rotor)} \nonumber \\
g_I &= \frac{3}{4} \textrm{ for } K_{-1} \textrm{ odd} \nonumber
\end{align}
We can derive the following equation for the 
molecular column density in H$_2$CO as derived from a measurement of
an ortho-H$_2$CO ($K_{-1}$ odd) K-doublet transition assuming
optically thin emission using Equation \ref{eq:ntotthin}: 
%
\begin{align}
N_{tot}(H_2CO) &= \frac{3h}{8 \pi^3 \mu^2 R_i} \frac{J_u(J_u + 1)}{K^2}
  \frac{Q_{rot}}{g_J g_K g_I} \notag \\
&\times\frac{\exp\left(\frac{E_u}{kT_{ex}}\right)}{\exp\left(\frac{h\nu}{kT_{ex}}\right)
  - 1} \left[\frac{\int T_R dv}{f\left(J_\nu(T_{ex}) -
    J_\nu(T_{bg})\right)}\right] \notag \\
&\simeq \frac{1.97\times 10^{12} J_u(J_u + 1) Q_{rot}}{R_i K^2 (2J_u +
  1)} \notag \\ 
&\times \frac{\exp\left(\frac{E_u}{kT_{ex}}\right)}{\exp\left(\frac{h\nu}{kT_{ex}}\right)
  - 1} \left[\frac{\int T_R dv(km/s)}{f\left(J_\nu(T_{ex}) -
      J_\nu(T_{bg})\right)}\right] \textrm{cm$^{-2}$}.
\label{eq:ntoth2co}
\end{align}

Note that the K-doublet transitions of H$_2$CO are rather unusual in
that, due to an unusual collisional excitation effect, these
transitions are often measured in \textit{absorption} against the
cosmic microwave background radiation.  For n($H_2$) $\lesssim$
10$^{5.5}$\,cm$^{-3}$, a collisional selection effect overpopulates the
lower energy states of the $1_{10}-1_{11}$ through $5_{14}-5_{15}$
transitions \citep{Evans1975,Garrison1975}.  This
overpopulation of the lower energy states results in a ``cooling'' of
the J$\leq$ 5 K-doublets to an excitation temperature that is lower
than that of the cosmic microwave background.  This causes the
J $\leq$ 5 K-doublet transitions to appear in absorption.  For volume
densities n($H_2$) $\gtrsim$ $10^{5.5}$\,cm$^{-3}$ and $T_K \simeq
40$\,K this collisional pump is quenched.  For these higher volume
densities the J $\leq$ 5 K-doublets are driven into emission over a wide range of 
kinetic temperatures and abundances \citep[see Figure\,1 in][]{Mangum2008}.

\section{Summary}

The most general form of the molecular spectral line column density is
given by Equation~\ref{eq:ntotfinal}:
\begin{multline}
N_{tot} = \frac{3h}{8 \pi^3 |{\mathbf \mu_{lu}}|^2}
  \frac{Q_{rot}}{g_u} \exp\left(\frac{E_u}{kT_{ex}}\right)\\
  \times\left[\exp\left(\frac{h\nu}{kT_{ex}}\right) - 1\right]^{-1}
  \int\tau_\nu dv.
\end{multline}
For absorption line measurements one should use the velocity
integrated opacity directly in Equation~\ref{eq:ntotfinal}.  For
emission lines, the velocity integrated opacity is usually converted
to an integrated intensity.  In the optically thin limit, the column
density becomes (Equation~\ref{eq:ntotthin}):
\begin{multline}
N^{thin}_{tot} = \left(\frac{3h}{8 \pi^3 S \mu^2
  R_i}\right)\left(\frac{Q_{rot}}{g_J g_K g_I}\right)
\frac{\exp\left(\frac{E_u}{kT_{ex}}\right)}{\exp\left(\frac{h\nu}{kT_{ex}}\right)
  - 1} \\
\times\frac{1}{\left(J_\nu(T_{ex}) - J_\nu(T_{bg})\right)}
  \int\frac{T_R dv}{f}.
\end{multline}
The optically thick column density is related to the optically thin
column density by Equation~\ref{eq:ntotthingenrat}:
\begin{equation}
N_{tot} = N^{thin}_{tot} \frac{\tau}{1 - \exp(-\tau)}.
\end{equation}
In this tutorial, we have carefully derived the terms in
Equations~\ref{eq:ntotfinal}, \ref{eq:ntotthin}, and
\ref{eq:ntotthingenrat}.  Note, though, that these equations assume
all energy levels are characterized by the same excitation temperature
(the ``CTEX LTE approximation''; See Section~\ref{Tex})).  We have
also explored the limits of the CTEX approximation
(Section~\ref{Tex}).  Non-LTE determinations of the column density
require a coupled radiative transfer and statistical equilibrium
calculation and can be performed with publically available tools.

\acknowledgments

JGM thanks Remy Indebetouw for many extremely fruitful discussions.  We also
thank the 2013 University of Arizona AST515 graduate ISM class, Brian
Svoboda, and Scott Schnee for reviewing the manuscript.  The referee,
Malcolm Walmsley, provided a fantastic review which resulted in many
important improvements to this manuscript.  The National
Radio Astronomy Observatory is a facility of the National Science
Foundation operated under cooperative agreement by Associated
Universities, Inc.

\bibliographystyle{apj}
%
%
\bibliography{ColumnDensityCalculation-MangumShirley}

\appendix

\section{Line Profile Functions}
\label{Lineprofile}

For a Gaussian profile the function $\phi(\nu)$ is given by
\begin{equation}
\phi(\nu) = \frac{1}{\sqrt{2\pi}\sigma}\exp\left[-\frac{(\nu -
    \nu_0)^2}{2\sigma^2}\right],
\label{eq:gaussprofile}
\end{equation}
where $\nu_0$ and $\nu$ are the central and offset frequencies
  of the line profile for a particle with mass $M$ and velocity $v$,
\begin{equation}
2\sigma^2 = \frac{\nu^2_0}{c^2}\left(\frac{2kT_k}{M} + v^2\right),
\label{eq:sigma}
\end{equation}
and
\begin{equation}
\int\phi(\nu) d\nu = 1.
\label{eq:intgauss}
\end{equation}
The Gaussian profile has a FWHM given by (in both frequency
and velocity):
\begin{eqnarray}
\Delta\nu_D &=& \frac{2\nu_0}{c}\left[\ln{2}\left(\frac{2kT_k}{M} +
  v^2\right)\right]^2 \\
\Delta v_D &=& 2\left[\ln{2}\left(\frac{2kT_k}{M} +
  v^2\right)\right]^2,
\label{eq:profilewidth}
\end{eqnarray}
and a peak value given by:
\begin{eqnarray}
\phi(\nu)_{peak} &=& \frac{2\sqrt{\ln{2}}}{\sqrt{\pi}\Delta\nu_D}
\nonumber \\
                 &=& \frac{2\sqrt{\ln{2}}c}{\sqrt{\pi}\nu_0 \Delta v_D}.
\label{eq:phipeak}
\end{eqnarray}

If one uses peak values instead of integrating over a Gaussian profile
to derive column densities, one must make the following correction:
\begin{equation}
(N_{tot})_{Gauss} = 2\sqrt{\frac{\ln{2}}{\pi}}(N_{tot})_{peak}.
\label{eq:colcor}
\end{equation}

\section{Integrated Fluxes Versus Brightness Temperatures}
\label{Intflux}

All calculations in this document assume the use of integrated
brightness temperatures ($\int T_B \Delta v$).  If one uses integrated
fluxes ($\int S_\nu \Delta v$), the total molecular column density
assuming optically-thin emission (Equation \ref{eq:ntotthinrayleighnobg}) is
modified by using the relationship between flux density and brightness
temperature:
\begin{equation}
S_\nu = \frac{2kT_B \nu^2}{c^2} \Omega_s,
\label{eq:fluxdens}
\end{equation}
and becomes
\begin{eqnarray}
N_{tot} &=& \left(\frac{D}{2 R_s}\right)^2 \frac{3c^2}{16\pi^3 S \mu^2
  \nu^3} \left(\frac{Q_{rot}}{g_J g_K g_{nuclear}}\right)
  \exp\left(\frac{E_u}{kT}\right) \int S_\nu \Delta v \nonumber \\
        &=& \frac{3c^2}{16\pi^3 \Omega_s S \mu^2 \nu^3}
    \left(\frac{Q_{rot}}{g_J g_K g_{nuclear}}\right)
    \exp\left(\frac{E_u}{kT}\right) \int S_\nu \Delta v.
\label{eq:ntotfluxden}
\end{eqnarray}

\section{Integrated Intensity Uncertainty}
\label{IntUncert}

For cases where you do not have a calculation from a fit to the
integrated intensity of a spectral line, one can use the following
estimate given a measurement of the baseline RMS and line profile
properties.
\begin{eqnarray}
\int T dv &=& \Delta v_c \sum_{n=1}^N T_n, \\
&\equiv& I. \nonumber
\label{eq:tdv}
\end{eqnarray}
where $\Delta v_c$ is the spectral velocity channel width,
$T_n$ is a spectral channel value, and the line spans N channels.  The
statistical uncertainty of the integrated line intensity is given by:
\begin{eqnarray}
\sigma^2_I &=& \sum_{n=1}^N \sigma^2_{T_n} \left[\frac{\partial I}{\partial T}\right]^2
+ \sigma^2_{\Delta v_c} \left[\frac{\partial I}{\partial (\Delta
    v_c)}\right]^2 \nonumber \\
&=& \sum_{n=1}^N \sigma^2_{T_n} (\Delta v_c)^2 \nonumber \\
&=& N\sigma^2_T (\Delta v_c)^2,
\label{eq:sigI}
\end{eqnarray}
where we have used the fact that we know the velocity
channel width ($\sigma_{\Delta v_c} = 0$) and assuming that all of the
channel noise values are equal.  We can then write
Equation~\ref{eq:sigI} as follows:
\begin{eqnarray}
\sigma_I &=& \sqrt{N} \sigma_T \Delta v_c \nonumber \\
&=&  \sigma_T\sqrt{\Delta v_{line} \Delta v_c},
\label{eq:sigIFinal}
\end{eqnarray}
where we have used the fact that the spectral line width
$\Delta v_{line} = N \Delta v_c$ to get the last expression for
$\sigma_I$.

\section{Hyperfine Ratios and Optical Depth}
\label{HyperTau}

For molecules with measurable hyperfine splitting one can derive an
estimate of the optical depth in a given transition.  Starting with
the form of the radiative transfer equation which involves the
observable \textsl{Source Radiation Temperature T$_R$}
(Equation~\ref{eq:tr}), we can write the ratio of two hyperfine
transitions as follows:
\begin{equation}
\frac{T_R(m)}{T_R(s)} = \frac{f_m\left[J_\nu(T_{ex,m}) -
    J_\nu(T_{bg})\right]\left[1 -
    \exp(-\tau_{\nu,m})\right]}{f_s\left[J_\nu(T_{ex,s}) -
    J_\nu(T_{bg})\right]\left[1 - \exp(-\tau_{\nu,s})\right]},
\label{eq:trrathyper}
\end{equation}
where ``m'' and ``s'' refer to the main and satellite hyperfine
components of a given rotational transition\footnote{Equivalently, one
can use molecular isotopologues with the additional assumptions that
one knows the isotopic ratio and there is no fractionation between
that atomic ratio and the molecular ratio.}.  If we further assume
that the hyperfine levels are in LTE ($J_\nu(T_{ex,m}) = J_\nu(T_{ex,s})$)\footnote{But note that
  anomalous hyperfine excitation has been observed in some
  molecules (see Section~\ref{Hyperfine}), notably the HCN J=$1-0$ transition
  \citep{Loughnane2012}.  This would invalidate the equal-excitation
  temperature assumption.} and that the two hyperfine
components emit over the same volume of gas ($f_m = f_s$), 
Equation~\ref{eq:trrathyper} becomes:
\begin{equation}
\frac{T_R(m)}{T_R(s)} = \frac{1 -
    \exp(-\tau_{\nu,m})}{1 - \exp(-\tau_{\nu,s})}.
\label{eq:trrathyper2}
\end{equation}
In equilibrium the optical depths of the hyperfine transitions within
a given rotational transition should be proportional to their relative
intensities:
\begin{equation}
\frac{\tau_{\nu,s}}{\tau_{\nu,m}} = \frac{R_{i,s}}{R_{i,m}}.
\label{eq:taurat}
\end{equation}
Inserting Equation~\ref{eq:taurat} into Equation~\ref{eq:trrathyper2}
results in a simple equation which allows for the calculation of the
optical depth of a hyperfine transition (and, therefore, the
rotational transition under study) using just the measured intensities
of two of its component hyperfine transitions:
\begin{equation}
\frac{T_R(m)}{T_R(s)} = \frac{1 -
    \exp(-\tau_{\nu,m})}{1 - \exp(-\frac{R_{i,s}}{R_{i,m}}\tau_{\nu,m})}.
\label{eq:trrathyper3}
\end{equation}
Taking the strongest and second-strongest transitions of the NH$_3$
(1,1) transition, from Table~\ref{tab:sixjji11} we find that:
\begin{equation}
\frac{T_R(m)}{T_R(s)}(NH_3 (1,1)) = \frac{1 -
    \exp(-\tau_{\nu,m})}{1 - \exp(-\frac{5}{18}\tau_{\nu,m})}.
\label{eq:trrathypernh3}
\end{equation}

\section{Excitation and Kinetic Temperature}
\label{AmonTexTk}

This section is drawn from Appendix A of \cite{Mangum1992}.
If the metastable ($J=K$; $A_{ij} < 10^{-7}$\,s$^{-1}$) states in
NH$_3$ are coupled only through collisions and the populations in the
upper non-metastable ($J \neq K$; $A_{ij} \sim 0.01$\,s$^{-1}$) states
in each K--ladder can be neglected, the populations in the metastable
states are related through the Boltzmann equation.  In molecular clouds, though,
$\Delta K = 1$ collisions across K-ladders will deplete metastable
states in favor of their next lower J metastable states.  Therefore,
for example, collisional de-excitation of the (2,2) transition will
result in an increase in the population of the (2,1) state, followed
by quick radiative relaxation of the (2,1) state into the (1,1)
state.  This implies that an excitation temperature,
$T_{ex}(J^\prime,K^\prime;J,K)$ relating the populations in the
(J$^\prime$,K$^\prime$) and (J,K) states, n(J$^\prime$,K$^\prime$) and
n(J,K), may be derived.  From the Boltzmann equation
\begin{equation}
\frac{n(J^\prime,K^\prime)}{n(J,K)} =
  \frac{g(J^\prime,K^\prime)}{g(J,K)}
  \exp\left[-\frac{\Delta E(J^\prime,K^\prime;J,K)}{T_{ex}(J^\prime,K^\prime;J,K)}\right],
\label{eq:boltzmannamon}
\end{equation}
and the ratio of level (J$^\prime$,K$^\prime$) and (J,K)
column densities (assuming $h\nu \ll kT_{ex}(J^\prime,K^\prime;J,K)$)
for the (J,K) and (J\arcmin,K\arcmin) transitions
\begin{equation}
\frac{N(J^\prime,K^\prime)}{N(J,K)} = \frac{J^\prime(J^\prime + 1)
K^2 \tau(J^\prime,K^\prime) \Delta v(J^\prime,K^\prime)}{J(J +
  1)(K^\prime)^2 \tau(J,K) \Delta v(J,K)},
\label{eq:nfrac}
\end{equation}
and the fact that in a homogeneous molecular cloud
\begin{equation}
\frac{n(J^\prime,K^\prime)}{n(J,K)} =
\frac{N(J^\prime,K^\prime)}{N(J,K)},
\label{eq:homofrac}
\end{equation}
we find that 
\begin{equation}
\frac{g(J^\prime,K^\prime)}{g(J,K)}\exp\left[-\frac{\Delta
    E(J^\prime,K^\prime;J,K)}{T_{ex}(J^\prime,K^\prime;J,K)}\right] = 
\frac{J^\prime(J^\prime + 1) K^2 \tau(J^\prime,K^\prime) \Delta
    v(J^\prime,K^\prime)}{J(J + 1)(K^\prime)^2 \tau(J,K) \Delta v(J,K)}.
\label{eq:nfrac2}
\end{equation}
Using
\begin{equation}
\tau(J,K) = \left[\sum^{F,F^\prime} R_{F,F^\prime}/\sum^{F,F^\prime}
R_m\right] \tau(J,K,m),
\label{eq:taufrac}
\end{equation}
where $R$ is the relative intensity for a quadrupole
($F$,$F_1$) or main ($m$) hyperfine component and
\begin{eqnarray}
I_{JK} &\equiv& \left[\sum^{F,F^\prime}
  R_{F,F^\prime}/\sum^{F,F^\prime} R_m\right], \nonumber \\
 &=& \frac{1.0}{\frac{5}{12} + \frac{1}{12}}, \nonumber \\
 &=& 2.000 \textrm{ for the (1,1) transition (see
  Tables~\ref{tab:sixjji11} or \ref{tab:hyper11})}, \nonumber \\ 
 &=& \frac{1.0}{\frac{56}{135} + \frac{25}{108} + \frac{3}{20}},
  \nonumber \\
 &=& 1.256 \textrm{ for the (2,K) transitions (see
    Tables~\ref{tab:sixjji2k} or \ref{tab:hyper2k})}, \nonumber \\
 &=& \frac{54}{43}, \nonumber \\
 &=& \frac{1.0}{\frac{45}{112} + \frac{121}{432} + \frac{40}{189}},
  \nonumber \\
 &=& \frac{216}{193}, \nonumber \\
 &=& 1.119 \textrm{ for the (3,K) transitions (see
    Table~\ref{tab:sixjji3k})}, \nonumber \\
 &=& \frac{1.0}{\frac{968}{2475} + \frac{361}{1200} + \frac{35}{144}},
  \nonumber \\
 &=& \frac{2200}{2057}, \nonumber \\
 &=& 1.070 \textrm{ for the (4,K) transitions (see
    Table~\ref{tab:sixjji4k})}, \nonumber 
\label{eq:i}
\end{eqnarray}
for the (1,1) and (2,2) transitions, we can relate the
total optical depth $\tau$(J,K) to the optical depth in the main
hyperfine component $\tau$(J,K,m), noting that for NH$_3$ g(J,K) =
$2J_u + 1$, and solving Equation \ref{eq:nfrac2} for
$T_{ex}(J^\prime,K^\prime;J,K)$ we find that
\begin{eqnarray}
T_{ex}(J^\prime,K^\prime;J,K) &=& 
-\Delta E(J^\prime,K^\prime;J,K) \nonumber \\
 & & \times\left\{\ln\left[\frac{(2J + 1) J^\prime (J^\prime + 1) K^2 I_{J^\prime
      K^\prime}
    \tau(J^\prime,K^\prime,m) \Delta v(J^\prime,K^\prime)}
    {(2J^\prime + 1) J (J + 1) (K^\prime)^2 I_{JK} \tau(J,K,m) \Delta
      v(J,K)}\right]\right\}^{-1}.
\label{eq:texamon}
\end{eqnarray}
Using
\begin{equation}
\frac{T_B(J,K,m)}{T_B(J^\prime,K^\prime,m)} = \frac{1 -
  \exp\left[-\tau(J,K,m)\right]}{1 -
  \exp\left[-\tau(J^\prime,K^\prime,m)\right]},
\label{eq:tbrat}
\end{equation}
which assumes equal excitation temperatures and beam
filling factors in the (J,K) and (J$^\prime$,K$^\prime$) transitions.
Solving Equation \ref{eq:tbrat} for $\tau(J^\prime,K^\prime,m)$ yields
\begin{equation}
\tau(J^\prime,K^\prime,m) = 
-\ln\left[1 - \frac{T_B(J^\prime,K^\prime,m)}{T_B(J,K,m)}
\left\{1 - \exp\left[-\tau(J,K,m)\right]\right\}\right].
\label{eq:tauamon}
\end{equation}
Substituting Equation \ref{eq:tauamon} into Equation~\ref{eq:texamon}:
\begin{multline}
T_{ex}(J^\prime,K^\prime;J,K) = -\Delta E(J^\prime,K^\prime;J,K) \\
 \times\Biggl\{\ln\Biggl[\frac{(2J + 1) J^\prime (J^\prime + 1) K^2
    I_{J^\prime K^\prime} \Delta v(J^\prime,K^\prime)}
    {(2J^\prime + 1) J (J + 1) (K^\prime)^2 I_{JK} \tau(J,K,m) \Delta
      v(J,K)}
 \ln\left(1 - \frac{T_B(J^\prime,K^\prime,m)}{T_B(J,K,m)}
     \left\{1 - \exp\left[-\tau(J,K,m)\right]\right\}\right)\Biggr]\Biggr\}^{-1}.
\label{eq:texamon2}
\end{multline}
For (J$^\prime$,K$^\prime$) = (2,2) and (J,K) = (1,1),
Equation~\ref{eq:texamon2} becomes:
\begin{equation}
T_{ex}(2,2;1,1) = -41.5 \Biggl\{\ln\Biggl[-\frac{0.283 \Delta
     v(2,2)}{\tau(1,1,m) \Delta v(1,1)}
     \ln\left(1 - \frac{T_B(2,2,m)}{T_B(1,1,m)} 
     \left\{1 -
     \exp\left[-\tau(1,1,m)\right]\right\}\right)\Biggr]\Biggr\}^{-1}.
\label{eq:texamon3}
\end{equation}

To derive the gas kinetic temperature from $T_{ex}(2,2;1,1)$, one uses
statistical equilibrium (noting that only collisional processes are allowed
between the different K--ladders), detailed balance, and the Boltzmann equation
to calculate T$_K$ from $T_{ex}(J^\prime,K^\prime;J,K)$.  Assuming that the
populations in the (1,1) and (2,2) transitions are much greater than that in
the higher lying levels of para--NH$_3$ and that the population of the
non--metastable (2,1) level is negligible in comparison to that in the (1,1)
level, we can use this ``three--level model'' of NH$_3$ to analytically 
derive an expression relating $T_{ex}(2,2;1,1)$ and T$_K$
\begin{equation}
1 + \frac{C(2,2;2,1)}{C(2,2;1,1)} = 
\left\{\frac{g(1,1)}{g(2,2)} \exp\left[\frac{\Delta
    E(2,2;1,1)}{T_{ex}(2,2;1,1)}\right]\right\}\left\{\frac{g(2,2)}{g(1,1)}
    \exp\left[-\frac{\Delta E(2,2;1,1)}{T_K}\right]\right\},
\label{eq:texamon4}
\end{equation}
where $C(J^\prime,K^\prime;J,K)$ is the collisional
excitation rate at temperature $T_K$ between levels
(J$^\prime$,K$^\prime$) and (J,K).  Equation \ref{eq:texamon4} can be
re--written as
\begin{equation}
T_{ex}(2,2;1,1)\left\{1 + \left(\frac{T_K}{41.5}\right)
\ln\left[1 + \frac{C(2,2;2,1)}{C(2,2;1,1)}\right]\right\} - T_K = 0.
\label{eq:tkamon}
\end{equation}
Solutions of Equation \ref{eq:tkamon} give $T_K$ for a
measured $T_{ex}(2,2;1,1)$.

\section{NH$_3$ Frequency and Relative Intensity Calculation Tables}
\label{amontabs}

\begin{deluxetable}{cccccccc}
\tablewidth{0pt}
\tablecaption{J--I$_N$ Hyperfine Intensities for
  NH$_3$(1,1)\label{tab:sixjji11}}
\tablehead{
\colhead{$F^\prime_1 \rightarrow F_1$\tablenotemark{a}} &
\colhead{$J^\prime \rightarrow J$\tablenotemark{a}} &
\colhead{a} & \colhead{b} & \colhead{c} &
\colhead{$\frac{(2F^\prime_1 + 1)(2F_1 + 1)}{(2I_N + 1)}$}  
& \colhead{6j} & \colhead{R$_i(F_1,J)$}
}
\startdata
(0,1) & (1,1) & 1 & 1 & 1 & 1 & $-\frac{1}{3}$ & $\frac{1}{9}$ \\
(2,1) & (1,1) & 1 & 1 & 2 & 5 & $\frac{1}{6}$  & $\frac{5}{36}$ \\
(2,2) & (1,1) & 1 & 2 & 1 & $\frac{25}{3}$ & $-\frac{1}{2\sqrt{5}}$ &
                                                 $\frac{5}{12}$ \\
(1,1) & (1,1) & 1 & 1 & 1 & 3 & $\frac{1}{6}$  & $\frac{1}{12}$ \\
(1,2) & (1,1) & 1 & 1 & 2 & 5 & $\frac{1}{6}$  & $\frac{5}{36}$ \\
(1,0) & (1,1) & 1 & 1 & 1 & 1 & $-\frac{1}{3}$ & $\frac{1}{9}$
\enddata
\tablenotetext{a}{$I_N = 1$.}
\end{deluxetable}

\begin{deluxetable}{cccccccc}
\tablewidth{0pt}
\tablecaption{J--I$_N$ Hyperfine Intensities for
  NH$_3$(2,K)\label{tab:sixjji2k}} 
\tablehead{
\colhead{$F^\prime_1 \rightarrow F_1$\tablenotemark{a}} &
\colhead{$J^\prime \rightarrow J$\tablenotemark{a}} &
\colhead{a} & \colhead{b} & \colhead{c} &
\colhead{$\frac{(2F^\prime_1 + 1)(2F_1 + 1)}{(2I_N + 1)}$} &
\colhead{6j} & \colhead{R$_i(F_1,J)$}
}
\startdata
(1,2) & (2,2) & 1 & 2 & 2 & 5 & $-\frac{1}{10}$ & $\frac{1}{20}$ \\
(3,2) & (2,2) & 1 & 2 & 3 & $\frac{35}{3}$ & $\frac{1}{15}$  & $\frac{7}{135}$ \\
(3,3) & (2,2) & 1 & 3 & 2 & $\frac{49}{3}$ & $-\frac{2\sqrt{2}}{3\sqrt{35}}$ &
                                                 $\frac{56}{135}$ \\
(2,2) & (2,2) & 1 & 2 & 2 & $\frac{25}{3}$ & $\frac{1}{6}$  & $\frac{25}{108}$ \\
(1,1) & (2,2) & 1 & 1 & 2 & 3 & $-\frac{1}{2\sqrt{5}}$  & $\frac{3}{20}$ \\
(2,3) & (2,2) & 1 & 2 & 3 & $\frac{35}{3}$ & $\frac{1}{15}$  &
$\frac{7}{135}$ \\
(2,1) & (2,2) & 1 & 2 & 2 & 5 & $-\frac{1}{10}$ & $\frac{1}{20}$
\enddata
\tablenotetext{a}{$I_N = 1$.}
\end{deluxetable}

\begin{deluxetable}{cccccccc}
\tablewidth{0pt}
\tablecaption{J--I$_N$ Hyperfine Intensities for
  NH$_3$(3,K)\label{tab:sixjji3k}}
\tablehead{
\colhead{$F^\prime_1 \rightarrow F_1$\tablenotemark{a}} &
\colhead{$J^\prime \rightarrow J$\tablenotemark{a}} &
\colhead{a} & \colhead{b} & \colhead{c} &
\colhead{$\frac{(2F^\prime_1 + 1)(2F_1 + 1)}{(2I_N + 1)}$} &
\colhead{6j} & \colhead{R$_i(F_1,J)$}
}
\startdata
(2,3) & (3,3) & 1 & 3 & 3 & $\frac{35}{3}$ & $-\frac{1}{21}$ & $\frac{5}{189}$ \\
(4,3) & (3,3) & 1 & 3 & 4 & 21 & $\frac{1}{28}$  & $\frac{3}{112}$ \\
(4,4) & (3,3) & 1 & 4 & 3 & 27 & $-\frac{\sqrt{5}}{4\sqrt{21}}$ & $\frac{45}{112}$ \\
(3,3) & (3,3) & 1 & 3 & 3 & $\frac{49}{3}$ & $\frac{11}{84}$  & $\frac{121}{432}$ \\
(2,2) & (3,3) & 1 & 2 & 3 & $\frac{25}{3}$ & $-\frac{2\sqrt{2}}{3\sqrt{35}}$  & $\frac{40}{189}$ \\
(3,4) & (3,3) & 1 & 3 & 4 & 21 & $\frac{1}{28}$  & $\frac{3}{112}$ \\
(3,2) & (3,3) & 1 & 3 & 3 & $\frac{35}{3}$ & $-\frac{1}{21}$ & $\frac{5}{189}$
\enddata
\tablenotetext{a}{$I_N = 1$.}
\end{deluxetable}

\begin{deluxetable}{cccccccc}
\tablewidth{0pt}
\tablecaption{J--I$_N$ Hyperfine Intensities for
  NH$_3$(4,K)\label{tab:sixjji4k}} 
\tablehead{
\colhead{$F^\prime_1 \rightarrow F_1$\tablenotemark{a}} &
\colhead{$J^\prime \rightarrow J$\tablenotemark{a}} &
\colhead{a} & \colhead{b} & \colhead{c} &
\colhead{$\frac{(2F^\prime_1 + 1)(2F_1 + 1)}{(2I_N + 1)}$} &
\colhead{6j} & \colhead{R$_i(F_1,J)$}
}
\startdata
(3,4) & (4,4) & 1 & 4 & 4 & 21 & $-\frac{1}{36}$ & $\frac{7}{432}$ \\
(5,4) & (4,4) & 1 & 4 & 5 & 33 & $\frac{1}{45}$  & $\frac{11}{675}$ \\
(4,4) & (4,4) & 1 & 4 & 4 & 27 & $\frac{19}{180}$  & $\frac{361}{1200}$ \\
(5,5) & (4,4) & 1 & 5 & 4 & $\frac{121}{3}$ & $-\frac{2\sqrt{2}}{5\sqrt{33}}$ & $\frac{968}{2475}$ \\
(3,3) & (4,4) & 1 & 3 & 4 & $\frac{49}{3}$ & $-\frac{\sqrt{5}}{4\sqrt{21}}$  & $\frac{35}{144}$ \\
(4,5) & (4,4) & 1 & 4 & 5 & 33 & $\frac{1}{45}$  & $\frac{11}{675}$ \\
(4,3) & (4,4) & 1 & 4 & 4 & 21 & $-\frac{1}{36}$ & $\frac{7}{432}$
\enddata
\tablenotetext{a}{$I_N = 1$.}
\end{deluxetable}

\begin{deluxetable}{ccccccccc}
\tablewidth{0pt}
\tabletypesize{\small}
\tablecaption{F$_1$--I$_H$ Hyperfine Frequencies and Intensities for
  NH$_3$(1,1)\label{tab:sixjfi11}} 
\tablehead{
\colhead{$F^\prime \rightarrow F$\tablenotemark{a}} & 
\colhead{$F^\prime_1 \rightarrow F_1$\tablenotemark{a}} & 
\colhead{$\Delta\nu_{HF}$\tablenotemark{b} (kHz)} &
\colhead{a} & \colhead{b} & \colhead{c} &
\colhead{$\frac{(2F^\prime + 1)(2F + 1)}{(2I_H + 1)}$} &  
\colhead{6j} & \colhead{R$_i(F,F_1)$}
}
\startdata
($\frac{1}{2},\frac{1}{2}$) & (0,1) & $-$1568.487
      & $\frac{1}{2}$ & $\frac{1}{2}$ & 1 & 2
      & $\frac{1}{\sqrt{6}}$ & $\frac{1}{3}$ \\

($\frac{1}{2},\frac{3}{2}$) & (0,1) & $-$1526.950
      & $\frac{1}{2}$ & 1 & $\frac{3}{2}$ & 4
      & $-\frac{1}{\sqrt{6}}$ & $\frac{2}{3}$ \\

($\frac{3}{2},\frac{1}{2}$) & (2,1) & $-$623.306
      & $\frac{1}{2}$ & 2 & $\frac{3}{2}$ & 4
      & $\frac{2}{\sqrt{3}}$ & $\frac{1}{3}$ \\

($\frac{5}{2},\frac{3}{2}$) & (2,1) & $-$590.338
      & $\frac{1}{2}$ & 2 & $\frac{5}{2}$ & 12
      & $-\frac{1}{2\sqrt{5}}$ & $\frac{3}{5}$ \\

($\frac{3}{2},\frac{3}{2}$) & (2,1) & $-$580.921
      & $\frac{1}{2}$ & $\frac{3}{2}$ & 2 & 8
      & $\frac{1}{2\sqrt{30}}$ & $\frac{1}{15}$ \\

($\frac{1}{2},\frac{1}{2}$) & (1,1) & $-$36.536
      & $\frac{1}{2}$ & $\frac{1}{2}$ & 1 & 2
      & $-\frac{1}{3}$ & $\frac{2}{9}$ \\

($\frac{3}{2},\frac{1}{2}$) & (1,1) & $-$25.538
      & $\frac{1}{2}$ & 1 & $\frac{3}{2}$ & 4
      & $-\frac{1}{6}$ & $\frac{1}{9}$ \\

($\frac{5}{2},\frac{3}{2}$) & (2,2) & $-$24.394
      & $\frac{1}{2}$ & 2 & $\frac{5}{2}$ & 12
      & $-\frac{1}{10\sqrt{3}}$ & $\frac{1}{25}$ \\

($\frac{3}{2},\frac{3}{2}$) & (2,2) & $-$14.977
      & $\frac{1}{2}$ & $\frac{3}{2}$ & 2 & 8
      & $-\frac{3}{10\sqrt{2}}$ & $\frac{18}{50}$ \\

($\frac{1}{2},\frac{3}{2}$) & (1,1) & $+$5.848
      & $\frac{1}{2}$ & 1 & $\frac{3}{2}$ & 4
      & $-\frac{1}{6}$ & $\frac{1}{9}$ \\

($\frac{5}{2},\frac{5}{2}$) & (2,2) & $+$10.515
      & $\frac{1}{2}$ & $\frac{5}{2}$ & 2 & 18
      & $\frac{\sqrt{7}}{15}$ & $\frac{14}{25}$ \\

($\frac{3}{2},\frac{3}{2}$) & (1,1) & $+$16.847
      & $\frac{1}{2}$ & $\frac{3}{2}$ & 1 & 8
      & $\frac{\sqrt{5}}{6\sqrt{2}}$ & $\frac{5}{9}$ \\

($\frac{3}{2},\frac{5}{2}$) & (2,2) & $+$19.932
      & $\frac{1}{2}$ & 2 & $\frac{5}{2}$ & 12
      & $-\frac{1}{10\sqrt{3}}$ & $\frac{1}{25}$ \\

($\frac{1}{2},\frac{3}{2}$) & (1,2) & $+$571.792
      & $\frac{1}{2}$ & 2 & $\frac{3}{2}$ & 4
      & $\frac{1}{2\sqrt{3}}$ & $\frac{1}{3}$ \\

($\frac{3}{2},\frac{3}{2}$) & (1,2) & $+$582.790
      & $\frac{1}{2}$ & $\frac{3}{2}$ & 2 & 8
      & $\frac{1}{2\sqrt{30}}$ & $\frac{1}{15}$ \\

($\frac{3}{2},\frac{5}{2}$) & (1,2) & $+$617.700
      & $\frac{1}{2}$ & 2 & $\frac{5}{2}$ & 12
      & $-\frac{1}{2\sqrt{5}}$ & $\frac{3}{5}$ \\

($\frac{1}{2},\frac{1}{2}$) & (1,0) & $+$1534.050
      & $\frac{1}{2}$ & $\frac{1}{2}$ & 1 & 2
      & $\frac{1}{\sqrt{6}}$ & $\frac{1}{3}$ \\

($\frac{3}{2},\frac{1}{2}$) & (1,0) & $+$1545.049
      & $\frac{1}{2}$ & 1 & $\frac{3}{2}$ & 4
      & $-\frac{1}{\sqrt{6}}$ & $\frac{2}{3}$

\enddata
\tablenotetext{a}{$F = I_H + F_1$ and $F_1 = J + I_N$, where
  $I_N = 1$ and $I_H = \frac{1}{2}$.}
\tablenotetext{b}{Frequency offset in kHz relative to 23694.495487 kHz
  \citep[][ Table~I (calculated)]{Kukolich1967}.}
\end{deluxetable}

\begin{deluxetable}{ccccccccc}
\tablewidth{0pt}
\tablecaption{F$_1$--I$_H$ Hyperfine Frequencies and Intensities for
  NH$_3$(2,2)\label{tab:sixjfi22}} 
\tablehead{
\colhead{$F^\prime \rightarrow F$\tablenotemark{a}} & 
\colhead{$F^\prime_1 \rightarrow F_1$\tablenotemark{a}} & 
\colhead{$\Delta\nu_{HF}$\tablenotemark{b} (kHz)} &
\colhead{a} & \colhead{b} & \colhead{c} &
\colhead{$\frac{(2F^\prime + 1)(2F + 1)}{(2I_H + 1)}$} &  
\colhead{6j} & \colhead{R$_i(F,F_1)$}
}
\startdata
($\frac{3}{2},\frac{3}{2}$) & (1,2) & $-$2099.027
      & $\frac{1}{2}$ & $\frac{3}{2}$ & 2 & 8
      & $\frac{1}{2\sqrt{30}}$ & $\frac{1}{15}$ \\

($\frac{3}{2},\frac{5}{2}$) & (1,2) & $-$2058.265
      & $\frac{1}{2}$ & 2 & $\frac{5}{2}$ & 12
      & $-\frac{1}{2\sqrt{5}}$ & $\frac{3}{5}$ \\

($\frac{1}{2},\frac{3}{2}$) & (1,2) & $-$2053.464
      & $\frac{1}{2}$ & 2 & $\frac{3}{2}$ & 4
      & $\frac{1}{2\sqrt{3}}$ & $\frac{1}{3}$ \\

($\frac{7}{2},\frac{5}{2}$) & (3,2) & $-$1297.079
      & $\frac{1}{2}$ & 3 & $\frac{7}{2}$ & 24
      & $-\frac{1}{\sqrt{42}}$ & $\frac{4}{7}$ \\

($\frac{5}{2},\frac{3}{2}$) & (3,2) & $-$1296.096
      & $\frac{1}{2}$ & 3 & $\frac{5}{2}$ & 12
      & $\frac{1}{\sqrt{30}}$ & $\frac{6}{15}$ \\

($\frac{5}{2},\frac{5}{2}$) & (3,2) & $-$1255.335
      & $\frac{1}{2}$ & $\frac{5}{2}$ & 3 & 18
      & $\frac{1}{3\sqrt{70}}$ & $\frac{1}{35}$ \\

($\frac{3}{2},\frac{1}{2}$) & (1,1) & $-$44.511
      & $\frac{1}{2}$ & 1 & $\frac{3}{2}$ & 4
      & $-\frac{1}{6}$ & $\frac{1}{9}$ \\

($\frac{5}{2},\frac{3}{2}$) & (2,2) & $-$41.813
      & $\frac{1}{2}$ & 2 & $\frac{5}{2}$ & 12
      & $-\frac{1}{10\sqrt{3}}$ & $\frac{1}{25}$ \\

($\frac{7}{2},\frac{5}{2}$) & (3,3) & $-$41.444
      & $\frac{1}{2}$ & 3 & $\frac{7}{2}$ & 24
      & $-\frac{1}{14\sqrt{6}}$ & $\frac{1}{49}$ \\

($\frac{5}{2},\frac{5}{2}$) & (2,2) & $-$1.051
      & $\frac{1}{2}$ & $\frac{5}{2}$ & 2 & 18
      & $\frac{\sqrt{7}}{15}$ & $\frac{14}{25}$ \\

($\frac{3}{2},\frac{3}{2}$) & (2,2) & $-$1.051
      & $\frac{1}{2}$ & $\frac{3}{2}$ & 2 & 8
      & $-\frac{3}{10\sqrt{2}}$ & $\frac{9}{25}$ \\

($\frac{7}{2},\frac{7}{2}$) & (3,3) & $+$0.309
      & $\frac{1}{2}$ & $\frac{7}{2}$ & 3 & 32
      & $\frac{3\sqrt{3}}{28\sqrt{2}}$ & $\frac{27}{49}$ \\

($\frac{5}{2},\frac{5}{2}$) & (3,3) & $+$0.309
      & $\frac{1}{2}$ & $\frac{5}{2}$ & 3 & 18
      & $-\frac{\sqrt{10}}{21}$ & $\frac{20}{49}$ \\

($\frac{3}{2},\frac{3}{2}$) & (1,1) & $+$1.054
      & $\frac{1}{2}$ & $\frac{3}{2}$ & 1 & 8
      & $\frac{\sqrt{5}}{6\sqrt{2}}$ & $\frac{5}{9}$ \\

($\frac{1}{2},\frac{1}{2}$) & (1,1) & $+$1.054
      & $\frac{1}{2}$ & $\frac{1}{2}$ & 1 & 2
      & $-\frac{1}{3}$ & $\frac{2}{9}$ \\

($\frac{3}{2},\frac{5}{2}$) & (2,2) & $+$39.710
      & $\frac{1}{2}$ & 2 & $\frac{5}{2}$ & 12
      & $-\frac{1}{10\sqrt{3}}$ & $\frac{1}{25}$ \\

($\frac{5}{2},\frac{7}{2}$) & (3,3) & $+$42.045
      & $\frac{1}{2}$ & 3 & $\frac{7}{2}$ & 24
      & $-\frac{1}{14\sqrt{6}}$ & $\frac{1}{49}$ \\

($\frac{1}{2},\frac{3}{2}$) & (1,1) & $+$46.614
      & $\frac{1}{2}$ & 1 & $\frac{3}{2}$ & 4
      & $-\frac{1}{6}$ & $\frac{1}{9}$ \\

($\frac{5}{2},\frac{5}{2}$) & (2,3) & $+$1254.584
      & $\frac{1}{2}$ & $\frac{5}{2}$ & 3 & 18
      & $\frac{1}{3\sqrt{70}}$ & $\frac{1}{35}$ \\

($\frac{3}{2},\frac{5}{2}$) & (2,3) & $+$1295.345
      & $\frac{1}{2}$ & 3 & $\frac{5}{2}$ & 12
      & $\frac{1}{\sqrt{30}}$ & $\frac{6}{15}$ \\

($\frac{5}{2},\frac{7}{2}$) & (2,3) & $+$1296.328
      & $\frac{1}{2}$ & 3 & $\frac{7}{2}$ & 24
      & $-\frac{1}{\sqrt{42}}$ & $\frac{4}{7}$ \\

($\frac{3}{2},\frac{1}{2}$) & (2,1) & $+$2053.464
      & $\frac{1}{2}$ & 2 & $\frac{3}{2}$ & 4
      & $\frac{1}{2\sqrt{3}}$ & $\frac{1}{3}$ \\

($\frac{5}{2},\frac{3}{2}$) & (2,1) & $+$2058.265
      & $\frac{1}{2}$ & 2 & $\frac{5}{2}$ & 12
      & $-\frac{1}{2\sqrt{5}}$ & $\frac{3}{5}$ \\

($\frac{3}{2},\frac{3}{2}$) & (2,1) & $+$2099.027
      & $\frac{1}{2}$ & $\frac{3}{2}$ & 2 & 8
      & $\frac{1}{2\sqrt{30}}$ & $\frac{1}{15}$

\enddata
\tablenotetext{a}{$F = I_H + F_1$ and $F_1 = J + I_N$, where
  $I_N = 1$ and $I_H = \frac{1}{2}$.}
\tablenotetext{b}{Frequency offset in kHz relative to 23722633.335\,kHz
  \cite[][ Table~II (calculated)]{Kukolich1967}} 
\end{deluxetable}

%
%
\begin{deluxetable}{ccccccccc}
\tablewidth{0pt}
\tablecaption{F$_1$--I$_H$ Hyperfine Frequencies and Intensities for
  NH$_3$(3,3)\label{tab:sixjfi33}} 
\tablehead{
\colhead{$F^\prime \rightarrow F$\tablenotemark{a}} & 
\colhead{$F^\prime_1 \rightarrow F_1$\tablenotemark{a}} & 
\colhead{$\Delta\nu_{HF}$\tablenotemark{b} (kHz)} &
\colhead{a} & \colhead{b} & \colhead{c} &
\colhead{$\frac{(2F^\prime + 1)(2F + 1)}{(2I_H + 1)}$} &  
\colhead{6j} & \colhead{R$_i(F,F_1)$}
}
\startdata

($\frac{7}{2},\frac{9}{2}$) & (2,3) & $-$2324.577
      & $\frac{3}{2}$ & 3 & $\frac{9}{2}$ & 20
      & $-\frac{1}{2\sqrt{14}}$ & $\frac{5}{14}$ \\

($\frac{5}{2},\frac{7}{2}$) & (2,3) & $-$2312.558
      & $\frac{3}{2}$ & 3 & $\frac{7}{2}$ & 12
      & $-\frac{1}{7}$ & $\frac{12}{49}$ \\

($\frac{1}{2},\frac{3}{2}$) & (2,3) & $-$2304.415
      & $\frac{3}{2}$ & 3 & $\frac{3}{2}$ & 2
      & $\frac{1}{2\sqrt{5}}$ & $\frac{1}{10}$ \\

($\frac{3}{2},\frac{5}{2}$) & (2,3) & $-$2301.989
      & $\frac{3}{2}$ & 3 & $\frac{5}{2}$ & 6
      & $-\frac{\sqrt{2}}{5\sqrt{3}}$ & $\frac{4}{25}$ \\

($\frac{7}{2},\frac{7}{2}$) & (2,3) & \ldots
      & $\frac{3}{2}$ & $\frac{7}{2}$ & 3 & 16
      & $\frac{1}{14\sqrt{2}}$ & $\frac{2}{49}$ \\

($\frac{5}{2},\frac{5}{2}$) & (2,3) & \ldots
      & $\frac{3}{2}$ & $\frac{5}{2}$ & 3 & 9
      & $-\frac{8}{105}$ & $\frac{64}{1225}$ \\

($\frac{3}{2},\frac{3}{2}$) & (2,3) & \ldots
      & $\frac{3}{2}$ & $\frac{3}{2}$ & 3 & 4
      & $\frac{1}{10}$ & $\frac{1}{25}$ \\

($\frac{7}{2},\frac{5}{2}$) & (2,3) & \ldots
      & $\frac{3}{2}$ & $\frac{5}{2}$ & 3 & 12
      & $-\frac{1}{14\sqrt{30}}$ & $\frac{1}{490}$ \\

($\frac{5}{2},\frac{3}{2}$) & (2,3) & \ldots
      & $\frac{3}{2}$ & $\frac{3}{2}$ & 3 & 6
      & $\frac{1}{10\sqrt{21}}$ & $\frac{1}{350}$ \\

($\frac{7}{2},\frac{5}{2}$) & (4,3) & $-$1690.763
      & $\frac{3}{2}$ & 4 & $\frac{7}{2}$ & 12
      & $-\frac{5}{28\sqrt{2}}$ & $\frac{150}{784}$ \\

($\frac{9}{2},\frac{7}{2}$) & (4,3) & $-$1689.154
      & $\frac{3}{2}$ & 4 & $\frac{9}{2}$ & 20
      & $\frac{\sqrt{11}}{12\sqrt{6}}$ & $\frac{110}{432}$ \\

($\frac{5}{2},\frac{3}{2}$) & (4,3) & $-$1682.925
      & $\frac{3}{2}$ & 4 & $\frac{5}{2}$ & 6
      & $\frac{1}{\sqrt{42}}$ & $\frac{1}{7}$ \\

($\frac{11}{2},\frac{9}{2}$) & (4,3) & $-$1679.029
      & $\frac{3}{2}$ & 4 & $\frac{11}{2}$ & 30
      & $-\frac{1}{3\sqrt{10}}$ & $\frac{1}{3}$ \\

($\frac{9}{2},\frac{9}{2}$) & (4,3) & \ldots
      & $\frac{3}{2}$ & $\frac{9}{2}$ & 4 & 25
      & $\frac{1}{6\sqrt{30}}$ & $\frac{5}{216}$ \\

($\frac{7}{2},\frac{7}{2}$) & (4,3) & \ldots
      & $\frac{3}{2}$ & $\frac{7}{2}$ & 4 & 16
      & $-\frac{\sqrt{5}}{21\sqrt{6}}$ & $\frac{40}{1323}$ \\

($\frac{5}{2},\frac{5}{2}$) & (4,3) & \ldots
      & $\frac{3}{2}$ & $\frac{5}{2}$ & 4 & 9
      & $\frac{1}{14\sqrt{2}}$ & $\frac{9}{392}$ \\

($\frac{5}{2},\frac{7}{2}$) & (4,3) & \ldots
      & $\frac{3}{2}$ & 3 & $\frac{7}{2}$ & 12
      & $\frac{1}{84\sqrt{2}}$ & $\frac{1}{1176}$ \\

($\frac{7}{2},\frac{9}{2}$) & (4,3) & \ldots
      & $\frac{3}{2}$ & 3 & $\frac{9}{2}$ & 20
      & $-\frac{1}{12\sqrt{210}}$ & $\frac{1}{1512}$ \\

($\frac{7}{2},\frac{5}{2}$) & (2,2) & \ldots
      & $\frac{3}{2}$ & 2 & $\frac{7}{2}$ & 12
      & $-\frac{1}{\sqrt{210}}$ & $\frac{6}{105}$ \\

($\frac{5}{2},\frac{3}{2}$) & (2,2) & $-$80.030
      & $\frac{3}{2}$ & 2 & $\frac{5}{2}$ & 6
      & $\frac{\sqrt{7}}{10\sqrt{6}}$ & $\frac{7}{100}$ \\

($\frac{3}{2},\frac{1}{2}$) & (2,2) & \ldots
      & $\frac{3}{2}$ & 2 & $\frac{3}{2}$ & 2
      & $-\frac{1}{2\sqrt{10}}$ & $\frac{1}{20}$ \\

($\frac{5}{2},\frac{3}{2}$) & (3,3) & $-$64.182
      & $\frac{3}{2}$ & 3 & $\frac{5}{2}$ & 6
      & $-\frac{1}{\sqrt{210}}$ & $\frac{3}{105}$ \\

($\frac{7}{2},\frac{5}{2}$) & (3,3) & \ldots
      & $\frac{3}{2}$ & 3 & $\frac{7}{2}$ & 12
      & $\frac{\sqrt{5}}{28\sqrt{2}}$ & $\frac{15}{392}$ \\

($\frac{9}{2},\frac{7}{2}$) & (3,3) & $-$61.951
      & $\frac{3}{2}$ & 3 & $\frac{9}{2}$ & 20
      & $-\frac{1}{4\sqrt{42}}$ & $\frac{5}{168}$ \\

($\frac{11}{2},\frac{9}{2}$) & (4,4) & $-$50.048
      & $\frac{3}{2}$ & 4 & $\frac{11}{2}$ & 30
      & $-\frac{1}{5\sqrt{66}}$ & $\frac{1}{55}$ \\

($\frac{9}{2},\frac{7}{2}$) & (4,4) & \ldots
      & $\frac{3}{2}$ & 4 & $\frac{9}{2}$ & 20
      & $\frac{\sqrt{77}}{180\sqrt{2}}$ & $\frac{77}{3240}$ \\

($\frac{7}{2},\frac{5}{2}$) & (4,4) & \ldots
      & $\frac{3}{2}$ & 4 & $\frac{7}{2}$ & 12
      & $-\frac{1}{4\sqrt{42}}$ & $\frac{1}{56}$ \\

($\frac{3}{2},\frac{3}{2}$) & (3,3) & $-$1.334
      & $\frac{3}{2}$ & $\frac{3}{2}$ & 3 & 4
      & $-\frac{1}{\sqrt{35}}$ & $\frac{4}{35}$ \\

($\frac{9}{2},\frac{9}{2}$) & (3,3) & \ldots
      & $\frac{3}{2}$ & $\frac{9}{2}$ & 3 & 25
      & $\frac{\sqrt{11}}{2\sqrt{210}}$ & $\frac{275}{840}$ \\

($\frac{7}{2},\frac{7}{2}$) & (3,3) & \ldots
      & $\frac{3}{2}$ & $\frac{7}{2}$ & 3 & 16
      & $-\frac{\sqrt{2}}{7\sqrt{3}}$ & $\frac{32}{147}$ \\

($\frac{5}{2},\frac{5}{2}$) & (3,3) & \ldots
      & $\frac{3}{2}$ & $\frac{5}{2}$ & 3 & 9
      & $\frac{17}{42\sqrt{10}}$ & $\frac{2601}{17640}$ \\

($\frac{5}{2},\frac{5}{2}$) & (4,4) & $+$0.445
      & $\frac{3}{2}$ & $\frac{5}{2}$ & 4 & 9
      & $-\frac{5}{6\sqrt{42}}$ & $\frac{25}{168}$ \\

($\frac{7}{2},\frac{7}{2}$) & (4,4) & \ldots
      & $\frac{3}{2}$ & $\frac{7}{2}$ & 4 & 16
      & $\frac{4\sqrt{2}}{9\sqrt{35}}$ & $\frac{512}{2835}$ \\

($\frac{9}{2},\frac{9}{2}$) & (4,4) & \ldots
      & $\frac{3}{2}$ & $\frac{9}{2}$ & 4 & 25
      & $-\frac{41}{90\sqrt{22}}$ & $\frac{1681}{7128}$ \\

($\frac{11}{2},\frac{11}{2}$) & (4,4) & \ldots
      & $\frac{3}{2}$ & $\frac{11}{2}$ & 4 & 36
      & $\frac{\sqrt{13}}{3\sqrt{165}}$ & $\frac{52}{165}$ \\

($\frac{1}{2},\frac{1}{2}$) & (2,2) & $+$1.067
      & $\frac{3}{2}$ & $\frac{1}{2}$ & 2 & 1
      & $-\frac{1}{2\sqrt{5}}$ & $\frac{1}{20}$ \\

($\frac{3}{2},\frac{3}{2}$) & (2,2) & \ldots
      & $\frac{3}{2}$ & $\frac{3}{2}$ & 2 & 4
      & $\frac{1}{5\sqrt{2}}$ & $\frac{2}{25}$ \\

($\frac{5}{2},\frac{5}{2}$) & (2,2) & \ldots
      & $\frac{3}{2}$ & $\frac{5}{2}$ & 2 & 9
      & $-\frac{11}{3\sqrt{7}}$ & $\frac{1089}{6300}$ \\

($\frac{7}{2},\frac{7}{2}$) & (2,2) & \ldots
      & $\frac{3}{2}$ & $\frac{7}{2}$ & 2 & 16
      & $\frac{\sqrt{3}}{2\sqrt{35}}$ & $\frac{12}{35}$ \\

($\frac{9}{2},\frac{11}{2}$) & (4,4) & $+$50.937
      & $\frac{3}{2}$ & 4 & $\frac{11}{2}$ & 30
      & $-\frac{1}{5\sqrt{66}}$ & $\frac{1}{55}$ \\

($\frac{7}{2},\frac{9}{2}$) & (4,4) & \ldots
      & $\frac{3}{2}$ & 4 & $\frac{9}{2}$ & 20
      & $\frac{\sqrt{77}}{180\sqrt{2}}$ & $\frac{77}{3240}$ \\

($\frac{5}{2},\frac{7}{2}$) & (4,4) & \ldots
      & $\frac{3}{2}$ & 4 & $\frac{7}{2}$ & 12
      & $-\frac{1}{4\sqrt{42}}$ & $\frac{1}{56}$ \\

($\frac{7}{2},\frac{9}{2}$) & (3,3) & $+$59.283
      & $\frac{3}{2}$ & 3 & $\frac{9}{2}$ & 20
      & $-\frac{1}{4\sqrt{42}}$ & $\frac{5}{168}$ \\

($\frac{5}{2},\frac{7}{2}$) & (3,3) & \ldots
      & $\frac{3}{2}$ & 3 & $\frac{7}{2}$ & 12
      & $\frac{\sqrt{5}}{28\sqrt{2}}$ & $\frac{15}{392}$ \\

($\frac{3}{2},\frac{5}{2}$) & (3,3) & \ldots
      & $\frac{3}{2}$ & 3 & $\frac{5}{2}$ & 6
      & $-\frac{1}{\sqrt{210}}$ & $\frac{3}{105}$ \\

($\frac{1}{2},\frac{3}{2}$) & (2,2) & \ldots
      & $\frac{3}{2}$ & 2 & $\frac{3}{2}$ & 2
      & $-\frac{1}{2\sqrt{10}}$ & $\frac{1}{20}$ \\

($\frac{3}{2},\frac{5}{2}$) & (2,2) & $+$82.164
      & $\frac{3}{2}$ & 2 & $\frac{5}{2}$ & 6
      & $\frac{\sqrt{7}}{10\sqrt{6}}$ & $\frac{7}{100}$ \\

($\frac{5}{2},\frac{7}{2}$) & (2,2) & \ldots
      & $\frac{3}{2}$ & 2 & $\frac{7}{2}$ & 12
      & $-\frac{1}{\sqrt{210}}$ & $\frac{6}{105}$ \\

\multicolumn{9}{c}{Continued on next page}
\enddata
\end{deluxetable}

\begin{deluxetable}{ccccccccc}
\tablewidth{0pt}
\tablenum{17}
\tablecaption{F$_1$--I$_H$ Hyperfine Frequencies and Intensities for
  NH$_3$(3,3) (continued)} 
\tablehead{
\colhead{$F^\prime \rightarrow F$\tablenotemark{a}} & 
\colhead{$F^\prime_1 \rightarrow F_1$\tablenotemark{a}} & 
\colhead{$\Delta\nu_{HF}$\tablenotemark{b} (kHz)} &
\colhead{a} & \colhead{b} & \colhead{c} &
\colhead{$\frac{(2F^\prime + 1)(2F + 1)}{(2I_H + 1)}$} &  
\colhead{6j} & \colhead{R$_i(F,F_1)$}
}
\startdata

($\frac{3}{2},\frac{5}{2}$) & (3,4) & $+$1682.036
      & $\frac{3}{2}$ & 4 & $\frac{5}{2}$ & 6
      & $\frac{1}{\sqrt{42}}$ & $\frac{1}{7}$ \\

($\frac{7}{2},\frac{9}{2}$) & (3,4) & $+$1688.264
      & $\frac{3}{2}$ & 4 & $\frac{9}{2}$ & 20
      & $\frac{\sqrt{11}}{12\sqrt{6}}$ & $\frac{55}{216}$ \\

($\frac{5}{2},\frac{7}{2}$) & (3,4) & $+$1689.873
      & $\frac{3}{2}$ & 4 & $\frac{7}{2}$ & 12
      & $-\frac{5}{28\sqrt{2}}$ & $\frac{75}{392}$ \\

($\frac{9}{2},\frac{11}{2}$) & (3,4) & $+$1678.140
      & $\frac{3}{2}$ & 4 & $\frac{11}{2}$ & 30
      & $-\frac{1}{3\sqrt{10}}$ & $\frac{1}{3}$ \\

($\frac{9}{2},\frac{7}{2}$) & (3,4) & \ldots
      & $\frac{3}{2}$ & $\frac{7}{2}$ & 4 & 20
      & $-\frac{1}{12\sqrt{210}}$ & $\frac{1}{1512}$ \\

($\frac{7}{2},\frac{5}{2}$) & (3,4) & \ldots
      & $\frac{3}{2}$ & $\frac{5}{2}$ & 4 & 12
      & $\frac{1}{84\sqrt{2}}$ & $\frac{1}{1176}$ \\

($\frac{9}{2},\frac{9}{2}$) & (3,4) & \ldots
      & $\frac{3}{2}$ & $\frac{9}{2}$ & 4 & 25
      & $\frac{1}{6\sqrt{30}}$ & $\frac{5}{216}$ \\

($\frac{7}{2},\frac{7}{2}$) & (3,4) & \ldots
      & $\frac{3}{2}$ & $\frac{7}{2}$ & 4 & 16
      & $-\frac{\sqrt{5}}{21\sqrt{6}}$ & $\frac{40}{1323}$ \\

($\frac{5}{2},\frac{5}{2}$) & (3,4) & \ldots
      & $\frac{3}{2}$ & $\frac{5}{2}$ & 4 & 9
      & $\frac{1}{14\sqrt{2}}$ & $\frac{9}{392}$ \\

($\frac{5}{2},\frac{3}{2}$) & (3,2) & $+$2301.723
      & $\frac{3}{2}$ & 3 & $\frac{5}{2}$ & 6
      & $-\frac{\sqrt{2}}{5\sqrt{3}}$ & $\frac{4}{25}$ \\

($\frac{3}{2},\frac{1}{2}$) & (3,2) & $+$2304.148
      & $\frac{3}{2}$ & 3 & $\frac{3}{2}$ & 2
      & $\frac{1}{2\sqrt{5}}$ & $\frac{1}{10}$ \\

($\frac{7}{2},\frac{5}{2}$) & (3,2) & $+$2312.291
      & $\frac{3}{2}$ & 3 & $\frac{7}{2}$ & 12
      & $\frac{1}{7}$ & $\frac{12}{49}$ \\

($\frac{9}{2},\frac{7}{2}$) & (3,2) & $+$2324.310
      & $\frac{3}{2}$ & 3 & $\frac{9}{2}$ & 20
      & $-\frac{1}{2\sqrt{14}}$ & $\frac{5}{14}$ \\

($\frac{5}{2},\frac{7}{2}$) & (3,2) & \ldots
      & $\frac{3}{2}$ & 2 & $\frac{7}{2}$ & 12
      & $-\frac{1}{14\sqrt{30}}$ & $\frac{1}{490}$ \\

($\frac{3}{2},\frac{5}{2}$) & (3,2) & \ldots
      & $\frac{3}{2}$ & 2 & $\frac{5}{2}$ & 6
      & $\frac{1}{10\sqrt{21}}$ & $\frac{1}{350}$ \\

($\frac{3}{2},\frac{3}{2}$) & (3,2) & \ldots
      & $\frac{3}{2}$ & $\frac{3}{2}$ & 3 & 4
      & $\frac{1}{10}$ & $\frac{1}{25}$ \\

($\frac{5}{2},\frac{5}{2}$) & (3,2) & \ldots
      & $\frac{3}{2}$ & $\frac{5}{2}$ & 3 & 9
      & $-\frac{8}{105}$ & $\frac{64}{1225}$ \\

($\frac{7}{2},\frac{7}{2}$) & (3,2) & \ldots
      & $\frac{3}{2}$ & $\frac{7}{2}$ & 3 & 16
      & $\frac{1}{14\sqrt{2}}$ & $\frac{2}{49}$

\enddata
\tablenotetext{a}{$F = I_H + F_1$ and $F_1 = J + I_N$, where
  $I_N = 1$ and $I_H = \frac{3}{2}$.}
\tablenotetext{b}{Frequency offset in kHz relative to 23870129.183\,kHz
  \cite[][ Table~III (calculated)]{Kukolich1967}.} 
\end{deluxetable}

\begin{deluxetable}{ccccccccc}
\tablewidth{0pt}
\tablecaption{F$_1$--I$_H$ Hyperfine Frequencies and Intensities for
  NH$_3$(4,4)\label{tab:sixjfi44}} 
\tablehead{
\colhead{$F^\prime \rightarrow F$\tablenotemark{a}} & 
\colhead{$F^\prime_1 \rightarrow F_1$\tablenotemark{a}} & 
\colhead{$\Delta\nu_{HF}$\tablenotemark{b} (kHz)} &
\colhead{a} & \colhead{b} & \colhead{c} &
\colhead{$\frac{(2F^\prime + 1)(2F + 1)}{(2I_H + 1)}$} &  
\colhead{6j} & \colhead{R$_i(F,F_1)$}
}
\startdata
($\frac{7}{2},\frac{9}{2}$) & (3,4) & $-$2485.883
      & $\frac{1}{2}$ & 4 & $\frac{9}{2}$ & 40
      & $-\frac{1}{6\sqrt{2}}$ & $\frac{5}{9}$ \\

($\frac{5}{2},\frac{7}{2}$) & (3,4) & $-$2481.291
      & $\frac{1}{2}$ & 4 & $\frac{7}{2}$ & 24
      & $\frac{1}{2\sqrt{14}}$ & $\frac{3}{7}$ \\

($\frac{7}{2},\frac{7}{2}$) & (3,4) & \ldots
      & $\frac{1}{2}$ & $\frac{7}{2}$ & 4 & 32
      & $\frac{1}{12\sqrt{14}}$ & $\frac{1}{63}$ \\

($\frac{9}{2},\frac{7}{2}$) & (5,4) & $-$1924.144
      & $\frac{1}{2}$ & 5 & $\frac{9}{2}$ & 40
      & $\frac{1}{3\sqrt{10}}$ & $\frac{4}{9}$ \\

($\frac{11}{2},\frac{9}{2}$) & (5,4) & $-$1922.222
      & $\frac{1}{2}$ & 5 & $\frac{11}{2}$ & 60
      & $-\frac{1}{\sqrt{110}}$ & $\frac{6}{11}$ \\

($\frac{9}{2},\frac{9}{2}$) & (5,4) & \ldots
      & $\frac{1}{2}$ & $\frac{9}{2}$ & 5 & 50
      & $\frac{1}{15\sqrt{22}}$ & $\frac{1}{99}$ \\

($\frac{9}{2},\frac{9}{2}$) & (4,4) & \ldots
      & $\frac{1}{2}$ & $\frac{9}{2}$ & 4 & 50
      & $\frac{\sqrt{22}}{45}$ & $\frac{44}{81}$ \\

($\frac{7}{2},\frac{7}{2}$) & (4,4) & $-$1.448
      & $\frac{1}{2}$ & $\frac{7}{2}$ & 4 & 32
      & $-\frac{\sqrt{35}}{36\sqrt{2}}$ & $\frac{35}{81}$ \\

($\frac{9}{2},\frac{7}{2}$) & (4,4) & \ldots
      & $\frac{1}{2}$ & 4 & $\frac{9}{2}$ & 40
      & $-\frac{1}{18\sqrt{10}}$ & $\frac{1}{81}$ \\

($\frac{7}{2},\frac{9}{2}$) & (4,4) & \ldots
      & $\frac{1}{2}$ & 4 & $\frac{9}{2}$ & 40
      & $-\frac{1}{18\sqrt{10}}$ & $\frac{1}{81}$ \\

($\frac{11}{2},\frac{11}{2}$) & (5,5) & \ldots
      & $\frac{1}{2}$ & $\frac{11}{2}$ & 5 & 72
      & $\frac{\sqrt{65}}{66\sqrt{2}}$ & $\frac{65}{121}$ \\

($\frac{9}{2},\frac{9}{2}$) & (5,5) & $+$0.526
      & $\frac{1}{2}$ & $\frac{9}{2}$ & 5 & 50
      & $-\frac{3\sqrt{3}}{55}$ & $\frac{54}{121}$ \\

($\frac{11}{2},\frac{9}{2}$) & (5,5) & \ldots
      & $\frac{1}{2}$ & 5 & $\frac{11}{2}$ & 60
      & $-\frac{1}{22\sqrt{15}}$ & $\frac{1}{121}$ \\

($\frac{9}{2},\frac{11}{2}$) & (5,5) & \ldots
      & $\frac{1}{2}$ & 5 & $\frac{11}{2}$ & 60
      & $-\frac{1}{22\sqrt{15}}$ & $\frac{1}{121}$ \\

($\frac{7}{2},\frac{7}{2}$) & (3,3) & \ldots
      & $\frac{1}{2}$ & $\frac{7}{2}$ & 3 & 32
      & $\frac{3\sqrt{3}}{28\sqrt{2}}$ & $\frac{27}{49}$ \\

($\frac{5}{2},\frac{5}{2}$) & (3,3) & \ldots
      & $\frac{1}{2}$ & $\frac{5}{2}$ & 3 & 18
      & $-\frac{\sqrt{10}}{21}$ & $\frac{20}{49}$ \\

($\frac{7}{2},\frac{5}{2}$) & (3,3) & \ldots
      & $\frac{1}{2}$ & 3 & $\frac{7}{2}$ & 24
      & $-\frac{1}{14\sqrt{6}}$ & $\frac{1}{49}$ \\

($\frac{5}{2},\frac{7}{2}$) & (3,3) & \ldots
      & $\frac{1}{2}$ & 3 & $\frac{7}{2}$ & 24
      & $-\frac{1}{14\sqrt{6}}$ & $\frac{1}{49}$ \\

($\frac{9}{2},\frac{11}{2}$) & (4,5) & $+$1931.301
      & $\frac{1}{2}$ & 5 & $\frac{11}{2}$ & 60
      & $-\frac{1}{\sqrt{110}}$ & $\frac{6}{11}$ \\

($\frac{7}{2},\frac{9}{2}$) & (4,5) & $+$1923.222
      & $\frac{1}{2}$ & 5 & $\frac{9}{2}$ & 40
      & $\frac{1}{3\sqrt{10}}$ & $\frac{4}{9}$ \\

($\frac{9}{2},\frac{9}{2}$) & (4,5) & \ldots
      & $\frac{1}{2}$ & $\frac{9}{2}$ & 5 & 50
      & $\frac{1}{15\sqrt{22}}$ & $\frac{1}{99}$ \\

($\frac{7}{2},\frac{5}{2}$) & (4,3) & $+$2480.878
      & $\frac{1}{2}$ & 4 & $\frac{7}{2}$ & 24
      & $\frac{1}{2\sqrt{14}}$ & $\frac{3}{7}$ \\

($\frac{9}{2},\frac{7}{2}$) & (4,3) & $+$2485.470
      & $\frac{1}{2}$ & 4 & $\frac{9}{2}$ & 40
      & $-\frac{1}{6\sqrt{2}}$ & $\frac{5}{9}$ \\

($\frac{7}{2},\frac{7}{2}$) & (4,3) & \ldots
      & $\frac{1}{2}$ & $\frac{7}{2}$ & 4 & 32
      & $\frac{1}{12\sqrt{14}}$ & $\frac{1}{63}$

\enddata
\tablenotetext{a}{$F = I_H + F_1$ and $F_1 = J + I_N$, where
  $I_N = 1$ and $I_H = \frac{1}{2}$.}
\tablenotetext{b}{Frequency offset in kHz relative to 24139416.34\,kHz
  \cite[][ Table~II (calculated)]{Kukolich1970}} 
\end{deluxetable}

\begin{deluxetable}{cccc}
\tablewidth{0pt}
\tablecaption{Hyperfine Intensities for NH$_3$(1,1)\label{tab:hyper11}}
\tablehead{
\colhead{$F^\prime \rightarrow F$} & 
\colhead{$F^\prime_1 \rightarrow F_1$} & 
\colhead{$J^\prime \rightarrow J$} &
\colhead{R$_i(F_1,J)$R$_i(F,F_1)$\tablenotemark{a,b}}
}
\startdata
($\frac{1}{2},\frac{1}{2}$) & (0,1) & (1,1) & $\frac{1}{27}$ \\

($\frac{1}{2},\frac{3}{2}$) & (0,1) & (1,1) & $\frac{2}{27}$ \\

($\frac{3}{2},\frac{1}{2}$) & (2,1) & (1,1) & $\frac{5}{108}$ \\

($\frac{5}{2},\frac{3}{2}$) & (2,1) & (1,1) & $\frac{1}{12}$ \\

($\frac{3}{2},\frac{3}{2}$) & (2,1) & (1,1) & $\frac{1}{108}$ \\

($\frac{1}{2},\frac{1}{2}$) & (1,1) & (1,1) & $\frac{2}{108}$ \\

($\frac{3}{2},\frac{1}{2}$) & (1,1) & (1,1) & $\frac{1}{108}$ \\

($\frac{5}{2},\frac{3}{2}$) & (2,2) & (1,1) & $\frac{1}{60}$ \\

($\frac{3}{2},\frac{3}{2}$) & (2,2) & (1,1) & $\frac{3}{20}$ \\

($\frac{1}{2},\frac{3}{2}$) & (1,1) & (1,1) & $\frac{1}{108}$ \\

($\frac{5}{2},\frac{5}{2}$) & (2,2) & (1,1) & $\frac{7}{30}$ \\

($\frac{3}{2},\frac{3}{2}$) & (1,1) & (1,1) & $\frac{5}{108}$ \\

($\frac{3}{2},\frac{5}{2}$) & (2,2) & (1,1) & $\frac{1}{60}$ \\

($\frac{1}{2},\frac{3}{2}$) & (1,2) & (1,1) & $\frac{5}{108}$ \\

($\frac{3}{2},\frac{3}{2}$) & (1,2) & (1,1) & $\frac{1}{108}$ \\

($\frac{3}{2},\frac{5}{2}$) & (1,2) & (1,1) & $\frac{1}{12}$ \\

($\frac{1}{2},\frac{1}{2}$) & (1,0) & (1,1) & $\frac{1}{27}$ \\

($\frac{3}{2},\frac{1}{2}$) & (1,0) & (1,1) & $\frac{2}{27}$

\enddata
\tablenotetext{a}{Compare with \cite{Kukolich1967} Table~IX after
  scaling $R_i$ by $(2I_H + 1)(2I_N + 1) = 6$ (\cite{Kukolich1967}
  lists unnormalized line strengths in their Table~IX).}
\tablenotetext{b}{Note that the sum of the relative intensities
  $\sum_i R_i = 1.0$.}
\end{deluxetable}

\begin{deluxetable}{cccc}
\tablewidth{0pt}
\tablecaption{Hyperfine Intensities for NH$_3$(2,2)\label{tab:hyper2k}}
\tablehead{
\colhead{$F^\prime \rightarrow F$} & 
\colhead{$F^\prime_1 \rightarrow F_1$} & 
\colhead{$J^\prime \rightarrow J$} &
\colhead{R$_i(F_1,J)$R$_i(F,F_1)$\tablenotemark{a,b}}
}
\startdata
($\frac{3}{2},\frac{3}{2}$) & (1,2) & (2,2) & $\frac{1}{300}$ \\

($\frac{3}{2},\frac{5}{2}$) & (1,2) & (2,2) & $\frac{3}{100}$ \\

($\frac{1}{2},\frac{3}{2}$) & (1,2) & (2,2) & $\frac{1}{60}$ \\

($\frac{7}{2},\frac{5}{2}$) & (3,2) & (2,2) & $\frac{4}{135}$ \\

($\frac{5}{2},\frac{3}{2}$) & (3,2) & (2,2) & $\frac{14}{675}$ \\

($\frac{5}{2},\frac{5}{2}$) & (3,2) & (2,2) & $\frac{1}{675}$ \\

($\frac{3}{2},\frac{1}{2}$) & (1,1) & (2,2) & $\frac{1}{60}$ \\

($\frac{5}{2},\frac{3}{2}$) & (2,2) & (2,2) & $\frac{1}{108}$ \\

($\frac{7}{2},\frac{5}{2}$) & (3,3) & (2,2) & $\frac{8}{945}$ \\

($\frac{5}{2},\frac{5}{2}$) & (2,2) & (2,2) & $\frac{7}{54}$ \\

($\frac{3}{2},\frac{3}{2}$) & (2,2) & (2,2) & $\frac{1}{12}$ \\

($\frac{7}{2},\frac{7}{2}$) & (3,3) & (2,2) & $\frac{8}{35}$ \\

($\frac{5}{2},\frac{5}{2}$) & (3,3) & (2,2) & $\frac{32}{189}$ \\

($\frac{3}{2},\frac{3}{2}$) & (1,1) & (2,2) & $\frac{1}{12}$ \\

($\frac{1}{2},\frac{1}{2}$) & (1,1) & (2,2) & $\frac{1}{30}$ \\

($\frac{3}{2},\frac{5}{2}$) & (2,2) & (2,2) & $\frac{1}{108}$ \\

($\frac{5}{2},\frac{7}{2}$) & (3,3) & (2,2) & $\frac{8}{945}$ \\

($\frac{1}{2},\frac{3}{2}$) & (1,1) & (2,2) & $\frac{1}{60}$ \\

($\frac{5}{2},\frac{5}{2}$) & (2,3) & (2,2) & $\frac{1}{675}$ \\

($\frac{3}{2},\frac{5}{2}$) & (2,3) & (2,2) & $\frac{14}{675}$ \\

($\frac{5}{2},\frac{7}{2}$) & (2,3) & (2,2) & $\frac{4}{135}$ \\

($\frac{3}{2},\frac{1}{2}$) & (2,1) & (2,2) & $\frac{1}{60}$ \\

($\frac{5}{2},\frac{3}{2}$) & (2,1) & (2,2) & $\frac{3}{100}$ \\

($\frac{3}{2},\frac{3}{2}$) & (2,1) & (2,2) & $\frac{1}{300}$

\enddata
\tablenotetext{a}{Compare with \cite{Kukolich1967} Table~IX after
  scaling $R_i$ by $(2I_H + 1)(2I_N + 1) = 6$ (\cite{Kukolich1967}
  lists unnormalized line strengths in their Table~IX).}
\tablenotetext{b}{Note that the sum of the relative intensities
  $\sum_i R_i = 1.0$.}
\end{deluxetable}

\end{document}